\def\preprintmode{1}
\providecommand{\preprintmode}{1}
\ifnum\preprintmode=1
  \documentclass[acmsmall,screen,nonacm]{acmart}
\else
  \documentclass[acmsmall,screen,review]{acmart}
\fi
\AtBeginDocument{%
  }

\usepackage{microtype}
\usepackage{algorithmic}
\usepackage{graphicx}
\usepackage{textcomp}
\usepackage[scaled]{beramono}
\usepackage[utf8]{inputenc}
\usepackage[T1]{fontenc}
\usepackage[dvipsnames]{xcolor}
\usepackage{listings}
\usepackage{alloy}
\usepackage{smv}
\usepackage{amsmath}
\usepackage{mathtools}
\usepackage[noend]{algorithm2e}
\usepackage{booktabs}
\usepackage{multirow}
\usepackage{utfsym}
\usepackage{stmaryrd}
\usepackage{subcaption}

\usepackage{amsmath, amssymb}
\usepackage{enumitem}
\usepackage{tikz}
\usetikzlibrary{positioning,fit,arrows.meta,calc,backgrounds}
\usepackage{mathrsfs}

\usepackage{fvextra}

\newcommand{\microsize}{\fontsize{2pt}{2.2pt}\selectfont}

\DefineVerbatimEnvironment{MicroVerbatim}{Verbatim}{
  breaklines,
  breakanywhere,
  breaksymbolleft={},
  breaksymbolright={},
  fontsize=\microsize,
  fontfamily=tt,
  frame=single,
  framesep=2pt,
  xleftmargin=0pt,
  xrightmargin=0pt
}

\newcommand{\hpard}[1]{\tool{HyperPardinus}}
\newcommand{\hsmv}[1]{\tool{HyperSMV}}
\newcommand{\tool}[1]{\textsf{#1}}
\newcommand{\proc}[1]{\ensuremath{\mathtt{#1}}}

\DeclarePairedDelimiter{\verts}{\lvert}{\rvert}
\let\sem\bbrackets
\newcommand{\mydots}{\makebox[1em][c]{.\hfil.\hfil.}}

\providecommand{\sem}[1]{\llbracket #1 \rrbracket}

\setcopyright{acmlicensed}
\copyrightyear{2018}
\acmYear{2018}
\acmDOI{XXXXXXX.XXXXXXX}
\acmISBN{978-1-4503-XXXX-X/2018/06}

\ifnum\preprintmode=1
  \renewcommand\footnotetextcopyrightpermission[1]{}
\fi

\newsavebox{\oksym}
\newsavebox{\noksym}
\newsavebox{\inclsym}
\newsavebox{\totsym}
\newsavebox{\oomsym}

\begin{document}

\title{Model checking of hyperproperties for high-level relational models}

\author{Nuno Macedo}
\authornote{Both authors contributed equally to this research.}
\email{nmacedo@di.uminho.pt}
\orcid{0000-0002-4817-948X}
\affiliation{%
  \institution{Universidade do Minho \& INESC TEC}
  \city{Braga}
  \country{Portugal}
}

\author{Hugo Pacheco}
\authornotemark[1]
\email{hpacheco@di.uminho.pt}
\orcid{0000-0003-0720-7744}
\affiliation{%
  \institution{Universidade do Minho \& INESC TEC}
  \city{Braga}
  \country{Portugal}
}

\begin{abstract}
Many properties related to security or concurrency must be encoded as so-called hyperproperties, temporal properties that allow reasoning about multiple traces of a system. However, despite recent advances on model checking hyperproperties, there is still a lack of higher-level specification languages that can effectively support software engineering practitioners in verifying properties of this class at early stages of system design.

\tool{Alloy} is a lightweight formal method with a high-level specification language that is supported by automated analysis procedures, making it particularly well-suited for the verification of design models at early development stages. It does not natively support, however, the verification of hyperproperties.

This work proposes \tool{HyperPardinus}, a new model finding procedure that extends \tool{Pardinus} -- the temporal logic backend of the \tool{Alloy} language -- to automatically verify hyperproperties over relational models by relying on existing low-level model checkers for hyperproperties. It then conservatively extends \tool{Alloy} to support the specification and automatic verification of hyperproperties over design models, as well as the visualization of (counter-)examples at a higher-level of abstraction. Evaluation shows that our approach enables modeling and finding (counter-)examples for complex hyperproperties with alternating quantifiers, making it feasible to address relevant scenarios from the state of the art.
\end{abstract}




\maketitle

\section{Introduction}
Many relevant system properties related to information flow, concurrency or robustness cannot be expressed by looking at individual execution traces of a system, and require reasoning about the relationship between multiple executions. This so-called class of \emph{hyperproperties}~\cite{hyperproperties} generalizes traditional trace properties to sets of traces and has seen increased applications across different research areas.

There are many widely used tools for checking specific hyperproperties for particular domains~\cite{surveyFMSecurity}, but work on checking properties specified in general hyperlogics is more recent.
For that purpose, classical temporal logics have been extended with operators that allow universal and existential quantification over traces, such as the extension of \textit{Linear-time Temporal Logic} (LTL) known as HyperLTL. A recent overview on the landscape of hyperlogics, their expressive power and algorithms is given in~\cite{hyperpropertiesSOK}. The decidability of model checking for HyperLTL has led to the development of several model checking tools for hyperlogics~\cite{MCHyper, HyperQube,AutoHyper,AutoHyperQ}. Other recent research focused on the deductive verification of imperative programs with respect to general hyperproperties, often requiring user guidance~\cite{hypra,Beutner24,ItzhakySV24,CorrensonNFW24}.

Research on model checking HyperLTL properties -- the focus of this work -- remains vibrant with algorithmic advances and novel model checking procedures, but has unfortunately not yet reached the point where general high-level modeling tools can support the verification of properties written in hyperlogics.
All existing model checkers for hyperlogics~\cite{MCHyper,AutoHyper,HyperQube} work with compact, machine-readable formats common in the model checking community for describing automata, circuits or state machines, such as fragments of the SMV language~\cite{SMV}. They support HyperLTL formulas written over such representation, which requires users to reason at a very low level of abstraction. Moreover, automatic verification of hyperproperties involving alternations is particularly challenging~\cite{AutoHyper,BMCloop}, and many approaches, such as \tool{MCHyper}~\cite{MCHyper} or the approach proposed by Lamport and Schneider~\cite{LamportS21} based on \tool{TLA}${}^+$~\cite{Lamport94}, require user input to generate witnesses for existential quantifications. As a consequence, there is a considerable gap between the conceptual model of the system a user expects to verify, and the artefacts that have to be developed for these tools, hindering their adoption. In order to reach the wider software engineering community, there is a need for modeling and specification languages that act at a higher-level of abstraction, as well as verification tools that are able to handle such representations, either through novel procedures or translations into lower-level ones.

\tool{Alloy}~\cite{Jackson19} is a high-level formal specification language that has become popular due to its simplicity and flexibility, whose \tool{Analyzer} quickly provides intuitive feedback. As of version 6~\cite{MacedoBCCK16}\footnote{\url{github.com/AlloyTools/org.alloytools.alloy/releases/tag/v6.0.0}}, \tool{Alloy} supports models with mutable elements, whose values can change over time, and the specification and verification of LTL properties. The \tool{Alloy Analyzer} provides a GUI through which users write specifications, run analysis commands, and interpret feedback graphically, but actual analysis is decoupled from the frontend and performed by a backend \emph{relational model finder}, \tool{Pardinus}~\cite{pardinus22}. Such model finders provide a higher-level interface for the analysis of formal specifications,  themselves typically relying on lower-level solvers. In the case of \tool{Pardinus}, it supports the same relational temporal logic as \tool{Alloy 6}, stripped away of certain syntactic constructs, and provides two alternative analysis procedures: one reducing to SAT solving through the relational model finder \tool{Kodkod}~\cite{KodKod}, and another reducing to SMV and relying on off-the-shelf symbolic model checkers for LTL. Unfortunately, the whole \tool{Alloy} toolchain focuses on the analysis of LTL properties for individual traces, rendering the analysis of arbitrary hyperproperties impossible.

\begin{figure}
    \centering
    \includegraphics[width=0.75\linewidth]{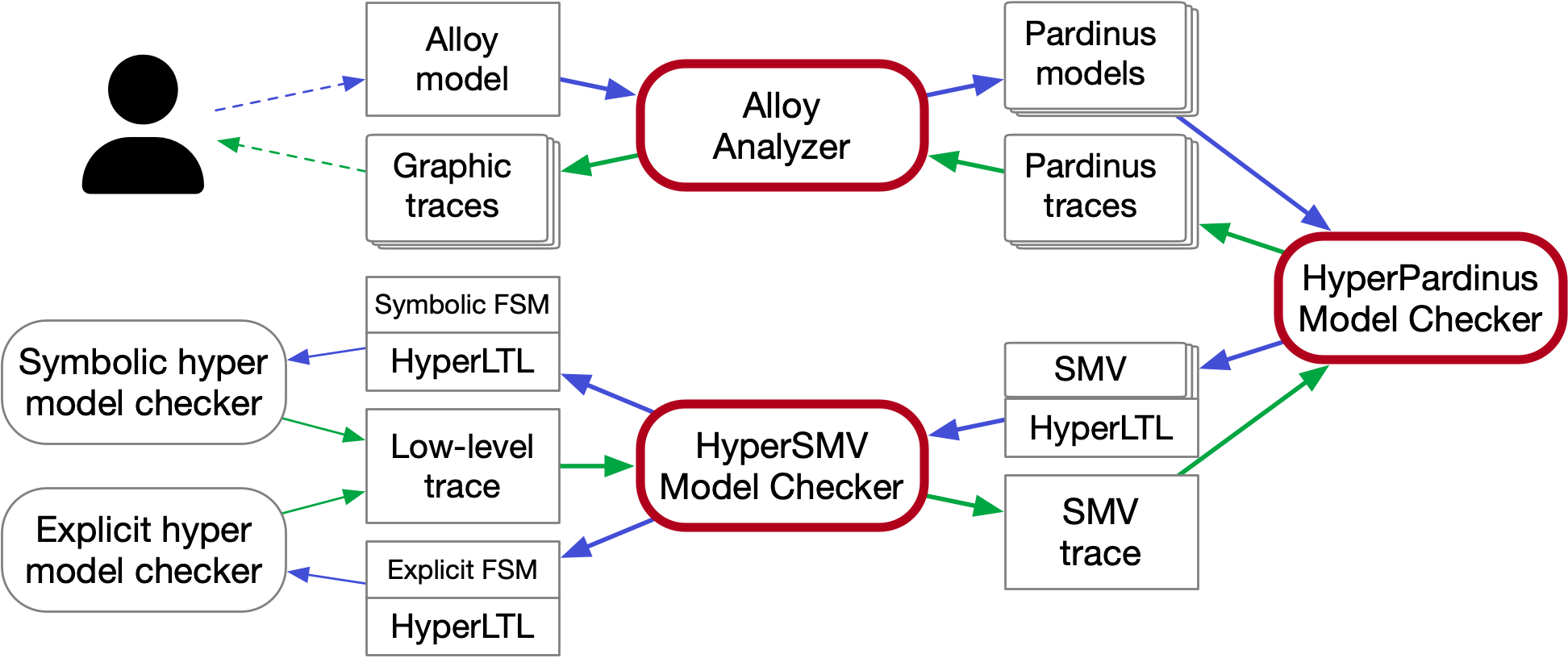}
    \caption{Overview of the proposed toolchain. Rectangles are interchangeable formats, rounded rectangles software components, \textcolor{BrickRed}{red} identifies contributions.}
    \label{fig:overview}
\end{figure}

In this paper, we extend \tool{Pardinus} and the surrounding ecosystem to allow for the automatic verification of hyperproperties and the inspection of (counter-)examples for models written in a high-level specification language\footnote{The tools and benchmarks are publicly available at \url{github.com/haslab/HyperPardinus-benchmarks}.}, whose overview is shown in Fig.~\ref{fig:overview}. Specifically, our main contributions are:

\begin{itemize}
\item \hpard{}, a high-level model finding tool for \emph{hyper relational temporal logic}, a hyper generalization of the underlying logic of \tool{Alloy 6} supporting a superset of HyperLTL properties with flexible occurrences of trace quantifiers. As far as we know, this is the first model checking tool for a first-order hyperlogic.

\item \hsmv{}, a SMV model checking tool that integrates with both explicit-state and symbolic state-of-the-art model checkers for HyperLTL, significantly improving the expressiveness of supported SMV models and the scalability of the verification process.

\item A minimal extension to \tool{Alloy 6} and its \tool{Analyzer}, to support analysis with \hpard{} and visualization of multiple trace (counter-)examples, allowing the verification of hyperproperties for high-level models.

\item An evaluation of our toolchain by modeling and verifying case studies that showcase hyperproperties from diverse domains, including low-level models from the literature and high-level models specifically crafted to evaluate its ability to handle more complex models.

\end{itemize}

The rest of the paper is organized as follows. 
Section~\ref{sec:motivation} motivates the approach with an example. Section~\ref{sec:relwork} discusses related work on hyperproperties. Section~\ref{sec:hpard} formalizes the notion of hyper model finding problem, and Section~\ref{sec:mfind} presents the \hpard{} model finder for such problems. Section~\ref{sec:smvtrans} presents the lower-level model checker \hsmv{}. Section~\ref{sec:halloy} presents the extended \tool{Alloy Analyzer} that supports analysis through \hpard{}. Section~\ref{sec:eval} presents the evaluation of \hpard{} and \hsmv{} against the state of the art. Section~\ref{sec:conclusions} concludes the paper and points direction for future work.

\section{Motivating example}
\label{sec:motivation}

\begin{figure*}[!t]
\centering
\begin{alloyfig}
sig Reviewer, Article {}
one sig Agent in Reviewer {}      // a designated reviewer whose view is used in the NI/GNI properties
enum Decision { Reject, Major, Accept }

§trace sig CMS§ { // a CMS includes static assignment of reviewers to papers and dynamic paper reviews
  assigns : Reviewer some -> some Article,
  var reviews : Article -> Reviewer -> lone Decision,
  var decisions : Article -> lone Decision }

pred cms[s:CMS] {                                          // determine valid traces of a CMS
  some Article - Agent.(s.assigns)                         // there are articles not assigned to agent
  no s.reviews and no s.decisions                          // initial state
  always some r:Reviewer, a:Article, d:Decision |          // possible events at each state
    review[s,r,a,d] or decide[s,a,d] or stutter[s] 
  eventually s.decisions.Decision = Article }              // force eventual decisions

pred review[s:CMS, r:Reviewer, a:Article, d:Decision] {   // reviewer submits a review
  a in r.(s.assigns)                                      // guard: a is assigned to r
  no r.(a.(s.reviews))                                    // guard: r has not submitted a review for a
  s.reviews' = s.reviews + a->r->d                        // effect: add the review for this article
  s.decisions' = s.decisions }                            // frame condition: decisions unchanged

pred decide[s:CMS, a:Article, d:Decision] {             // the editor makes a decision
  a not in s.decisions.Decision                         // guard: editor has not decided on a
  a.(s.reviews).Decision = s.assigns.a                  // guard: all a reviews submitted
  d in criteria[s,a]                                    // guard: d is according to specified criteria
  s.decisions' = s.decisions + a->d                     // effect: add the decision for a
  s.reviews' = s.reviews }                              // frame condition: reviews unchanged

pred stutter[s:CMS] {                               // the system stutters (does not change the state)
  s.reviews' = s.reviews and s.decisions' = s.decisions }

fun low[s1, s2:CMS] : set Article {                  // articles are low security if assigned to agent
  Agent.(s1.assigns + s2.assigns) } 
  
fun high[s1, s2:CMS] : set Article {            // articles are high security if not assigned to agent
  Article - low[s1,s2] } 
  
pred same_assigns[s1, s2:CMS, arts:set Article] { // whether arts have the same assignments in two CMS
  s1.assigns :> arts = s2.assigns :> arts }  
\end{alloyfig}
\caption{The \texttt{CMS} system in extended \tool{Alloy}}
\label{fig:easychair}
\end{figure*}

Conference management systems (CMSs) such as EasyChair or HotCRP are expected to enforce certain security policies, and there has been quite some work on using hyperproperties to model and verify their security. Famously, \tool{CoCon}~\cite{CoCon} has found bugs in production systems using a theorem prover to check confidentiality properties, expressed in a variant of \textit{nondeducibility}. 
This has made CMSs a target of work on hyperproperties, such as in the verification of CMS workflows written in a multi-agent framework~\cite{ccs17EasyChair,csf18CMS}, where special emphasis is placed on workflows with loops and different restrictions on agent behavior.

Following this inspiration, we start by defining a high-level \tool{Alloy} model of a CMS that will illustrate our approach. Later, we will use this same example to highlight the limitations of the state of the art.
This abstract version considers assignment of articles to reviewers, and we wish to analyze whether the process of submitting reviews and reaching decisions on submissions ensures fine-grained confidentiality policies. 

\subsection{Modeling a CMS}

The \a{CMS} model presented in Fig.~\ref{fig:easychair} is a possible specification of a CMS in our extended \tool{Alloy}, whose syntax is mostly conformant with regular \tool{Alloy}. In an \tool{Alloy} model, structure is introduced by the declaration of \emph{signatures} (keyword \a{sig})---defining the entities of the model---and \emph{fields} declared within signatures---representing the relationships between entities. In ll.~1--3, signatures are declared for the main entities of our model: reviewers, articles and decisions. During solving, signatures are assigned abstract, uninterpreted, \textit{atoms} according to a scope defined by the user. For this model, the user will have to set a scope for reviewers and articles, since decisions are declared as an exact enumeration signature. Additionally, we introduce a distinguished reviewer, \a{Agent} (a sub-signature with multiplicity \a{one}), to define security properties in his perspective. Signature \a{CMS} then encapsulates the content of a CMS: it \a{assigns} a set of articles to reviewers, registers \a{reviews} which are reviewer-proposed decisions for articles, and notifies final \a{decisions} made by the editor on articles. In \tool{Alloy}, fields encode relations with arbitrary arity and can have multiplicity constraints imposed. Declaring \a{assigns} (l.~6) within \a{CMS} with type \a{Reviewer some -> some Article} results in a ternary relation between \a{CMS}s, \a{Reviewer}s and \a{Article}s, with the additional constraints that there are no articles without reviewers nor reviewers without articles assigned (literally, for all \a{Reviewer}s there are \a{some} \a{Article}s in \a{assigns}, and vice-versa); in contrast, \a{reviews} (l.~7) is a quaternary relation between \a{CMS}s, \a{Article}s, \a{Reviewer}s and \a{Decision}s, where a reviewer submits at most one decision per article (\a{lone}). Our traces start at a point where articles have already been assigned to reviewers, so \a{assigns} is declared as a static relation (the default); in contrast, \a{reviews} and \a{decisions} will change as the systems progresses, and thus are declared as mutable (\a{var}), one of the temporal features introduced in \tool{Alloy 6}. This signature is explicitely marked as a {\color{BrickRed}\a{trace}}, making it prone to be quantified in a hyperproperty, as we shall see in the next section.

Once the structure of the model is defined, the next step is to specify valid \a{CMS} traces. In our model this is encoded by the auxiliary \a{pred}icate \a{cms} (ll.~10--15). \tool{Alloy} is based on relational logic, and includes set operators such as union (\a{+}) and difference (\a{-}), and relational join (\a{.}) that allows composition of relations with arbitrary arity. For example, expression \a{Agent.assigns} is a unary relation (i.e., a set) containing all articles related with \a{Agent} through \a{assigns}, while \a{Article - Agent.assigns} represents all articles not assigned to \a{Agent}. Atomic formulas are created through inclusion tests (\a{in}) or cardinality tests, which are then combined through Boolean operators, quantifications, and temporal operators. The \a{cms} constraint in l.~11 thus states that there must exist \a{some} articles that are not assigned to \a{Agent}, as the security properties will become vacuous when the \a{Agent}'s view of the system is empty. The one in l.~12 restricts the initial state to be empty: properties over mutable elements outside a temporal operator apply to the first state. The one in ll.~13--14 states how the system is allowed to evolve: at any state (\a{always}), either a review is submitted, the editor makes a decision, or the system stutters (does not change its state). These events are specified as auxiliary predicates that specify how succeeding states can be related, through formulas that compare the current state with the next using primed expressions (\a{'}). Finally, we force a decision to be \a{eventually} reached on every article (l.~15).

At this point, this could be a regular \tool{Alloy} model, and we could execute commands to animate and validate this model, such as the one below that asks for a valid \a{CMS} with up to 3 elements in each signature. In \tool{Alloy} all execution traces are infinite, represented as a finite prefix of the trace with a looping segment, and by default are obtained through bounded model checking (BMC).
\begin{alloy}
run example { some s:CMS | cms[s] } for 3 but 1 CMS
\end{alloy}

To postulate security properties over our model, we also need to specify which information is considered confidential. The \a{Agent} in our model will serve precisely this role.
We assume that articles assigned to the \a{Agent} are low-security (i.e., public), and dually that all other articles are high-security (i.e., secret). This security policy is encoded in \a{fun}ctions \a{low} and \a{high}, which declaratively select the joint public/secret view of the \a{Agent} in two \a{CMS}s.
The policy extends to associated article information (assignments, reviews and decisions), as encoded in the remaining \a{pred}icates. For example, \a{same_assigns} (ll.~39--40) defines when two \a{CMS}s have the same assignments for a set of articles (operator \a{:>} restricts the range of a relation to a certain set).

\subsection{Specifying \& verifying hyperproperties}

The model from Fig.~\ref{fig:easychair} is written in the temporal dialect of \tool{Alloy}. Still, it is impossible for \tool{Alloy} to test hyperproperties relating multiple traces, as it can only test whether a temporal property holds in each individual trace.
To allow higher-order quantifications, we introduced a new annotation {\color{BrickRed}\a{trace}} for signatures, highlighted in {\color{BrickRed}red}:
quantifications over such signatures are lifted into quantifications over traces. {\color{BrickRed}\a{CMS}} is marked as one such signature in Fig.~\ref{fig:easychair} (l.~5).

\begin{figure*}[!t]
\centering
\begin{alloyfig}
assert NI { 
  §all s1,s2:CMS§ | cms[s1] and cms[s2] and aligned[s1,s2,low[s1,s2]] implies 
    (same_assigns[s1,s2,low[s1,s2]] and always same_reviews[s1,s2,low[s1,s2]]) implies 
      always same_decisions[s1,s2,low[s1,s2]]
  }

assert GNI { 
  §all s1,s2:CMS§ | cms[s1] and cms[s2] and aligned[s1,s2,low[s1,s2]] implies §some s3:CMS§ { 
    cms[s3]
    same_assigns[s1,s3,low[s1,s3]] and always same_reviews[s1,s3,low[s1,s3]]
    always same_decisions[s1,s3,low[s1,s3]]
    same_assigns[s2,s3,high[s2,s3]] and always same_reviews[s2,s3,high[s2,s3]]
  } }
\end{alloyfig}
\caption{Hyperproperties in extended \tool{Alloy} over the \texttt{CMS} system}
\label{fig:hyperprops}
\end{figure*}

Consider, for instance, a classical security hyperproperty, known as \textit{noninterference} (NI), postulated as a relation on traces: any two traces that have the same low inputs must also have the same low outputs. It declaratively encodes the fact that the system appears deterministic to a low user (our \a{Agent}), in the sense that low outputs can only depend on low inputs. This is encoded for \a{CMS} as shown in ll.~1--4 of Fig.~\ref{fig:hyperprops}. 
We consider article assignments and reviews to be inputs of the system, and the editor's decision to be the outputs\footnote{Note that this abstract system naturally supports reactive behavior, e.g., a reviewer can submit his review after viewing his assignments.}.

Hyperproperties that use temporal operators to compare events occurring in multiple traces require such traces to be aligned at least in the occurrence of low events; if the system is allowed to stutter, this still allows an arbitrary number of high events to occur between low events (see~\cite{LamportS21} for a deeper discussion). Thus, \a{NI} literally asserts that for any two \a{CMS} traces that behave according to \a{cms} and are aligned on low events, if the low inputs are the same at every point of the traces, so must be the low outputs.

\begin{figure*}
    \centering
    \includegraphics[width=0.48\textwidth]{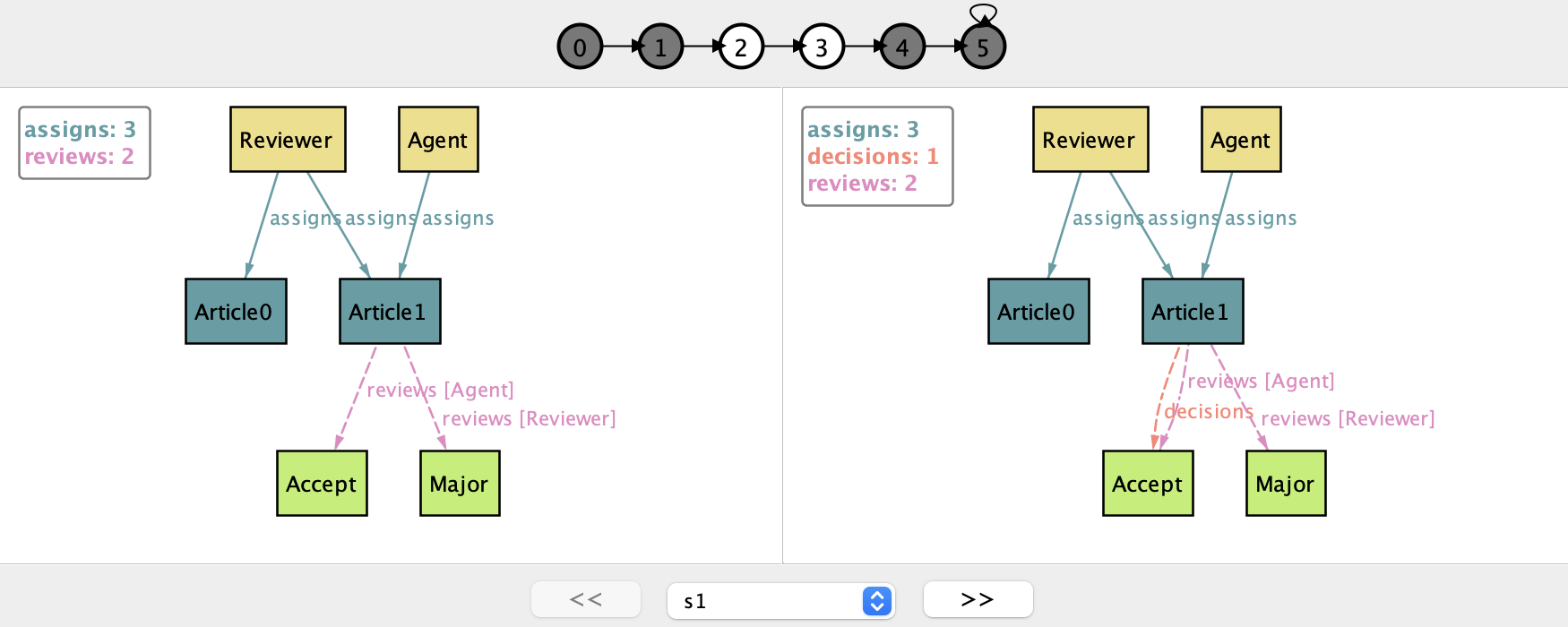} $\,$
    \includegraphics[width=0.48\textwidth]{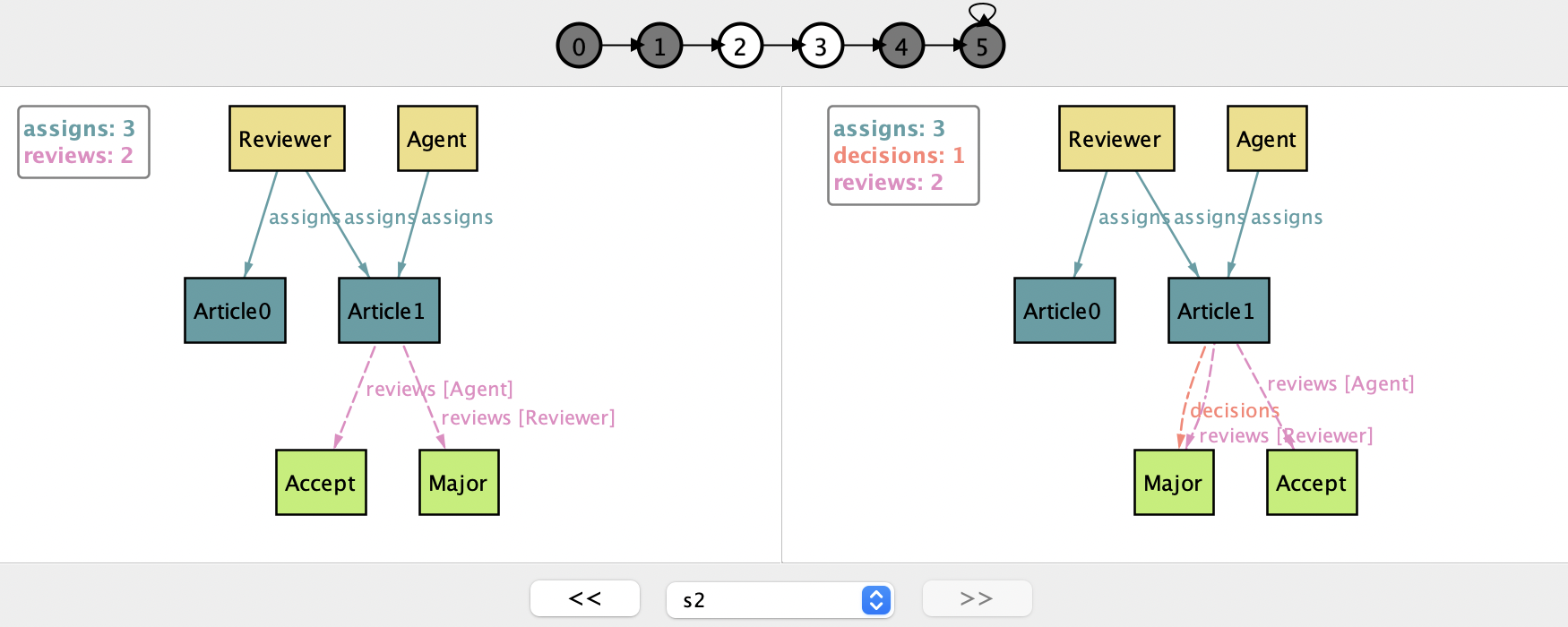}
    \caption{Possible counter-example for the \a{NI} property of \texttt{CMS}}
    \label{fig:cex}
\end{figure*}

To check the \a{NI} assertion, we can write a command with the desirable scopes, such as the following for up to 3 reviewers and articles. 
\begin{alloy}
check NI for 3 Reviewer, 3 Article
\end{alloy}
Whether this property holds for \a{CMS} depends on the behavior of the editor when making a decision. To explore the consequences of alternative designs, namely which information the editor has available to make a decision, in \a{CMS} (l.~26) this behavior was left as an auxiliary \a{fun}ction \a{criteria[s:CMS, a:Article]} that determines the possible decision an editor may take for an article. For the \a{NI} property to hold, an idealized editor would have to make the same decision with the same information, hence looking deterministic to the \a{Agent}, who knows all reviews and decisions of articles assigned to him. Let us model an optimistic editor that always chooses the best decision from those submitted by reviewers to the given article \a{a}, which could be expressed as follows.
\begin{alloy}
fun criteria[s:CMS, a:Article] : set Decision {
  max[Reviewer.(a.(s.reviews))] }
\end{alloy}
Here, expression \a{a.(s.reviews)} retrieves decisions submitted by reviewers for article \a{a} in CMS \a{s}. \a{Reviewer.(a.(s.reviews))} then retrieves all submitted decisions regardless of the reviewer; elements of an enumeration signature are ordered in the order they are declared, so we can then select the \a{max}imum decision.

Our tool will report the \a{check} above to be valid, within the specified scope. However, we stated that any non-determinism on the low output would invalidate \a{NI}. To observe this, we could allow an editor to arbitrarily select a decision from those submitted, changing criteria to the following
\begin{alloy}
fun criteria[s:CMS, a:Article] : set Decision {
  Reviewer.(a.(s.reviews)) }
\end{alloy}
Our extension of \tool{Alloy} will now quickly report \a{NI} to be invalid and provide a counter-example, which for \a{NI} is composed by traces \a{s1} and \a{s2}. Figure~\ref{fig:cex} shows how such a counter-example can be inspected in the visualizer of the \tool{Alloy Analyzer}. Observing this counter-example, the \a{Agent} knows the two decisions for \a{Article1} (his own \a{Accept} and \a{Reviewer}'s \a{Major}); nonetheless, the fact that the editor selected differently among the two in the two traces \a{s1} and \a{s2} is sufficient to invalidate \a{NI}.

A more complex confidentiality hyperproperty is \textit{generalized noninterference} (GNI), which relaxes NI to allow non-deterministic behavior, by postulating that any observable low behavior must be compatible with any sequence of high inputs. It can be formalized by stating that, for any two traces of the system, there must exist a third trace that combines the low information from the first and the high information from the second. This could be encoded for \a{CMS} as shown in ll.~6--10 of Fig.~\ref{fig:hyperprops}. Our extension of \tool{Alloy} will report \a{GNI} as valid for both \a{criteria} functions defined above.

Intuitively, this shows that non-determinism is the sole explanation for differences in the output. To break \a{GNI}, the editor would have to consult secret information -- for instance, to enforce a quota of accepted articles -- to make the decision on an article. To observe this, we could let the editor select the decision not only from article \a{a}, but from all articles in the system. If \a{criteria} is changed to the following, our tool will then report \a{GNI} to be invalid.

\begin{alloy}
fun criteria[s:CMS, a:Article] : set Decision {
  Reviewer.(Article.(s.reviews)) }
\end{alloy}

As this example illustrates, our lightweight extension to the \tool{Alloy} language enables the specification of properties in a first-order hyperlogic over high-level relational models. This significantly increases expressiveness compared to the languages supported by current state-of-the-art tools, as shall become clear in Section~\ref{sub:qualitative_evaluation}. At the same time, the added expressiveness can impose considerable analysis costs. The remainder of the paper presents a pipeline for analyzing these abstract representations using state-of-the-art lower-level solvers.
\section{Related work}
\label{sec:relwork}

\subsection{Model checking}

HyperLTL~\cite{hyperltl}, as the most studied logic for expressing hyperproperties, has seen the development of various general verification methods in recent years (see \cite{hyperpropertiesSOK} for an  extensive review). This vibrant landscape includes various recent model checking tools, which are of particular interest to this paper.

The seminal \tool{MCHyper}~\cite{MCHyper} supports the verification of alternation-free HyperLTL via model composition, 
but requires explicit witnesses to resolve existential quantifications. Strategy-based verification, where players of a game represent distinct traces of a joint property, can be used to support quantifier alternations~\cite{HyPro}.
\tool{HyperQB}~\cite{HyperQube} supports BMC for HyperLTL by reducing to QBF (\textit{Quantified Boolean Formula}) solving, but is incomplete as it is limited to finite, non-looping, prefixes of infinite traces. The same authors~\cite{BMCloop} propose a method with loops for one quantifier alternation, which is not integrated into \tool{HyperQB}. Models are provided in SMV but counter-examples provided at the QBF level.
\tool{AutoHyper}~\cite{AutoHyper,AutoHyperQ} 
is an explicit-state automata-based model checker which supports complete verification of \emph{Hyper Quantified Propositional Temporal Logic} (HyperQPTL), in which trace and propositional quantifiers can be freely interleaved. Models are provided in a non-declarative subset of SMV or a low-level representation of an explicit-state system; counter-examples are generated at the level of models.

In terms of expressiveness, these tools~\cite{MCHyper,HyperQube,AutoHyper} support HyperLTL or HyperQPTL restricted to prenex form (trace quantifiers appear in outermost positions of the formula).
Our approach flexibly blends trace quantifications with \textit{First-Order Logic} (FOL), and is a natural hyper generalization of \textit{First-Order LTL} (FOLTL).

The visualization of hyperproperty counter-examples is a challenge on itself. \tool{HYPERVIS}~\cite{visualHyper} is one of the first to address this, providing explanations for hyperproperty violations as minimal causative sub-formulas~\cite{explainhyper}.
Nonetheless, note that all of the mentioned model checkers work with low-level models, while ours support specification at a much higher level of abstraction. This becomes even more evident when counter-examples of tools like \tool{HyperQB} are the raw transcript of the underlying QBF solver. In contrast, \hpard{} is fully integrated into the \tool{Analyzer}, which provides high-level counter-examples via its graphical user interface.

\subsection{Deductive verification}

Beyond model checking, significant effort has been dedicated to the deductive verification of imperative programs with respect to hyperproperties. These approaches generally navigate a trade-off between the expressiveness of the property language and the degree of automation.

On the highly expressive end, \tool{Hypra}~\cite{hypra} introduces \emph{Hyper Hoare Logic}, a logic capable of verifying general hyperproperties. However, this expressiveness comes at the cost of automation, often requiring manual proof guidance or detailed invariants from the user. To improve automation, \tool{ForEx}~\cite{Beutner24} targets a specific subset of properties known as \emph{hyperliveness}, by restricting the specification language to a less liberal \textit{$\forall$-$\exists$ Hoare Logic}.

Another class of approaches reframes the problem of hyperlogic verification to game-based interpretations.
To capture asynchronous behaviors, \tool{HyPA}~\cite{BeutnerF22} enables the verification of \emph{Observation-based HyperLTL} (OHyperLTL). This approach aligns traces based on observable events rather than strict ordering, allowing more natural verification of properties where traces proceed at different speeds.
\tool{HyHorn}~\cite{ItzhakySV24} focuses on a subset of OHyperLTL with one quantifier alternation over LTL safety properties. By reducing verification to the satisfiability of \emph{Constrained Horn Clauses} (CHCs), and using off-the-shelf CHC solvers, it has shown to outperform \tool{HyPA} in many benchmarks.
In contrast, \tool{Hy$\exists$n$\forall$}~\cite{CorrensonNFW24} considers properties in a subset of OHyperLTL with arbitrary quantification over LTL safety properties, and employs symbolic execution to automate the search for violations (or ``hyperbugs'') in asynchronous programs.

These tools, much like traditional deductive verification, focus primarily in reasoning at the level of the operational semantics of imperative programs, often requiring user-provided invariants. We take a fundamentally different approach, and focus in the specification and automated verification of hyperproperties for high-level, declarative models expressed directly in first-order logic.

\section{Hyper relational model finding}
\label{sec:hpard}

The analysis procedures of the \tool{Alloy Analyzer} are powered by a relational temporal model finder, \tool{Pardinus}~\cite{pardinus22} (itself an extension of the static model finder \tool{Kodkod}~\cite{KodKod}). This section formalizes hyper relational model finding problems as an extension of the non-hyper counterpart, allowing for the animation and verification of hyperproperties over relational models.

A (non-hyper) model finding \emph{problem} $P = \langle \mathcal{A}, \mathcal{D}, \phi \rangle$ is composed of a universe $\mathcal{A}$ of uninterpreted atoms; the declaration $\mathcal{D}$ of a set of free relations $\mathcal{R}$ with upper- and lower-bounds restricting their possible values; and a relational temporal logic constraint $\phi$ over $\mathcal{R}$ variables that is expected to hold for the generated instances\footnote{Model finders search for instances of a problem, so checking the validity of a property $\psi$ amounts to creating a problem that searches for (counter-)examples that break it. This is done by negating $\psi$ in the problem's constraint $\phi$.}. A \emph{trace} $I : \mathbb{N} \rightarrow \mathcal{R} \mapsto \mathbb{P}(\mathcal{A}^*)$ is an infinite sequence of states assigning to each relation a tuple set. A trace $I$ is an instance of a problem $P$, $I \models P$, iff the relation bounds and the constraint hold, whose semantics are detailed next. A hyper model finding problem builds on this formalism for non-hyper problems. It considers a set of non-hyper problems $P_i$ over which traces will be quantified, and an additional problem $H$ with a hyper relational temporal logic constraint $\Phi$ whose trace quantifications range over $P_i$ problems; thus, $\Phi$ may refer to both the relations declared in $H$ and those from models $P_i$.

\begin{definition}[Hyper model finding problem]
For a domain of discourse $\mathcal{A}$, a \emph{hyper model finding problem} is composed of
\begin{itemize}
\item a set of problems $P_i = \langle \mathcal{A}, \mathcal{D}_i, \phi_i \rangle$, where $\mathcal{D}_i$ declares a set of relations $\mathcal{R}_i$ bounded by tuples from $\mathcal{A}$ and $\phi_i$ is a (non-hyper) relational temporal logic constraint over $\mathcal{R}_i$;
\item a problem $H = \langle \mathcal{A}, \mathcal{D}, \Phi \rangle$, where $\mathcal{D}$ declares a set of relations $\mathcal{R}$ bounded by tuples from $\mathcal{A}$ and $\Phi$ is a hyper relational temporal logic constraint over $\mathcal{R} \cup \bigcup \mathcal{R}_i \cup \bigcup \{P_i\}$.
\end{itemize}
\end{definition}

The verification of the NI hyperproperty for the CMS model from Section~\ref{sec:motivation}, for instance, would result in two non-hyper models, each a copy of \a{CMS} trace signature and the restriction of their behaviour by predicate \a{cms}, and a hyper model finding problem with the hyperproperty defined in assertion \a{NI}. The translation of the higher-level \tool{Alloy} model into this is addressed in Section~\ref{sec:halloy}.

\subsection{Relation bounds} 

\paragraph{Syntax}
The declaration of each free relation $r \in \mathcal{R}$ bounds the possible values that $r$ may take, namely through a lower-bound denoting tuples that \emph{must} belong to $r$, and an upper-bound, denoting tuples that \emph{may} belong to $r$. In \tool{Pardinus}, bounds are either tuple sets or an expression referring to other relations~\cite{pardinus22}. We have extended the upper-bounds to support \emph{multiplicity constraints} over arbitrary $n$-ary relations. These are available at the \tool{Alloy} level (as can be seen in Fig.~\ref{fig:easychair}, ll.~6--8), but are expanded before reaching \tool{Pardinus}; by supporting them natively, we are able to generate finer state machines for the backends\footnote{\tool{Pardinus} already had preliminary, undocumented, support for this, but we have extended it to richer multiplicities of arbitrary arity.}. 

\begin{definition}[Relation bounds]
The declaration of an $n$-ary relation $r \in \mathcal{R}$ has the shape $\square \ r :_{n} L, U$, where $\square$ is either \texttt{static} or \texttt{mutable}, denoting whether the value of $r$ can change over time, and $L$ and $U$ are the lower- and upper-bounds of $r$, defined as:

{  \centering\footnotesize
  \begin{displaymath}
    \begin{array}{lcl}
L & \bnfeq & \mathbb{P}(\mathcal{A}^n) \mid  \Gamma \\
U & \bnfeq & \mathbb{P}(\mathcal{A}^n) \mid \color{BrickRed}{conjunction} \\
conjunction & \bnfeq & conjunction \wedge arrow \mid arrow\\
arrow & \bnfeq & arrow \ mult \ \rightarrow \ mult \ arrow \mid mult \ \Gamma\\
mult & \bnfeq & \ma{set} \mid \ma{lone} \mid \ma{some} \mid \ma{one} \mid \ma{no}
\end{array}%
\end{displaymath}}%
where $\Gamma$ is any relational expression over variables with constant bounds.
\end{definition}

Roughly, multiplicity bounds specify a relation between atoms of the domain and the range of a relation. For instance, a multiplicity constraint \a{A lone -> one B} defines an injective function from \a{A} to \a{B}: it must relate every \a{A} atom with exactly one \a{B} atom, and every \a{B} atom with at most one \a{A} atom. For a relation of arity $>2$, this process is applied inductively, and the conjunction of multiple multiplicity constraints is supported. The syntax and semantics of relational expressions $\Gamma$ are presented in the next section. For instance, the \texttt{CMS} model from Section~\ref{sec:motivation} could be represented by the following declarations, assuming a scope of exactly 2 reviewers and 2 articles that would result in the universe of atoms $\mathcal{A} = \{ \mathtt{A_0}, \mathtt{A_1}, \mathtt{R_0}, \mathtt{R_1}, \mathtt{D_0}, \mathtt{D_1}, \mathtt{D_0}\}$.

{  \centering\footnotesize
\begin{displaymath}
\begin{array}{lcll}
\mathtt{static}  \ \ma{Article} &:_1&    \{ \mathtt{(A_0)}, \mathtt{(A_1)} \}                  &  \{ \mathtt{(A_0)}, \mathtt{(A_1)} \} \\
\mathtt{static}  \ \ma{Reviewer} &:_1&   \{ \mathtt{(R_0)}, \mathtt{(R_1)} \}                  &  \{ \mathtt{(R_0)}, \mathtt{(R_1)} \} \\
\mathtt{static}  \ \ma{Agent} &:_1&      \{ \mathtt{(R_0)} \}                                  &  \{ \mathtt{(R_0)}  \} \\
\mathtt{static}  \ \ma{Decision} &:_1&   \{ \mathtt{(D_0)}, \mathtt{(D_1)}, \mathtt{(D_2)} \}  &  \{ \mathtt{(D_0)}, \mathtt{(D_1)}, \mathtt{(D_2)} \} \\
\mathtt{static}  \ \ma{assigns} &:_2&    \{\}                                                  &  \ma{Reviewer some -> some Article} \\
\mathtt{mutable} \ \ma{reviews} &:_3&    \{\}                                                  &  \ma{Article -> Reviewer -> lone Decision} \\
\mathtt{mutable} \ \ma{decisions} &:_2&  \{\}                                                  &  \ma{Article -> lone Decision}
\end{array}%
\end{displaymath}}

\paragraph{Semantics}

For a trace $I$, $I \models \mathcal{D}$ holds iff for every $r \in \mathcal{R}$ and $i \in \mathbb{N}$, $L(r) \subseteq I(i)(r) \subseteq U(r)$ (that is, the bounds are valid in each state of the trace). If a bound is a tuple set $\mathbb{P}(\mathcal{A}^n)$, then $\subseteq$ is just subset relation. If a bound is an expression $\Gamma$ over relations with constant bounds, we interpret the bound as a \emph{set of valuations} that satisfy the imposed multiplicity constraints; and a relation valuation is valid if it belongs to that set. These sets of valuations are formalized as follows.

Given a universe of atoms $\mathcal{A}$, a valuation for an $n$-ary relation $r$ is a tuple set in $\mathbb{P}(\mathcal{A}^n)$. A multiplicity bound for such an $n$-ary has $n$ unary relational expressions $\Gamma_i$ over relations with constant bounds at the leaves. Thus, it is possible to statically evaluate the value of all \emph{domains} $\Gamma_i$ into unary tuple sets, which we denote by $\sem{\Gamma_i} \in \mathbb{P}(\mathcal{A}^1)$ (following the semantics for relational expressions from the next section); the domain of $r$ is the Cartesian product of all its unary domains. The set of all possible valuations of $r$ is the powerset $\mathcal{V}(r) \;\triangleq\; \mathbb{P}(\sem{\Gamma_1} \times \ldots \times \sem{\Gamma_n})$. We call a \emph{valuation set} of $r$ any $V \subseteq \mathcal{V}(r)$. 
We recover the domain of a valuation set by taking the union of all the tuples in its valuations as $\bigcup V$.

To each multiplicity annotation $m \in \mathtt{mult}$ we associate a predicate
$\mu_m$ to constrain the cardinality of a valuation $v$:
\begin{equation*}
  \begin{array}{r@{\;\;}l@{\qquad}r@{\;\;}l}
    \mu_{\ma{set}}(v)  &\triangleq \top
    & \mu_{\ma{some}}(v) &\triangleq |v| \geq 1 \\[2pt]
    \mu_{\ma{one}}(v)  &\triangleq |v| = 1
    & \mu_{\ma{no}}(v)   &\triangleq |v| = 0 \\[2pt]
    \mu_{\ma{lone}}(v) &\triangleq |v| \leq 1
  \end{array}
\end{equation*}

The \emph{multiplicity restriction} of a valuation set $V$ by $m$ keeps
only the valuations satisfying $\mu_m$:
\begin{equation*}
  V \upharpoonright m \;\triangleq\;
  \{\, v \in V \mid \mu_m(v) \,\}
\end{equation*}
Whenever $\mu_m$ is satisfiable on $\mathcal{V}(r)$ for a non-empty domain (i.e., for every multiplicity except \a{no} on a non-empty $\sem{\Gamma_1} \times \ldots \times \sem{\Gamma_n}$), then $\bigcup( \mathcal{V}(r) \upharpoonright m) = \sem{\Gamma_1} \times \ldots \times \sem{\Gamma_n}$.

Given valuation sets $V_1 \subseteq \mathcal{V}(r_1)$ and
$V_2 \subseteq \mathcal{V}(r_2)$ with domains $\sem{\Gamma_1} \times \ldots \times \sem{\Gamma_k}$ and $\sem{\Gamma_{k+1}} \times \ldots \times \sem{\Gamma_n}$, respectively, we define their \emph{matched Cartesian product}
$V_1 \otimes V_2 \subseteq \mathcal{V}(r_1 \times r_2)$ as the set of
valuations where every tuple from $V_1$ has a match in $V_2$, and vice-versa.

\begin{equation*}
  V_1 \otimes V_2 \;\triangleq\;
  \left\{\, v \subseteq \bigcup V_1 \times \bigcup V_2 \;\middle|\;
  \begin{array}{l}
    \forall\, \ma{(}a_1\ma{,}\ldots\ma{,}a_{k}\ma{)} \in \bigcup V_1 \mid  \{ \ma{(}b_{k+1}\ma{,}\ldots\ma{,}b_n\ma{)} \mid \ma{(}a_1\ma{,}\ldots\ma{,}b_n\ma{)} \in v \} \in V_2 \\[2pt]
    \forall\, \ma{(}b_{k+1}\ma{,}\ldots\ma{,}b_m\ma{)} \in \bigcup V_2 \mid  \{ \ma{(}a_1\ma{,}\ldots\ma{,}a_{k}\ma{)} \mid \ma{(}a_1\ma{,}\ldots\ma{,}b_n\ma{)} \in v \} \in V_1 
  \end{array}
  \,\right\}
\end{equation*}

Finally, the semantic function $\sem{\cdot}$ maps each multiplicity expression to a valuation set. For an arrow $a$ with underlying relation of arity $n$,
$\sem{a} \subseteq \mathcal{V}(\mathcal{U}^n)$.

\begin{align*}
  \sem{m\,\Gamma}
    &\;\triangleq\; \mathbb{P}(\Gamma) \upharpoonright m
    \label{eq:sem-leaf} \\
  \sem{a_1\; m_1 \to m_2\; a_2}
    &\;\triangleq\; \sem{a_1} \upharpoonright {m_1}
                    \;\otimes\;
                    \sem{a_2} \upharpoonright {m_2}
\end{align*}

A conjunction whose top-level constructor is an arrow $a$ inherits the
arrow's semantics; an explicit intersection is interpreted as the
set intersection of the two valuation sets:
\begin{equation*}
  \sem{c \wedge a} \;\triangleq\; \sem{c} \,\cap\, \sem{a},
\end{equation*}

As an example, let us calculate the set of possible valuations for relation binary relation \a{assigns} of \texttt{CMS}, whose upper-bound is \a{Reviewer some -> some Article}, meaning it has two unary domains, \a{Reviewer} and \a{Article}.

{  \centering\footnotesize
\begin{align*}
& \sem{\ma{Reviewer some -> some Article}} \\
=&\sem{\ma{Reviewer}} \upharpoonright {\ma{some}}  \;\otimes\; \sem{\ma{Article}} \upharpoonright \ma{some} \\
=&\sem{\ma{Reviewer}} \upharpoonright {\ma{some}}  \;\otimes\; \{\{\},\{(\mathtt{A_0})\},\{(\mathtt{A_1})\},\{(\mathtt{A_0}),(\mathtt{A_1})\}\} \upharpoonright \ma{some}\\
=&\sem{\ma{Reviewer}} \upharpoonright {\ma{some}}  \;\otimes\; \{\{(\mathtt{A_0})\},\{(\mathtt{A_1})\},\{(\mathtt{A_0}),(\mathtt{A_1})\}\} \\
=&\{\{\},\{(\mathtt{R_0})\},\{(\mathtt{R_1})\},\{(\mathtt{R_0}),(\mathtt{R_1})\}\} \upharpoonright {\ma{some}}  \;\otimes\; \{\{(\mathtt{A_0})\},\{(\mathtt{A_1})\},\{(\mathtt{A_0}),(\mathtt{A_1})\}\} \\
=&\{\{(\mathtt{R_0})\},\{(\mathtt{R_1})\},\{(\mathtt{R_0}),(\mathtt{R_1})\}\}   \;\otimes\; \{\{(\mathtt{A_0})\},\{(\mathtt{A_1})\},\{(\mathtt{A_0}),(\mathtt{A_1})\}\} \\
=&\{\{(\mathtt{R_0},\mathtt{A_0}),(\mathtt{R_1},\mathtt{A_1})\},
  \{(\mathtt{R_0},\mathtt{A_1}),(\mathtt{R_1},\mathtt{A_0})\}, \\
 & \ \{(\mathtt{R_0},\mathtt{A_0}),(\mathtt{R_0},\mathtt{A_1}),(\mathtt{R_1},\mathtt{A_0})\},
  \{(\mathtt{R_0},\mathtt{A_0}),(\mathtt{R_0},\mathtt{A_1}),(\mathtt{R_1},\mathtt{A_1})\}, 
  \{(\mathtt{R_0},\mathtt{A_0}),(\mathtt{R_1},\mathtt{A_0}),(\mathtt{R_1},\mathtt{A_1})\}, 
  \{(\mathtt{R_0},\mathtt{A_1}),(\mathtt{R_1},\mathtt{A_0}),(\mathtt{R_1},\mathtt{A_1})\}, \\
 & \ \{(\mathtt{R_0},\mathtt{A_0}),(\mathtt{R_0},\mathtt{A_1}),(\mathtt{R_1},\mathtt{A_0}),(\mathtt{R_1},\mathtt{A_1})\}\}
\end{align*} }

\subsection{Hyper relational temporal logic constraints} 

\paragraph{Syntax} Hyper model finding supports constraints specified in a extension of HyperLTL with first-order and relational operators, as well as more flexible trace quantification placement.

\begin{definition}[Hyper relational temporal logic syntax] A hyper relational temporal logic formula $\Phi$ is defined as follows, where $x \in \mathcal{R} \cup \bigcup \mathcal{R}_i \cup bv(\Phi) \cup \bigcup bv(\phi_i)$ are free or bound relational variables, and $\pi \in \bigcup \{ P_i \}$ are trace variables.

{\centering\scriptsize
\begin{displaymath}
\begin{array}{lcl}
  \Phi     & \bnfeq & \textrm{\a{not} } \Phi \mid \Phi \textrm{ \a{and} } \Phi \mid {\color{BrickRed}{\textrm{\a{all} } \pi \textrm{ \a{:} } P \textrm{ \a{|} } \Phi}} \mid \phi \\
  \phi     & \bnfeq & \ma{true} \mid \Gamma \textrm{ \a{in} } \Gamma \mid \textrm{\a{some} } \Gamma \mid \textrm{\a{lone} } \Gamma \mid \textrm{\a{not} } \phi \mid \phi \textrm{ \a{and} } \phi \mid \textrm{\a{all} } x \textrm{ \a{:} } \Gamma \textrm{ \a{|} } \phi  \\
           & \mid & \textrm{\a{after} } \phi \mid \phi \textrm{ \a{until} } \phi  \mid \textrm{\a{before} } \phi \mid  \phi \textrm{ \a{since} } \phi \\
  \Gamma   & \bnfeq & x \mid {\color{BrickRed}{\Gamma[\pi]}} \mid \textrm{\a{none}} \mid \textrm{\a{iden}} \mid \ma{\~}\Gamma \mid \ma{\^}\Gamma \mid \Gamma\ \ma{\+}\ \Gamma \mid \Gamma\ \ma{\&}\ \Gamma \mid \Gamma\ \ma{-}\ \Gamma \mid \Gamma\ \ma{->}\ \Gamma \mid \Gamma\ \ma{.}\ \Gamma \mid \ma{\{} x_1 \ma{:} \Gamma_1 \ma{,} \mydots \ma{,} x_n \ma{:} \Gamma_n\ \ma{|}\ \phi \ma{\}}\mid \Gamma\ma{'}
\end{array}
\end{displaymath}}
\end{definition}

The fragment not highlighted in {\color{BrickRed}{red}} corresponds to the non-hyper relational temporal logic fragment 
and allows the definition of properties in relational temporal logic, an extension of first-order temporal logic with relational and transitive closure operators. Here, for the sake of brevity, we focus on a kernel of the language into which all other operators can be translated. For expressions, this includes the converse (\a{~}), union (\a{+}), intersection (\a{&}), difference (\a{-}), product (\a{->}), join (\a{.}) and transitive closure (\a{^}) relational operators, as well as the temporal prime operator (\a{'}) and relations by comprehension. For formulas, this includes atomic inclusion tests (\a{in}) and cardinality tests (\a{some} and \a{lone}), and combinations with negation (\a{not}), conjunction (\a{and}) and (first-order) universal quantifications (\a{all}), as well as future (\a{until} and \a{after}) and past (\a{since} and \a{before}) temporal operators. In relational model finding everything is seen as a relation, including quantified variables. This allows the same operators to be used for relations and variables, since variables are simply singleton tuple sets (a set with exactly one tuple of arity 1). This is the reason why there is no set membership, but only the inclusion operator \a{in}. The hyper fragment introduces trace quantifications and a trace selector to determine under which trace variable an expression is evaluated. Notice that trace quantifications can be combined by Boolean connectives, but not nested inside temporal operators nor first-order quantifications.

\paragraph{Semantics}
The semantics of formulas and expressions are shown in Fig.~\ref{fig:formulasemantics}, with entries highlighted in {\color{BrickRed}{red}} denoting the hyper fragment. They are evaluated under a set of traces, given by a context $\sigma$ for each trace variable, the current trace variable $\pi$ being evaluated, and an index $i$ of the current state being evaluated. The validity of a formula follows standard semantics for relational logic and LTL with future and past operators. First-order quantification produces static variables, and for a tuple $t$, we write $\pi \mapsto \mathbb{N} \mapsto x \mapsto \{t\}$ to assign $t$ to $x$ at every state of trace variable $\pi$. The value of a relational expression $\Gamma$  -- a tuple set -- is denoted by $\sem{\Gamma}^{\pi,i}_{\sigma}$ (the universe $\mathcal{A}$ is given implicitly). The hyper fragment introduces a new trace into $\sigma$ for each quantification over a model $P$, while context selection changes the trace variable being evaluated.
A trace instance of a hyper problem $H$ is a valuation $I$ for the variables declared in $H$ such that the bounds and constraint hold. 

\begin{definition}
A trace $I : \mathbb{N} \rightarrow \mathcal{R} \mapsto \mathbb{P}(\mathcal{A}^*)$ is an \emph{instance} of an hyper model finding problem over problems $P_i$ and $H = \langle \mathcal{A}, \mathcal{D}, \Phi \rangle$, denoted $I \models H$, iff $I \models \mathcal{D}$ and $\pi_0 \mapsto I, \pi_0, 0 \models \Phi$, for some distinguished trace variable $\pi_0$ not occurring in $\Phi$. A problem $H$ is satisfiable, $\models H$, if it has any instance $I$, and unsatisfiable, $\not \models H$, otherwise. 
\end{definition}

Notice that if $\Phi$ has no hyper constructs, the semantics degenerates into that of (non-hyper) relational temporal logic problems evaluated under a single trace variable. So, for a non-hyper problem $P$, $I \models P$, iff $I \models \mathcal{D}$ and $\pi_0 \mapsto I, \pi_0, 0 \models \phi$.

\begin{figure}
  \centering\scriptsize
  \begin{displaymath}
    \begin{array}{lcl}
      \sigma, \pi, i \models \textrm{\a{not} } \Phi       & \equiv & \sigma, \pi, i \not\models \Phi\\
      \sigma, \pi, i \models \Phi_1 \textrm{ \a{and} } \Phi_2 & \equiv & \sigma, \pi, i \models \Phi_1 \wedge \sigma, \pi, i \models \Phi_2\\
      {\color{BrickRed}{\sigma, \pi, i \models \textrm{\a{all} } \pi_1 \textrm{ \a{:} } P \textrm{ \a{|} } \Phi}}  & \equiv & \forall I \cdot I \models P \Rightarrow \sigma \oplus \pi_1 \mapsto I, \pi, i \models \Phi\\
\end{array}
  \end{displaymath}  
\begin{displaymath}
    \begin{array}{lcl}
      \sigma, \pi, i \models \Gamma_1 \textrm{ \a{in} } \Gamma_2  & \equiv & \sem{\Gamma_1}^{\pi,i}_{\sigma} \subseteq \sem{\Gamma_2}^{\pi,i}_{\sigma}\\
      \sigma, \pi, i \models \textrm{\a{some} } \Gamma      & \equiv & \verts*{\sem{\Gamma}^{\pi,i}_{\sigma}} \ge 1\\
      \sigma, \pi, i \models \textrm{\a{lone} } \Gamma      & \equiv &\verts*{\sem{\Gamma}^{\pi,i}_{\sigma}} \le 1\\
      \sigma, \pi, i \models \textrm{\a{not} } \phi       & \equiv & \sigma, \pi, i \not\models \phi\\
      \sigma, \pi, i \models \phi_1 \textrm{ \a{and} } \phi_2 & \equiv & \sigma, \pi, i \models \phi_1 \wedge \sigma, \pi, i \models \phi_2\\
      \sigma, \pi, i \models \textrm{\a{all} } x \textrm{ \a{:} } \Gamma \textrm{ \a{|} } \phi  & \equiv & \forall t \in \sem{\Gamma}^{\pi,i}_{\sigma} \sigma \oplus \pi \mapsto \mathbb{N} \mapsto x \mapsto \{t\}, \pi, i \models \phi\\
      \sigma, \pi, i \models \textrm{\a{after} } \phi                       & \equiv & \sigma, \pi, {i+1} \models \phi\\
      \sigma, \pi, i \models \phi_1 \textrm{ \a{until} } \phi_2                 & \equiv & \exists i \leq l \cdot \sigma, \pi, l \models \phi_2  \wedge \forall i \leq j < l \cdot \sigma, \pi, j \models \phi_1 \\
      \sigma, \pi, i \models \textrm{\a{before} } \phi                  & \equiv & 0 < i \wedge \sigma, \pi, {i-1} \models \phi\\
      \sigma, \pi, i \models \phi_1 \textrm{ \a{since} } \phi_2                 & \equiv & \exists 0 \leq l \leq i \cdot \sigma, \pi, l \models \phi_2 \wedge \forall l < j \leq i \cdot \sigma, \pi, j \models \phi_1 
\end{array}
  \end{displaymath}  
    \begin{displaymath}
    \begin{array}{lcl}
      \sem{x}^{\pi,i}_{\sigma} & = & \sigma(\pi)(i)(x)\\
      {\color{BrickRed}{\sem{\Gamma[\pi_1]}^{\pi,i}_{\sigma}}} & = & \sem{\Gamma}^{\pi_1, i}_{\sigma}\\
      \sem{\textrm{\a{none}}}^{\pi,i}_{\sigma} & = & \{\}\\
      \sem{\textrm{\a{iden}}}^{\pi,i}_{\sigma} & = & \{\ma{(} a \ma{,} a \ma{)} \mid a \in {\mathcal{A}} \}\\
      \sem{\ma{\~}\Gamma}^{\pi,i}_{\sigma} & = & \{\ma{(} b \ma{,} a \ma{)} \mid \ma{(} a \ma{,} b \ma{)} \in \sem{\Gamma}^{\pi,i}_{\sigma} \}\\
      \sem{\ma{\^}\Gamma}^{\pi,i}_{\sigma} & = & \{\ma{(} a \ma{,} b \ma{)} \mid \exists c_1, \mydots ,c_n \cdot \ma{(} a \ma{,} c_1 \ma{)},\ma{(} c_1 \ma{,} c_2 \ma{)}, \mydots ,\ma{(} c_n \ma{,} b \ma{)} \in \sem{\Gamma}^{\pi,i}_{\sigma} \}\\
      \sem{\Gamma_1\ \ma{\+}\ \Gamma_2}^{\pi,i}_{\sigma} & = & \sem{\Gamma_1}^{\pi,i}_{\sigma} \cup \sem{\Gamma_2}^{\pi,i}_{\sigma}\\
      \sem{\Gamma_1\ \ma{\&}\ \Gamma_2}^{\pi,i}_{\sigma} & = & \sem{\Gamma_1}^{\pi,i}_{\sigma} \cap \sem{\Gamma_2}^{\pi,i}_{\sigma}\\
      \sem{\Gamma_1\ \ma{-}\ \Gamma_2}^{\pi,i}_{\sigma} & = & \sem{\Gamma_1}^{\pi,i}_{\sigma} \setminus \sem{\Gamma_2}^{\pi,i}_{\sigma}\\
      \sem{\Gamma_1\ \ma{->}\ \Gamma_2}^{\pi,i}_{\sigma} & = & \{\ma{(} a_1 \ma{,} \mydots \ma{,} a_n \ma{,} b_1  \ma{,} \mydots \ma{,} b_m\ma{)} \mid \ma{(} a_1 \ma{,} \mydots \ma{,} a_n \ma{)} \in \sem{\Gamma_1}^{\pi,i}_{\sigma} \wedge \ma{(} b_1 \ma{,} \mydots \ma{,} b_m \ma{)} \in \sem{\Gamma_2}^{\pi,i}_{\sigma} \}\\
      \sem{\Gamma_1\ \ma{.}\ \Gamma_2}^{\pi,i}_{\sigma} & = & \{\ma{(} a_1 \ma{,} \mydots \ma{,} a_{n-1} \ma{,} b_2  \ma{,} \mydots \ma{,} b_m\ma{)} \mid   \exists c \cdot \ma{(} a_1 \ma{,} \mydots \ma{,} c \ma{)} \in \sem{\Gamma_1}^{\pi,i}_{\sigma} \wedge \ma{(} c \ma{,} \mydots \ma{,} b_m \ma{)} \in \sem{\Gamma_2}^{\pi,i}_{\sigma} \}\\
      \sem{\ma{\{} x_1 \ma{:} \Gamma_1 \ma{,} \mydots \ma{,} x_n \ma{:} \Gamma_n\ \ma{|}\ \phi \ma{\}}}^{\pi,i}_{\sigma} &=& \\
      \multicolumn{3}{l}{\quad\quad\quad\{\ma{(} a_1,\mydots,a_n \ma{)} \mid \ma{(}a_1\ma{)} \in \sem{\Gamma_1}^{\pi,i}_{\sigma} \wedge \mydots \wedge \ma{(} a_n \ma{)} \in \sem{\Gamma_n}^{\pi,i}_{\sigma} \wedge\sigma \oplus \pi \mapsto \mathbb{N} \mapsto (x_1 \mapsto \ma{\{(} a_1 \ma{)\}}, \mydots, x_n \mapsto \ma{\{(} a_n \ma{)\}}), \pi, i \models \phi \}}\\
      \sem{\Gamma\ma{'}}^{\pi,i}_{\sigma} & = & \sem{\Gamma}^{\pi, i+1}_{\sigma}
    \end{array}
  \end{displaymath}
  \caption{Formula and expression semantics, $a$, $b$, $c$ are atoms from an implicitly defined $\mathcal{A}$.}
  \label{fig:formulasemantics}
\end{figure}

\paragraph{Finite prefix semantics} 
In model checking instance traces are always infinite, but some BMC techniques draw conclusions from finite prefixes of infinite traces. This requires the semantics to be adapted~\cite{BiereCCZ99}, depending on assumptions made about the unknown continuation of the prefix. \tool{HyperQB}~\cite{HyperQube} implements BMC for HyperLTL, and our approach considers the same variants of the semantics. For instance, for a prefix where $p$ holds at every state, a \emph{pessimistic} semantics will assume that $p$ may not hold in the future and consider \a{always p} trivially false; in contrast, an \emph{optimistic} semantics will assume $p$ will continue to hold and consider \a{always p} trivially true. Since certain identities (such as $\ma{always p} \equiv \ma{not eventually not p}$) no longer hold, an additional temporal operator, \a{releases}, is added to the kernel language. In those cases, results are still reported but they are inconclusive. The pessimistic semantics for a prefix of size $k$ is as follows:

{\centering\footnotesize
  \begin{displaymath}
    \begin{array}{lcl}
      \sigma, \pi, i \models_k \textrm{\a{after} } \phi                      & \equiv & i < k \wedge \sigma, \pi, {i+1} \models_k \phi\\
      \sigma, \pi, i \models_k \phi_1 \textrm{ \a{until} } \phi_2              & \equiv & \exists i \leq l \leq k \cdot \sigma, \pi, l \models_k \phi_2  \wedge \forall i \leq j < l \cdot \sigma, \pi, j \models_k \phi_1  \\
      \sigma, \pi, i \models_k \phi_1 \textrm{ \a{releases} } \phi_2           & \equiv & \exists i \leq l \leq k \cdot \sigma, \pi, l \models_k \phi_2  \wedge \forall i \leq j \leq l \cdot \sigma, \pi, j \models_k \phi_1
\end{array} 
  \end{displaymath}  }

The optimistic semantics is defined in a dual way.
In classical BMC, if the state machine representing the system is total (i.e., there are no deadlock states) then the outcome is conclusive when a finite trace is found under the pessimistic semantics. However, as we have seen, relational model finding problems do not have an explicit state machine. Instead, behavior is enforced declaratively in constraint $\phi$, making it hard to test for totality.\footnote{Our tool will actually be able to perform this test, but requires infinite trace analysis, which defeats the purpose of BMC; thus, it is disabled by default.}
Under the optimistic semantics the outcome is conclusive if no trace is found. 
\tool{HyperQB} actually introduces two other variants of bounded semantics for models without loops other than self-loops in \emph{halting} states. In these variants, which we also support, a sound conclusion can also be made for traces that reach a halting state.

\section{Model finding with \hpard{}}
\label{sec:mfind}

\hpard{} is our implementation of a hyper model finder, and is the first step of our analysis procedure. For a hyper model finding problem\footnote{As a backend solver, model finders typically do not provide a concrete syntax for the definition of problems, which are defined programmatically through an API.}, it generates a set of SMV models and a HyperLTL formula, lower level formats already consumable by existing HyperLTL model checkers and our own \hsmv{} presented in the next section. By relying on standard formats, our technique is expected to benefit from further advances in model checking techniques. 

\subsection{Hyper model finding algorithm}

\begin{figure}
\begin{smvfig}
FROZENVAR
  assigns : array 0..1 of array 0..1 of boolean;
VAR
  review   : array 0..1 of array 0..1 of 0..3;
  decision : array 0..1 of 0..3;
IVAR
    a : 0..2;
    r : 0..2;
    d : 1..3;
ASSIGN
    init(review[0][0]) := 0;
    next(review[0][0]) := case 
                            r = 0 & a = 0 & assigns[0][0] & review[0][0] = 0: d; 
                            TRUE: review[0][0]; 
                          esac;
    init(review[0][1]) := 0;
    next(review[0][1]) := case 
                            r = 1 & a = 0 & assigns[0][1] & review[0][1] = 0: d; 
                            TRUE: review[0][1]; 
                          esac;
    ...
    init(decision[0]) := 0;
    next(decision[0]) := case 
                           r = 2 & a = 0 & decision[0] = 0 & 
                           (assigns[0][0] -> review[0][0] != 0) & (assigns[0][1] -> review[0][1] != 0) & 
                           (review[0][0] = d | review[0][1] = d | review[1][0] = d | review[1][1] = d): d; 
                           TRUE: decision[0]; 
                         esac;
    ...
\end{smvfig}
\caption{Possible encoding of a CMS model in SMV for 2 reviewers and 2 articles}
\label{fig:cms_smv}
\end{figure}

Existing model checkers for HyperLTL follow the traditional approach of having a distinct notation for the state machine over which traces are quantified, and the hyperproperty to be checked. Most advanced ones (e.g., \tool{AutoHyper} and \tool{HyperQB}) support the definition of the state machine in (a fragment of) the SMV language. SMV models can be defined in two styles: one more imperative based on direct \s{ASSIGN} clauses, and another declarative based on \s{INIT}, \s{INVAR} and \s{TRANS} restrictions. The property to be verified is specified in an \s{LTLSPEC} clause.

Figure~\ref{fig:cms_smv} presents a possible (manual) encoding of a CMS similar to the one presented in \tool{Alloy} in Section~\ref{sec:motivation}. Since SMV has no first-order features, the scope of the model is fixed; here we consider 2 reviewers and 2 articles. SMV supports atomic types, and words/arrays of fixed length. Here, we define \s{assigns} as a $2\times2$ array of Booleans to represent papers assigned to reviewers; it is defined as a \s{FROZENVAR}, meaning its value will not change in the trace. Both \s{review} and \s{decision} are defined as arrays with 4 possible values (0 meaning unassigned); they are defined as variables (\s{VAR}) whose value is assigned by the transitions of the system. \s{a}, \s{r} and \s{d} are input variables (\s{IVAR}), whose value is not controlled by the state machine but rather assigned arbitrarily; they are used to represent inputs that guide the selection of events. To enforce the behavior of the system, we employ the imperative style using \s{ASSIGN} clauses. These determine the next value of each variable independently, making it cumbersome to model events with multiple effects. Here, a \s{review} may change value if it is still unassigned; a \s{decision} may change value if all assigned reviewers have submitted decisions and the decision is within those reviews (the second criteria defined in Section~\ref{sec:motivation}).

\tool{Pardinus} already provides a translation of (non-hyper) model finding problems into declarative SMV by expanding the relational fragment of the language into (propositional) LTL~\cite{pardinus22}. Although \tool{Pardinus} models do not support an explicit definition of the state machine, this translation identifies conjuncts with no temporal operators (referring to the initial state), and those only under \a{always} operator (denoting invariants and transitions if non-nested prime operators occur), into a declarative description of the state machine; the remaining conjuncts go to the \s{LTLSPEC} clause. 

\hpard{} builds on this procedure, as shown in Algorithm~\ref{alg:smvtrans} which translates a hyper model finding problem into a set of (SMV) state machines and a HyperLTL formula. We abstract the translation of relational temporal logic to LTL translation by a procedure \proc{RL2LTL} that is identical to the one provided by \tool{Pardinus} except that it propagates the trace selectors from relational expressions down to propositional variables. We also assume that a procedure annotates a non-hyper problem $P$ with a trace variable $\pi$, denoted by $P[\pi]$; this connects a problem $P$ with a particular hyper quantification $\pi : P$ in $H$.

A mismatch between existing HyperLTL model checkers and hyper model finding problems is that the former only support hyperproperties in prenex normal form. Quantifiers can be trivially moved to the outermost position if their domain is guaranteed to be non-empty. The algorithm starts by converting constraint $\Phi$ from $H$ into prenex form, and separating it between the quantifiers $\mathbf{Q}$ and the non-hyper fragment $\psi$; this non-hyper fragment of the outermost model $H$ is translated into a state machine $M$ and an LTL constraint $\psi^0$. 
Then, for each hyper quantification $\pi : P$, it annotates $P$ with the correct trace variable and converts it into a lower-level problem consisting of a state machine $M$ and a formula an LTL $\phi^0$. To guarantee that the prenex transformation was sound, the LTL problem is checked for non-emptiness, which can be performed by a standard SMV model checker. All $\phi^0$ formulas that could not be encoded in the state machine are added as premises of formula $\psi^0$. The final formula applies the hyper quantifiers $\mathbf{Q}$ to $\psi^0$, resulting in a HyperLTL constraint $\Phi^0$.

\begin{algorithm}[t]
\footnotesize
\KwIn{Hyper model finding problem with problems $P_i$ and $H = \langle \mathcal{A}, \mathcal{D}, \Phi \rangle$.}
\KwOut{State machine definitions $\mathscr{M}$, HyperLTL formula $\Phi^0$, or $\bot$}
$\mathbf{Q}, \psi \leftarrow \proc{prenex}(\Phi)$\tcp*{convert to prenex and extract quantifiers}
$M, \psi^0 \leftarrow \proc{RL2LTL}(\langle \mathcal{A}, \mathcal{D}, \psi \rangle)$\tcp*{translate non-hyper fragment into LTL}
$\mathscr{M} \leftarrow [ M ]$\; 
\ForEach{$\mathop{\mathrm{Q}} \pi:P \in \proc{reverse}(\mathbf{Q})$}{
  $M, \phi^0 \leftarrow \proc{RL2LTL}(P[\pi])$\tcp*{translate problem into LTL}
  \lIf*{$\proc{empty}(M,\phi^0)$}{\KwRet{$\bot$}}\tcp*{invalid state machine}
  \eIf{$Q = \forall$}{$\psi^0 \leftarrow \phi^0 \ \ma{implies (} \psi^0 \ma{)}$\;}{$\psi^0 \leftarrow \phi^0 \ \ma{and (}\psi^0\ma{)}$\;}
  $\mathscr{M} \leftarrow [M] + \mathscr{M}$\;
}
$\Phi^0 \leftarrow \mathbf{Q} \ \ma{|}\ \psi^0$\tcp*{prefix hyper quantifiers and problem constraints}
\KwRet{$\mathscr{M}, \Phi^0$}
\caption{Overview of the \hpard{}-to-SMV procedure}
\label{alg:smvtrans}
\end{algorithm}

\begin{theorem}
Let \proc{LTL} and \proc{HyperLTL} be model checking procedure for LTL and HyperLTL, respectively.
Assume that, for any $M, \phi^0 = \proc{RL2LTL}(P)$, $\models P \ \text{iff}\ \proc{LTL}(M,\neg \phi^0)$ finds counter-examples. Then, if $\mathscr{M}, \Phi^0$ is obtained from problems $P_i$ and $H$ through Algorithm~\ref{alg:smvtrans}, $\models H \ \text{iff}\ \proc{HyperLTL}(\mathscr{M}, \neg \Phi^0)$ finds counter-examples.
\end{theorem}
\begin{proof}[Proof (draft)] 
We assume that the \proc{RL2LTL} translation implements the relational temporal logic semantics from Fig.~\ref{fig:formulasemantics}; that is, $P$ is satisfiable iff \proc{LTL} finds a counter-example for $\neg \phi^0$ over a state machine $M$ (model checkers, unlike model finders, attempt to falsify properties). The conversion into prenex normal form is sound if the quantification ranges are non-empty; for trace quantifiers, this amounts to $\models P$, which due to the previous assumption can be performed through an emptiness check with an LTL model checker. After conversion to prenex normal form, the non-hyper subformula of $\Phi$ is translated with \proc{RL2LTL}, preserving the semantics of the relational temporal logic fragment. Hyper model finding problems range over problems $P$, while \proc{HyperLTL} ranges over state machines $M$; a problem $P$ may not fully be encoded as a state machine $M$, requiring the extra $\phi^0$ constraint produced by \proc{RL2LTL}; these are imposed as premises of the property passed to \proc{HyperLTL} in order to only consider valid traces of $P$ problems (as conjunction for existential quantifications, and implication for universal ones). 
\end{proof}

\subsection{Optimizations}

In the rest of this section, we describe further optimizations related to this translation.

\paragraph{Composing quantifications} Sequential quantifications of the same type (universal or existential) can be merged into a single quantification over the product of the individual models. This composition results in more complex models but fewer trace quantifications, which usually results in more efficient analysis. The product of two model finding problems $P_1 \times P_2$ is obtained by simply uniting the universes and declarations, and conjoining the constraints, apart from renamings to avoid clashes (in particular, in the self composition $P \times P$). 
Outermost existential quantifications can also be merged into the hyper problem $H$. So, for instance, a hyper problem with models $P_1$, $P_2$ and $H$, a constraint $\Phi$ of the shape $\ma{some x, y :} \ P_1 \ \ma{| all w, z :} \ P_2 \ \ma{|} \ \phi$, becomes $H \times P_1 \times P_1$ with formula $\ma{all wz :} \ P_2 \times P_2 \ \ma{|} \ \phi$.
As a side-effect, the generated trace instances will now also instantiate these outermost existential quantifications (such as \a{s1} and \a{s2} in the motivating example). 

\paragraph{Multiplicity constraints}
The default \proc{RL2LTL} translation from \tool{Pardinus} produces Boolean-encoded SMV models in which the presence of each tuple of an $n$-ary relation $R$ is recorded by an independent Boolean variable. This encoding represents the full space $\mathcal{V}(R) = \mathbb{P}(R)$ of valuations, and thus corresponds to an unconstrained declaration $\sem{\ma{set}\,R}$.
Our extension exploits the multiplicity constraints attached to field declarations to encode instead the smaller space $\sem{R}$, retaining only the valuations that respect the declared multiplicities. We realise this by replacing the per-tuple Booleans by SMV integer variables ranging over $\sem{R}$.

For instance, a binary relation $R : A \rightarrow \ma{one}\, B$ -- a total function from $A$ to $B$ -- has exactly $|\sem{R}| = |B|^{|A|}$ valuations, and can be encoded as $|A|$ integer variables of size $|B|$ (one per atom of $A$). This will be crucial for explicit backends such as \tool{AutoHyper}, whose scalability is dominated by the explored state space, and will also yield a more compact representation for symbolic backends such as \tool{HyperQB}: even after the customary translation to a Boolean encoding (``bit blasting''), each integer of size $|B|$ takes only $\lceil \log_2 |B| \rceil$ Booleans, so the $|A|$ integers together require $|A| \cdot \lceil \log_2 |B| \rceil$ Booleans, against the $|A| \cdot |B|$ Booleans generated by default by \proc{RL2LTL}.

In general, $|\sem{R}| \le |\mathcal{V}(R)| = 2^{|R|}$, whose worst case for an $n$-ary $R$ over domains of size $|D_1|,\ldots,|D_n|$ is $2^{|D_1|\cdots|D_n|}$. Assuming $|D_i| = k$, this is $2^{k^n}$ -- doubly exponential in the arity $n$. Materialising $\sem{R}$ explicitly is therefore only feasible for relations with tight multiplicity bounds. Our implementation accordingly manipulates relations symbolically, and applies reduction rules for the common multiplicity patterns; explicit enumeration is reserved as a last resort for more intricate constraints, typically over small domains.

\paragraph{Symmetry breaking} An important feature of \tool{Pardinus}, inherited from \tool{Kodkod}, is the detection of symmetries to avoid the generation of isomorphic solutions. The introduction of symmetry breaking predicates considerably reduces the search space, and has shown to largely improve the performance of solvers~\cite{KodKod}. We also explore this feature for hyper problems, but it can only be applied to the outermost model, since breaking symmetries in the inner models would be dependent on the valuations of the outermost relations. However, by applying this detection after merging quantifications we can actually apply symmetries over multiple traces. For instance, in the CMS example, articles are considered symmetric in the outermost quantification \a{s1}, so having the \a{Agent} assigned to \a{Article0} or \a{Article1} is indistinguishable. In \a{s2}, articles would be considered symmetric if they are not distinguishable in \a{s1}, but \a{s2} has no knowledge of \a{s1}. Composing \a{s1} and \a{s2} together enables symmetry breaking in the latter. Remark that state-of-the-art model checkers for HyperLTL also support merging quantifications but at much lower level; applying symmetry breaking over merged quantifications is one of the domain-specific advantages of starting the analysis from a higher-level model.

\section{Model checking with \hsmv{}}
\label{sec:smvtrans}

\hsmv{} is the second step of our analysis pipeline (Fig.~\ref{fig:hsmv-arch}).
It consumes the set of SMV models $\mathscr{M}$ and HyperLTL formula $\Phi^0$
produced by \hpard{}, and discharges the verification problem using one of two
backends: an infinite explicit-state backend (\tool{Exp}) and a finite
symbolic backend (\tool{Sym}).

\begin{figure}[t]
\centering
\resizebox{\linewidth}{!}{%
\begin{tikzpicture}[
  font=\small, >=Stealth,
  io/.style    = {draw, rounded corners=3pt, fill=gray!15,
                  inner sep=5pt, align=center},
  novel/.style = {draw, thick, fill=BrickRed!18,
                  inner sep=5pt, align=center},
  reimp/.style = {draw, fill=Orange!30,
                  inner sep=5pt, align=center},
  inh/.style   = {draw, dashed, fill=gray!10,
                  inner sep=5pt, align=center},
  smn/.style   = {draw, thick, fill=BrickRed!18,
                  inner sep=4pt, align=center, font=\scriptsize},
  smi/.style   = {draw, dashed, fill=gray!10,
                  inner sep=4pt, align=center, font=\scriptsize},
  arr/.style   = {->, thick},
  lin/.style   = {thick},
]

\def\Ex{-4.3}
\def\Sx{4.3}

\node[novel, font=\scriptsize, minimum width=2.4cm] (lN) at (\Ex, 0.5) {novel};
\node[font=\bfseries\scriptsize, anchor=south west, inner sep=0pt, yshift=3pt] at (lN.north west) {Legend:};
\node[reimp, font=\scriptsize, minimum width=2.4cm] (lR) at (\Ex, -0.1) {reimplemented};
\node[inh,   font=\scriptsize, minimum width=2.4cm] (lI) at (\Ex, -0.7) {inherited};

\node[io] (input)  at (0.0, 0.5) {SMV $\mathscr{M}$ + HyperLTL $\Phi^0$};
\node[io] (output) at (8.2, 0.5) {SMV verdict + counter-examples};

\node[novel] (ir)   at (0.0, -0.6) {Expression IR};
\node[novel] (splt) at (0.0, -1.6) {Split initial configurations};
\node[novel] (mvdd) at (0.0, -2.6) {MVDD \;/\; BDD};
\draw[arr] (input)--(ir);
\draw[arr] (ir)--(splt);
\draw[arr] (splt)--(mvdd);

\draw[lin] (mvdd.south) -- ++(0,-0.35) coordinate (m_down);

\node[reimp] (expst) at (\Ex,-4.0) {Explicit state system construction};
\node[novel] (fmov)  at (\Ex,-4.9) {Formula sub-expression extraction};
\node[reimp] (bisim) at (\Ex,-6.0) {Bisimulation};
\node[novel] (ap)    at (\Ex,-7.1) {AP identification};
\draw[arr] (expst)--(fmov);
\draw[arr] (fmov)--(bisim);
\draw[arr] (bisim)--(ap);

\draw[arr, rounded corners=4pt] (bisim.west) -- ([xshift=-0.6cm]fmov.west |- bisim.west) |- (fmov.west);

\node[draw=gray!55, dashed, fill=white, rounded corners=5pt,
      minimum width=10.3cm, minimum height=3.2cm] (AH) at (\Ex+0.65, -9.3) {};

\node[anchor=north west, font=\normalsize\sffamily, inner xsep=8pt, inner ysep=6pt]
      at (AH.north west) {AutoHyper};

\node[inh, text width=4.0cm] (nba) at (\Ex,-8.65)
      {NBA + $\exists$-intersection\\+ $\forall$-complementation};

\node[novel] (apc) at (\Ex+4.35,-8.65) {AP compaction};
\draw[arr] ([yshift=5pt]nba.east)  -- ([yshift=5pt]apc.west);
\draw[arr] ([yshift=-5pt]apc.west) -- ([yshift=-5pt]nba.east);

\node[smi] (nowit) at (\Ex-2.3,-10.25)
      {Automaton solver w/o witnesses\\ spot / forq / bait / forklift / rabit / roll};
\node[smn] (wit)   at (\Ex+2.3,-10.25)
      {Automaton solver w/ witnesses\\ forklift / rabit / roll};

\draw[arr] (ap.south) -- (nba.north);

\draw[lin] (nba.south) -- ++(0,-0.2) coordinate (a_down);
\draw[arr, rounded corners=3pt] (a_down) -| (nowit.north);
\draw[arr, rounded corners=3pt] (a_down) -| (wit.north);

\node[novel] (qbfc) at (\Sx,-4.0) {Formula sub-expression extraction};
\node[reimp] (smov) at (\Sx,-4.9) {QBF unrolling};
\node[inh] (qbfs) at (\Sx,-6.1)
      {QBF solver\\{\scriptsize quabs / cqesto / qute / qfun}};
\draw[arr] (qbfc)--(smov);
\draw[arr] (smov)--(qbfs);

\draw[arr, rounded corners=3pt] (m_down) -| (expst.north);
\draw[arr, rounded corners=3pt] (m_down) -| (qbfc.north);

\path (m_down) -- (m_down -| expst.north) 
  node[midway, above, font=\footnotesize\sffamily] {Exp};
\path (m_down) -- (m_down -| qbfc.north) 
  node[midway, above, font=\footnotesize\sffamily] {Sym};

\node[novel] (btrace) at (0,-11.6) {Backtracing witnesses};

\draw[lin] (wit.south) -- ++(0,-0.5) coordinate (w_drop);
\draw[arr, rounded corners=3pt] (w_drop) |- (btrace.west);

\draw[lin] (qbfs.south) -- ++(0,-0.5) coordinate (q_drop);
\draw[arr, rounded corners=3pt] (q_drop) |- (btrace.east);

\draw[arr] (btrace.south) -- ++(0,-0.6) coordinate (b_down);

\draw[lin] (nowit.south) -- ++(0,-1.0) coordinate (n_drop);
\draw[thick, rounded corners=3pt] (n_drop) |- (b_down);

\draw[arr, rounded corners=5pt] (b_down) -| (output.south);

\end{tikzpicture}%
}
\caption{Architecture of \hsmv{}.
  Thick red borders mark novel contributions; orange solid borders mark
  reimplemented components with significant extensions; dashed grey borders
  mark inherited components.
  Inside \tool{AutoHyper}, the witness track
  for counter-example generation during inclusion checking is novel.}
\label{fig:hsmv-arch}
\end{figure}

Applying existing HyperLTL model checkers to our higher-level examples turned
out to be far from straightforward. The SMV models and HyperLTL formulas that
arise from high-level \tool{Alloy} specifications are significantly larger and
more complex than the low-level benchmarks those tools were designed for:
they use the full declarative SMV language (including \s{INIT},
\s{INVAR} and \s{TRANS} clauses with integer-valued variables),
whereas \tool{AutoHyper} only accepts imperative \s{ASSIGN}-style models,
and \tool{HyperQB} struggled to cope with the size of the resulting QBF
instances.
Scaling these tools to our setting required significant reengineering of their
pipelines; \hsmv{} is the result of that effort.

\hsmv{} is designed to be a retrocompatible drop-in replacement: it accepts
the same input formats (SMV + HyperLTL) and calls the same underlying
solvers --- \tool{AutoHyper} directly for \tool{Exp}, and off-the-shelf QBF
solvers for \tool{Sym} --- while supporting a much more liberal class of SMV
models and applying several optimizations along the way, as described below
and summarized in Fig.~\ref{fig:hsmv-arch}.

Beyond the technical pipeline improvements, the two main usability
contributions of \hsmv{} are support for \emph{general SMV models} (removing
the restriction to imperative style) and \emph{SMV-level witnesses} (the
ability to translate solver counter-examples back to readable SMV traces),
which was a major hindrance in the existing tools and is essential for the
\tool{Alloy Analyzer} visualization described in Section~\ref{sec:halloy}.

\paragraph{Representing expressions as decision diagrams}
We first apply integer-level simplifications -- such as equality tests and
range checks -- to all non-hyper, non-temporal sub-expressions of SMV models
and HyperLTL formulas at an intermediate expression level, before any
decision diagram is constructed.
We then convert the simplified expressions into a shared internal
representation based on \emph{Multi-Valued Decision Diagrams} (MVDD),
using hash-consing to promote memory sharing of sub-expressions, and
implementing horizontal (conjunctive) and vertical (disjunctive) partitioning
for scalable processing.
As an optional preprocessing step, we can convert and specialize to \emph{Binary Decision Diagrams} (BDDs) via \tool{nuXmv}.

The SMV integer variables produced by the multiplicity-aware encoding of Section~\ref{sec:mfind} are carried into the MVDD as multi-valued nodes; this keeps the diagram shallower and avoids
inflating the node count without reducing the underlying state space.
The explicit-state backend (\tool{Exp}) manipulates this representation directly.
The symbolic backend (\tool{Sym}) bit-blasts multi-valued nodes
into Boolean gates at encoding time.

\subsection{Explicit-state analysis (\tool{Exp})}

Regarding our explicit backend, we have encountered two major challenges in trying to connect \tool{HyperPardinus} directly to \tool{AutoHyper}.
The first challenge is that \tool{AutoHyper} only supports imperative SMV models, being incompatible with those generated by \tool{Pardinus}\footnote{It is possible to convert a declarative SMV model to an  imperative one~\cite{TLAassigns}, but this is more complex than necessary for our purposes.}.
The second challenge concerns the generation of counter-examples. Succinctly, \tool{AutoHyper} converts the formula to a \textit{Non-deterministic B\"uchi Automata} (NBA) and iteratively intersects it with the automaton of each $\exists$ quantification; $\forall$ quantifiers are handled using automata complementation.
Since NBA complementation is a very expensive operation, the outermost $\forall$ quantification is checked via automaton inclusion.
This optimization is however disabled when computing witnesses, and \tool{AutoHyper} instead relies on the \tool{spot}~\cite{spot} automaton library to compute witnesses as any valid trace of the final NBA.

\paragraph{Constructing explicit-state systems}
The first step in \tool{Exp} is thus to convert each SMV model into an explicit-state system, which essentially corresponds to an NBA. The MVDD representation greatly eases the implementation of this pass. Conversion proceeds by enumerating all initial SMV states and recursively following all transitions. The above \tool{HyperPardinus} optimizations ensure that the input SMV models are already as compact as possible.

\paragraph{Reducing explicit-state systems}
The next pass of \tool{AutoHyper}, replicated in \tool{Exp}, is to compress models by computing bisimulation quotients.
This consists in projecting away state variables not mentioned by the formula, jointly with transitions that only depend on projected-away variables.
We refine quotients to observe the formula's non-temporal sub-expressions (namely APs, introduced below) rather than the raw variables, which is strictly coarser since these are functions of the variables.

\paragraph{Compressing formulas}
Remember that the solving process of \tool{AutoHyper} consists in converting the formula to an NBA, to then be intersected with the NBA of each quantification. To reduce the size of the formulas, arbitrary non-temporal sub-expressions can be treated as variables -- termed \emph{Atomic Propositions (APs)} -- in the resulting NBA. Since the choice of APs can greatly affect the performance of the conversion and the subsequent solving process, \tool{AutoHyper} allows users to directly annotate APs in hyper formulas. In \tool{Alloy} models -- and later at the level of abstraction of \tool{HyperPardinus} -- there are no such annotations, and \tool{HyperSMV} thus infers APs automatically. This inference is performed on formula expressions, before any decision diagram is built.
AP boundaries are placed syntactically following two rules:
1) a maximal non-temporal sub-expression mentioning a single trace becomes one AP;
2) families of multi-trace Boolean expressions corresponding to relational operations such as equality or comparisons (that the Alloy-to-SMV translation has previously expanded) are grouped into a single AP.
We also ensure that semantically identical APs are shared.

\paragraph{Compressing APs}
Compressing the formula is not sufficient, because \tool{AutoHyper} introduces new APs as it solves: intersecting the NBA with the automaton of a quantification $\pi_2 : M$ eliminates that trace variable, and a multi-trace AP such as $x[\pi_1] = x[\pi_2]$ is then left behind as one AP $x[\pi_1] = v$ for \emph{every} value $v$ that $x$ takes in $M$.
The number of APs thus grows with the size of the models rather than with the formula, even though the whole family is determined by the single value that $x$ takes in trace $\pi_2$: for an equality at most one
of them can hold, and for a comparison such as $x[\pi_1] \leq x[\pi_2]$ the residual tests $x[\pi_1] \leq v$ are ordered by implication. In either case, most combinations of these APs are impossible.
This increase in APs can make the subsequent complementations blow up, and we address it by tuning \tool{AutoHyper} in two ways.
First, we discard operations that no reachable value of the model can satisfy, and replace variables with a single reachable value by constants.
Second, much like on the \hsmv{} side, we recognize the families whose APs are determined by such a value -- equalities and comparisons between variables or against constants, and
their Boolean combinations -- and replace each family by $\lceil \log_2 n \rceil$ APs, one per bit of an SMV integer variable of size $n$, as in the bit blasting of Section~\ref{sec:mfind}, rewriting the operation bitwise, so that the APs are fixed in advance. Unrecognized shapes fall back to the original APs.
This rewriting is applied to the automaton rather than to the formula, so the LTL-to-NBA conversion still sees the original APs.

\paragraph{Calling \tool{AutoHyper} and computing witnesses}
We then call \tool{AutoHyper} over the reduced systems and formula.
When computing witnesses, \tool{AutoHyper} originally relied solely on automata complementation, defaulting to it whenever witnesses are requested.
This is because mapping witnesses over bisimulation-compressed models back to the original uncompressed ones is non-trivial.
Since complementation quickly becomes a bottleneck for larger models and more complex formulas, we extended \tool{AutoHyper} to produce SMV-level witnesses from the output of inclusion checkers \tool{RABIT}, \tool{forklift} and \tool{ROLL}, which natively provide automaton-level witnesses.

\paragraph{Backtracing witnesses}
Finally, we translate witness traces over reduced systems back to the original systems, and then to the SMV level. Since the reduced system is a projection of the original, this can be framed as a reachability problem over the original system that always has a solution.
Still, bisimulation entails that witness traces may not have the same length, since one state in the reduced witness trace may correspond to multiple transitions in the original system, namely the source states in the same equivalence class (that have been compressed to the same reduced state). Witnesses for zero or more outermost traces -- always with the same quantifier type -- are handled similarly under composition.

\subsection{Symbolic analysis (\tool{Sym})}

The main challenge for \tool{Sym} is that the \tool{HyperQB} pipeline was not coping with larger models.

\paragraph{Constructing QBF problems}
We reimplemented the \tool{HyperQB}~\cite{HyperQube} algorithm to convert models and formulas to QBF, represented as a Boolean circuit with $\forall$ and $\exists$ quantifiers in prenex form.
For a prefix length $k$, we define $k+1$ copies of the variables of each model and construct a circuit that unrolls the initial states, transitions and invariants.
We then unroll the formula over states $0..k$, assuming an optimistic or pessimistic semantics at step $k+1$.
For the halting semantics, we additionally test that the circuit corresponding to the transition from $k$ to $k+1$ is equivalent to the identity relation.

\paragraph{Scaling to larger models}
As for \tool{Exp}, the simplicity of this pass is greatly empowered by our MVDD representation.
In a similar fashion, \tool{HyperQB} builds on the \tool{PyNuSMV} library to construct a symbolic internal BDD representations of input SMV models, using it to first enumerate all initial states, and then find all reachable states. It then constructs circuits of the form $\bigwedge(vars(i) \rightarrow vars(i+1))$ for each transition from state $i$ to its next state $i+1$. This enumeration is comparable to the process of constructing an explicit-state system performed by \tool{AutoHyper}, and may also help explaining why both tools achieve comparable performance in our benchmarks later in Section~\ref{sec:eval}.
In contrast, we convert each MVDD/BDD to a circuit by applying a simple Shannon Expansion that mirrors its structure, preserving sharing. Larger expressions in the models and formula are naturally handled via partitioning.

\paragraph{Calling QBF solver and backtracing witnesses}
We then call an off-the-shelf (non-CNF) QBF solver, such as \tool{Quabs}, \tool{cqesto}, \tool{qute} or \tool{qfun}. 
The result of the solver will include a partial sequence of assignments, that we convert back to the original SMV model by mapping each QBF variable to a different variable of a different state in the witness trace. 
At the source SMV level, we present the validity of the formula, a witness SMV trace and whether the outcome is conclusive or not.

\subsection{Optimizations}

We now describe additional \tool{HyperSMV} optimizations that proved to be critical for pratical efficiency.

\paragraph{Moving formula sub-expressions}
The translation performed by \tool{HyperPardinus} moves into the SMV definition sub-expressions of the problems' constraints that represent initial conditions, invariants and transitions. The remaining constraints are moved into the HyperLTL property $\Phi^0$ to be checked.
To further simplify the formula, we extract additional sub-expressions that refer to only one trace and move them to the respective state machine.
For \tool{Exp}, this further reduces the size of the explicit-state system. We apply the same LTL-to-NBA conversion and intersection with the NBA of the model performed by \tool{AutoHyper} for $\exists$ quantifications, and iteratively move formula components and apply bisimulation until no further reduction is possible.
For \tool{Sym}, the performance gain is not so direct as both the state machine and the formula are encoded as Boolean expressions in the resulting QBF problem. 
However, the same expression may have different semantics depending on whether it appears in the state machine or in the formula, due to the bounded semantics; pushing sub-expressions from the formula into the state machine can therefore improve the conclusiveness of the analysis.
For example, a formula \am{some a:A | always $\phi[a]$ and eventually $\psi[a]$} will become false under the pessimistic semantics, but if $\phi$ is moved to the state machine of \ma{A} then \am!some a:{$A_\phi$} | eventually $\psi[a]$! may find a witness \ma{a} that satisfies $\psi$ in a finite prefix.
This optimization is sound only if the totality of the state machine is preserved — i.e., that every prefix has at least one possible continuation; this may be violated, for instance, when $\phi$ -- that will act as an invariant -- rules out all outgoing transitions from some state. In Section~\ref{sec:eval} we discuss how this is handled in our selected benchmarks.

\paragraph{Splitting initial configurations}
For larger SMV models, a significant bottleneck lies in the processing and state-space exploration of the full system; therefore, HyperSMV adopts a partitioning strategy~\cite{LerdaSisto99,HolzmannJoshiGroce11} by decomposing the initial state predicate into disjoint subsets. This effectively splits the input model into smaller submodels with restricted initial configurations that can be solved independently. We implement several splitting strategies to control granularity, such as explicitly computing all initial states, partitioning based only on variables referenced in the \s{INIT} predicate, or splitting only ``frozen'' variables that remain constant throughout a trace. While our current implementation checks these submodels sequentially, the approach is sound because the total set of traces is the union of the submodels' traces. Consequently, to refute a $\forall$ property or witness an $\exists$ property, it is sufficient to find a single submodel that respectively invalidates or satisfies the formula, allowing the \tool{Exp} backend to prune the search space and scale better for complex models.

\section{The extended \tool{Alloy Analyzer}}
\label{sec:halloy}

As seen in Section~\ref{sec:motivation}, \tool{Alloy} is based on the same relational linear temporal logic of \tool{Pardinus}, but introduces additional syntactic features to ease the design of software systems. For a thorough presentation of \tool{Alloy} see~\cite{Jackson16,MacedoBCCK16}.

When an \tool{Alloy} command is executed, the model is converted into a single \tool{Pardinus} problem $P$ -- $\mathcal{A}$ is derived from the command's scope, $\mathcal{D}$ from the signature hierarchy, and $\phi$ from conjoining the command's formula with other model restrictions. To generate a hyper model finding problem instead, we must identify portions of the \tool{Alloy} model that represent $P_i$ models for trace quantifications. The only extension implemented in the \tool{Alloy} language is a keyword {\color{BrickRed}\a{trace}} to identify signatures that represent a $P_i$ model; any quantification over such signatures results in a trace quantification $\pi_i \ \ma{:} \ P_i$ in the hyperproperty from $H$. Occurrences of such variables will result in context selections, and thus $\pi_i$ can only be used when composed with fields of $P_i$.
To obtain the constraint $\phi$ of each $P_i$ problem -- which defines its behavior declaratively -- we convert the original formula into negation normal form and extract top-level conjuncts where $\pi_i$ is the only trace variable mentioned. This extraction is performed syntactically over the whole formula without requiring any user annotation; moving conjuncts into individual models can only further constrain them, so no solutions are lost. It may even move into the models parts of the formula to be checked in the command, resulting in simpler hyperproperties.\footnote{Since moving constraints into a $P_i$ model is equivalent to restricting an initially unconstrained state machine, doing so can affect conclusiveness in the bounded semantics. User predictability is therefore essential. Unlike \tool{HyperSMV}, which simplifies formulas prior to extraction, this Alloy-level transformation is conservative. We preserve the structure of the user's formula and restrict extraction to top-level conjuncts where $\pi_i$ is the sole trace variable, to ensure that only the portions of the formula the user most naturally associates with $P_i$'s behavior are moved there.}
Additionally, we also search the conjuncts for any multiplicity constraints over the declared relations, which are then moved from constraint $\phi$ to the bound declarations $\mathcal{D}$.

The \tool{Alloy Analyzer} was extended to support this translation, launch \hpard{} (choosing the \tool{Sym} or \tool{Exp} is done through the GUI as in regular \tool{Alloy}), and present trace instances back to the user for satisfiable commands, as exemplified in Fig.~\ref{fig:cex}. The counter-example shows the declarations of the outermost model $H$. When multiple outermost existential quantifiers are present, their domains are combined into a single composite model, so a single instance is displayed that ranges over all of them simultaneously (see Section~\ref{sec:smvtrans}).

\section{Evaluation}
\label{sec:eval}

\newcommand{\benchnat}{\textsc{High}}
\newcommand{\benchrev}{\textsc{Low}}

This section intends to evaluate the feasibility of our approach by answering the following research questions:
\begin{description}[leftmargin=\parindent,labelindent=0pt]
\item[RQ1] Can relevant hyperproperties be specified over rich models in our extension to \tool{Alloy}? 
\item[RQ2] How do high-level \tool{Alloy} models improve usability compared to low-level SMV ones?
\item[RQ3] Are analyses and optimizations in the pipeline effective for high-level models?
\item[RQ4] How does the performance of \tool{HyperSMV} compare with state-of-the-art model checkers?
\end{description}

To answer these, we consider 2 sets of benchmarks, namely: 1) \tool{Alloy} models with hyperproperties written by the authors (\benchnat{}) and 2) SMV models from a benchmark suite that is shared by \tool{HyperQB}~\cite{hyperqb2024} and \tool{AutoHyper}~\cite{AutoHyper} (\benchrev{}), as well as their automatic translation into \tool{Alloy}\footnote{We developed a specialized \tool{HyperSMV} mode that performs a direct, syntactic translation from SMV to \tool{Alloy}. Although \tool{Alloy} is designed for high-level abstraction, it can also express these lower-level patterns, even if the resulting models are not considered idiomatic.}.
For \textbf{RQ1}, \benchnat{} explores the flexibility of writing high-level models natively, and \benchrev{} the support for examples from the state of the art.
For \textbf{RQ2}, we analyze the readability of specifications and the interpretability of feedback, benchmarking the \tool{Alloy} experience against standard SMV model checkers for demonstrative benchmarks.
For \textbf{RQ3}, we run \hpard{} for i) \benchnat{} models, ii) the \benchrev{} \tool{Alloy} models translated from SMV to analyze the overhead of the pipeline, and iii) \benchnat{} models with the \hpard{} optimizations disabled both altogether and selectively to assess their impact.
For \textbf{RQ4}, we compare \hsmv{} against \tool{AutoHyper} and \tool{HyperQB} with i) the original SMVs from \benchrev{}, and ii) the SMVs generated by \hpard{} for \benchnat{}. All executions had witness reporting enabled and were performed in a MacBook Pro 13-inch, M1, 2020 16GB with a timeout of $200$s.

\subsection{Benchmark models}

\benchnat{} includes the first 4 groups of models and consists of idiomatic examples written natively in \tool{Alloy} to showcase the full expressive power of the language.

\paragraph{Conference management system} This is our running example from Section~\ref{sec:motivation}, with the variants for \a{criteria} that render NI and GNI valid or invalid: NI and GNI hold for $\text{CMS}_\text{max}$, $\text{CMS}_\text{ndet}$ breaks NI, and $\text{CMS}_\text{any}$ breaks GNI as well. The scope determines the number of $A$rticles and $R$eviewers.

\paragraph{Robotic path synthesis} Path synthesis is a typical use case of hyperproperties. Here we consider different variants, such as finding a shortest path~\cite{HyperQube} ($\text{Robot}_\text{short}$), robust paths which solve multiple goals~\cite{HyperQube} ($\text{Robot}_\text{robust}$), robust paths against adversaries~\cite{BMCloop} ($\text{Robot}_\text{adv}$), and paths with minimal penalties ($\text{Robot}_\text{score}$). We consider also a simpler $\text{Robot}_\text{two}$ property that checks if at least two distinct paths exist.

Scope $P$osition determines the size of the board, and in $\text{Robot}_\text{score}$ the maximum penalty.
 Below is the property for $\text{Robot}_\text{adv}$, searching for a \a{rob}ot trace against all possible \a{adv}ersaries traces.
\begin{alloy}
some r:Rob | rob[r] and all a:Rob | adv[a] implies 
  always (not Robot.(r.robot) in Robot.(a.robot))
\end{alloy}
In $\text{Robot}_\text{score}$, we allow robots to step onto obstacles with a penalty. The hyperproperty searches for a path with a smaller penalty than any other path reaching a goal.
\begin{alloy}
some r:Rob | rob[r] and all o:Rob | rob[o] implies 
  always ((o.pos in o.goals and r.pos in r.goals) implies lte[r.penalty,o.penalty])
\end{alloy}

\paragraph{Mutation testing} Hyperproperties have been used in mutation-driven test case generation~\cite{FellnerBW21}. We consider whether a mutant of a non-deterministic system is \textit{possibly} or \textit{definitely} killed by a test case, and 3 different vending systems for which the properties hold or fail~\cite{FellnerBW21} (scope determines the maximum $Q$uantity of items). Below is the definition of definitely killable mutant with 3 trace quantifiers stating, for a sequence of inputs, the output of the test for the mutant must differ from all possible original system outputs.
\begin{alloy}
some s1:System | system[m1] and all m,s2:System |
  (mutant[m] and system[s2] and always (same_in[s1,m] and same_in[s1,s2])) implies 
    eventually not same_out[m,s2]
\end{alloy}

\paragraph{Linearizability} 
Verifying that a concurrent system adheres to a sequential specification is a hallmark challenge in formal methods; such properties are hard to express compositionally and more easily framed as hyperproperties. We consider the SNARK concurrent queue~\cite{DohertyDGFLMMSS04}, a data structure that is linearizable but not sequentially consistent. While sequential consistency requires a sequential execution that preserves the program order of all traces, linearizability is more flexible, allowing concurrent operations to be reordered to match a valid sequential execution. For this benchmark, the analysis complexity is determined by the scope of available $N$odes, possible $V$alues and concurrent $P$rocesses.

\paragraph{\benchrev{} benchmarks} This is a set of SMV models and HyperLTL properties from a benchmark suite jointly used by \tool{AutoHyper} and \tool{HyperQB}. These include the Bakery3, Bakery5, SNARK, NI, NRP, Mutant and Robot benchmarks, and were also directly translated to \tool{Alloy}.

\subsection{Qualitative evaluation} \label{sub:qualitative_evaluation}

\begin{figure*}
    \centering
    \includegraphics[width=0.9\textwidth]{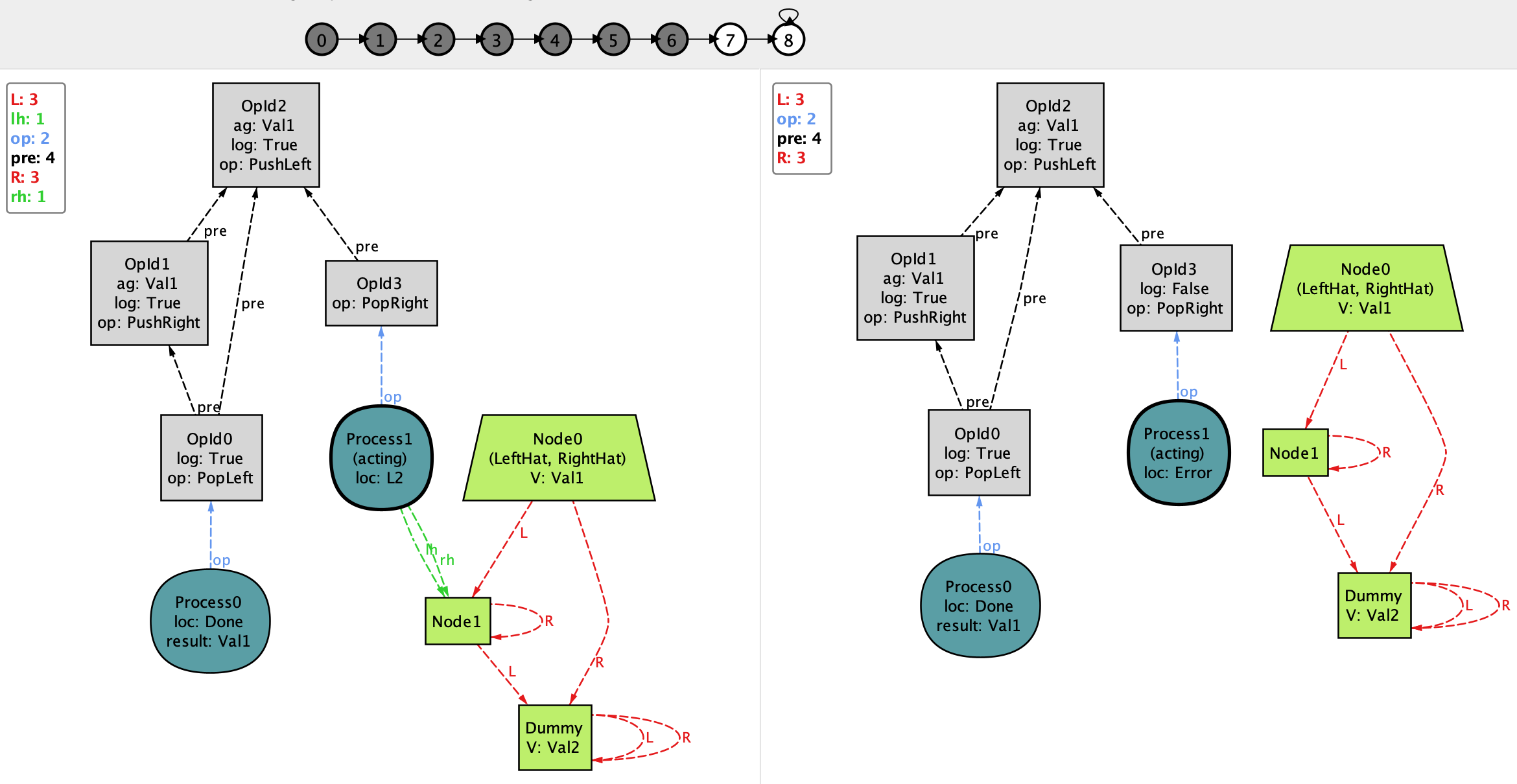} $\,$
    \caption{Last transition of a known bug for SNARK~\cite{DohertyDGFLMMSS04}, process 1 fails a pop even though the queue was never empty since it started, which is not linearizable; the precedences between the occurred operations are shown, required to test linearizability.}
    \label{fig:snark}
\end{figure*}

Regarding \textbf{RQ1}, our approach has shown to be suitable for modeling the selected systems at a high level of abstraction. This is patent, e.g., in the robot examples, where in \tool{Alloy} the movement is defined declaratively for any current position and is easily readable, while the original SMVs encoded actions for each explicit position, and were thus inscrutable and required dedicated auxiliary scripts to be generated for various map sizes. Moreover, we cover varied classes of quantifier alternations. 
The different level of abstraction is also patent in the returned instances, which will become clear for the SNARK examples. While \tool{Alloy} allows relational instances to be graphically visualized, SMV-level tools report traces with only Boolean or integer variables, and \tool{HyperQB} provides just the output of the QBF solver. 

Regarding \textbf{RQ2}, we have set to compare the experience of specifying models and interpreting the results using our extension of \tool{Alloy}. For that purpose, we have chosen two demonstrative high-level examples that were inspired by existing low-level benchmarks.

\paragraph{Robotic path synthesis} In this family of problems, the models encode valid movements of a robot in a grid map with obstacles; the hyperproperty then defines different goals for the path that is to be synthesized. Here, we compare the encoding found in the existing, SMV-based benchmarks, with our based on \tool{Alloy}.

Figure~\ref{fig:robot_smv_model} shows an excerpt (focusing on horizontal movement) of the SMV encoding of a $10\times10$ map with 26 obstacles provided in the \tool{HyperQB} benchmarks. SMV does not have first-order quantifications or auxiliary predicates, so any encoding of such a grid will require addressing each grid position explicitly. In the case of Fig.~\ref{fig:robot_smv_model}, the authors defined an auxiliary constraint for each movement direction, determining when they might occur according to (implicit) grid constraints (here, \s{LEFT} and \s{RIGHT} are shown). Actual movement is then defined by assignments in an imperative flavour. Such a model is clearly unmanageable, since any changes to the obstacles requires changing all the auxiliary formulas, and changes to the grid size extending the assignment clauses. Evidence of this is that the authors actually relied on Python scripts to generate such SMV files from a given dimension and list of obstacles. 
While other SMV encodings in a more declarative style could be explored (e.g., using \s{TRANS} constraints), they would still need to be defined for a concrete grid dimension.

Contrast this with the \tool{Alloy} model shown in Fig.~\ref{fig:robot_alloy_model} (again, an excerpt focusing on horizontal movement). It imposes a total order on a signature position that is navigated through relations \a{next}/\a{prev}; the size of the grid is simply defined by the scope on this signature. Then, each movement action is described by a declarative predicate with two pre-conditions (the position to which the robot exists, and is not an obstacle), and an effect (the robot position is updated). The obstacles are defined in a separate predicate as a binary relation between positions (i.e., sets of coordinates); any update to the obstacles requires only a local change in this predicate.

\begin{figure}
\centering

\begin{subfigure}{0.5\textwidth}

\begin{smvfig}[numbers=none,xleftmargin=0pt,xrightmargin=0pt,basicstyle=\ttfamily\tiny]
ASSIGN
  init(x_axis) := {0};
  next(x_axis) :=
    case
      (TOKEN=0) : x_axis;
      (x_axis=0 & RIGHT ) : {1};
      (x_axis=1 & LEFT & RIGHT) : {0,2};
      (x_axis=1 & LEFT & !RIGHT) : {0};
      (x_axis=1 & !LEFT & RIGHT) : {2};
      (x_axis=2 & LEFT & RIGHT) : {1,3};
      (x_axis=2 & LEFT & !RIGHT) : {1};
      (x_axis=2 & !LEFT & RIGHT) : {3};
      (x_axis=3 & LEFT & RIGHT) : {2,4};
      (x_axis=3 & LEFT & !RIGHT) : {2};
      (x_axis=3 & !LEFT & RIGHT) : {4};
      (x_axis=4 & LEFT & RIGHT) : {3,5};
      (x_axis=4 & LEFT & !RIGHT) : {3};
      (x_axis=4 & !LEFT & RIGHT) : {5};
      (x_axis=5 & LEFT & RIGHT) : {4,6};
      (x_axis=5 & LEFT & !RIGHT) : {4};
      (x_axis=5 & !LEFT & RIGHT) : {6};
      (x_axis=6 & LEFT & RIGHT) : {5,7};
      (x_axis=6 & LEFT & !RIGHT) : {5};
      (x_axis=6 & !LEFT & RIGHT) : {7};
      (x_axis=7 & LEFT & RIGHT) : {6,8};
      (x_axis=7 & LEFT & !RIGHT) : {6};
      (x_axis=7 & !LEFT & RIGHT) : {8};
      (x_axis=8 & LEFT & RIGHT) : {7,9};
      (x_axis=8 & LEFT & !RIGHT) : {7};
      (x_axis=8 & !LEFT & RIGHT) : {9};
      (x_axis=9 & LEFT & RIGHT) : {8,10};
      (x_axis=9 & LEFT & !RIGHT) : {8};
      (x_axis=9 & !LEFT & RIGHT) : {10};
      TRUE: x_axis;
    esac;
...
DEFINE
  GOAL := ((x_axis=7 & y_axis=5));
  LEFT := (!(x_axis=0)) & !(x_axis=7 & y_axis=6) 
        & !(x_axis=8 & y_axis=6) & !(x_axis=5 & y_axis=5) 
        & !(x_axis=4 & y_axis=4) & !(x_axis=6 & y_axis=4) 
        & !(x_axis=2 & y_axis=8) & !(x_axis=7 & y_axis=4) 
        & !(x_axis=5 & y_axis=9) & !(x_axis=3 & y_axis=6) 
        & !(x_axis=5 & y_axis=6) & !(x_axis=2 & y_axis=5) 
        & !(x_axis=7 & y_axis=3) & !(x_axis=3 & y_axis=9) 
        & !(x_axis=9 & y_axis=5) & !(x_axis=4 & y_axis=5) 
        & !(x_axis=4 & y_axis=1) & !(x_axis=6 & y_axis=0) 
        & !(x_axis=9 & y_axis=6) & !(x_axis=2 & y_axis=6) 
        & !(x_axis=4 & y_axis=6) & !(x_axis=4 & y_axis=8) 
        & !(x_axis=2 & y_axis=9) & !(x_axis=3 & y_axis=8) 
        & !(x_axis=8 & y_axis=4) & !(x_axis=2 & y_axis=3);    
  RIGHT:= (!(x_axis=9)) & !(x_axis=5 & y_axis=4) 
        & !(x_axis=7 & y_axis=6) & !(x_axis=7 & y_axis=5) 
        & !(x_axis=4 & y_axis=0) & !(x_axis=4 & y_axis=4) 
        & !(x_axis=6 & y_axis=6) & !(x_axis=6 & y_axis=4) 
        & !(x_axis=2 & y_axis=8) & !(x_axis=2 & y_axis=4) 
        & !(x_axis=3 & y_axis=6) & !(x_axis=1 & y_axis=6) 
        & !(x_axis=8 & y_axis=1) & !(x_axis=2 & y_axis=1) 
        & !(x_axis=5 & y_axis=6) & !(x_axis=2 & y_axis=5) 
        & !(x_axis=0 & y_axis=3) & !(x_axis=5 & y_axis=3) 
        & !(x_axis=3 & y_axis=9) & !(x_axis=1 & y_axis=8) 
        & !(x_axis=0 & y_axis=6) & !(x_axis=0 & y_axis=9) 
        & !(x_axis=2 & y_axis=6) & !(x_axis=1 & y_axis=9) 
        & !(x_axis=0 & y_axis=8) & !(x_axis=3 & y_axis=5) 
        & !(x_axis=0 & y_axis=5);
\end{smvfig}
\caption{SMV model}
\label{fig:robot_smv_model}
\end{subfigure}
\hfill
\begin{subfigure}{0.45\textwidth}
\begin{alloyfig}[numbers=none,xleftmargin=0pt,xrightmargin=0pt,basicstyle=\ttfamily\tiny]
pred moveleft[b:Board] {
  some next.(b.robot) 
  no next.(b.robot) & b.obstacles
  b.robot' = next.(b.robot) }

pred moveright[b:Board] {
  some prev.(b.robot) 
  no prev.(b.robot) & b.obstacles
  b.robot' = prev.(b.robot) }
...
pred board[b:Board] {
  let p0=first, p1=p0.next, p2=p1.next, p3=p2.next, 
      p4=p3.next, p5=p4.next, p6=p5.next, p7=p6.next, 
      p8=p7.next, p9=p8.next {
    b.goals = p7->p5
    b.obstacles =
      p1->p3 + p1->p5 + p1->p6 + p1->p8 + p1->p9 + 
      p2->p6 + p2->p8 + p2->p9 + p3->p1 + p3->p4 + 
      p3->p5 + p3->p6 + p3->p8 + p4->p5 + p4->p6 + 
      p4->p9 + p5->p0 + p5->p4 + p6->p3 + p6->p4 + 
      p6->p6 + p7->p4 + p7->p6 + p8->p5 + p8->p6 + 
      p9->p1 
    b.robot = p0->p0 } 

    always {
      moveleft[b] or moveright[b] or ... or stutter[b]
    } }
\end{alloyfig}
\caption{Alloy model}
\label{fig:robot_alloy_model}
\end{subfigure}

\caption{Excerpt of the robot model, showing the same map and right/left actions}
\end{figure}

\paragraph{Linearizability} Our largest \tool{Alloy} case study involves modeling the SNARK concurrent deque~\cite{DohertyDGFLMMSS04}. This data structure implements a double-ended queue using two ``hats'' (pointers to the leftmost and rightmost elements) and supports concurrent push and pop operations across multiple processes.
While intended to be linearizable, the original design contained a subtle bug that was later identified and corrected in~\cite{DohertyDGFLMMSS04}. The SNARK structure effectively illustrates the distinction between linearizability and sequential consistency. For example, if a process initiates a \texttt{pop} on an empty queue while another process concurrently completes a \texttt{push}, the \texttt{pop} might still return a failure. Such a trace is not sequentially consistent, as no sequential execution explains a failed \texttt{pop} on a non-empty queue; however, it is linearizable because concurrent operations can be reordered. While the specific bug found in~\cite{DohertyDGFLMMSS04} violates both properties, many valid linearizable executions do not satisfy sequential consistency.

Previous attempts to verify the SNARK structure~\cite{hyperqb2024} -- as modelled in \benchrev{} -- relied on lower-level models and substituted the target linearizability property with the significantly stricter requirement of sequential consistency. We contend that this substitution is fundamentally problematic. First, it forces developers to verify a rigid proxy that misaligns with the actual design goals of concurrent data structures. Second, because many valid, linearizable traces do not satisfy sequential consistency, previous work had to impose ad-hoc restrictions on the search space to suppress these unwanted counter-examples. Such artificial narrowing of the state space is counter-productive to verification, as it requires the user to manually guide the tool away from legitimate concurrent behaviors.

To our knowledge, this is the first time the actual linearizability of the SNARK has been verified using a model checker for hyperproperties. To express the valid reorderings of concurrent operations, each trace maintains structures analogous to logical clocks, recording the history of initiated operations (\a{log}) and those that have completed when each operation initiates (\a{pre}). This allows us to express linearizability as a global hyperproperty as follows:

\begin{alloy}
pred precedes[A:Con, B:Seq] {
    Process.(A.loc) in Done+Error implies (A.log = B.log and A.pre in B.pre) }

pred Linearizability {
  all A:Con | RunCon[A] implies 
    some B:Seq | RunSeq[B] and sameOps[A,B] and always precedes[A,B] }
\end{alloy}

That is, for any concurrent execution, there exists some sequential execution that, at a stable state (no processes executing), recorded the same completed operations in an order that respects linearizability (order must be preserved between operations terminated before another operation initiates). Predicate \a{sameOps} (omitted) forces each operation atom to represent the same operation in both traces.

\begin{figure}[ht]
\centering
\begin{MicroVerbatim}
V -5 -145 -159 173 187 -201 -215 -229 -243 -257 -271 19 -285 33 47 61 75 89 103 -117 -131 -4 -144 -158 -172 -186 -200 -214 228 242 256 270 -18 284 -32 -46 -60 -74 -88 -102 -116 -130 -3 143 157 171 -185 -199 213 227 241 255 269 -17 283 -31 -45 -59 -73 -87 -101 115 129 2 142 156 170 -184 198 212 -226 -240 -254 -268 16 -282 30 -44 58 -72 -86 100 114 128 -6 146 160 -174 -188 202 216 -230 -244 -258 -272 -20 -286 -34 -48 -62 -76 -90 104 118 132 1 -141 155 169 183 197 -211 -225 -239 -253 -267 15 -281 29 43 57 71 85 99 -113 -127 -10 150 -164 178 192 206 -220 234 248 262 276 -24 290 -38 -52 -66 -80 -94 -108 122 -136 -9 149 -163 -177 -191 -205 219 233 247 261 275 -23 289 -37 -51 -65 -79 -93 -107 -121 135 -8 -148 162 176 190 204 218 232 246 260 274 -22 288 -36 -50 -64 -78 -92 -106 -120 -134 -7 -147 -161 -175 -189 -203 -217 -231 -245 -259 -273 -21 -287 -35 -49 -63 -77 -91 -105 -119 -133 -14 154 168 182 -196 210 224 238 252 266 280 28 294 -42 56 -70 84 -98 112 126 140 -13 153 167 181 195 -209 -223 -237 -251 -265 -279 -27 -293 41 55 -69 -83 97 111 125 139 -12 152 166 180 194 208 222 236 250 264 278 -26 292 -40 -54 68 82 96 110 124 138 -11 -151 -165 -179 -193 -207 -221 -235 -249 -263 -277 -25 -291 -39 -53 -67 -81 -95 -109 -123 -137 0
\end{MicroVerbatim}

\vspace{4pt}

\begin{MicroVerbatim}
A: ({"AllNodes_0_0":0, "AllNodes_0_1":0, "AllNodes_1_0":0, "AllNodes_1_1":0, "AllNodes_2_0":2, "AllNodes_2_1":2, "First_ProcID":1, "LeftHat":1, "RightHat":1, "Second_ProcID":2, "newnode":2, "popRightFAIL":false, "proc1.line":0, "proc1.local_popL_LH":0, "proc1.local_popL_LHR":0, "proc1.local_popL_RH":0, "proc1.local_popR_LH":0, "proc1.local_popR_RH":0, "proc1.local_popR_RHL":0, "proc1.local_pushR_LH":0, "proc1.local_pushR_RH":0, "proc1.local_pushR_RHR":0, "proc2.line":0, "proc2.local_popL_LH":0, "proc2.local_popL_LHR":0, "proc2.local_popL_RH":0, "proc2.local_popR_LH":0, "proc2.local_popR_RH":0, "proc2.local_popR_RHL":0, "proc2.local_pushR_LH":0, "proc2.local_pushR_RH":0, "proc2.local_pushR_RHR":0} {"AllNodes_0_0":0, "AllNodes_0_1":0, "AllNodes_1_0":0, "AllNodes_1_1":0, "AllNodes_2_0":2, "AllNodes_2_1":2, "First_ProcID":1, "LeftHat":1, "RightHat":1, "Second_ProcID":2, "newnode":2, "popRightFAIL":false, "proc1.line":1, "proc1.local_popL_LH":0, "proc1.local_popL_LHR":0, "proc1.local_popL_RH":0, "proc1.local_popR_LH":0, "proc1.local_popR_RH":0, "proc1.local_popR_RHL":0, "proc1.local_pushR_LH":0, "proc1.local_pushR_RH":0, "proc1.local_pushR_RHR":0, "proc2.line":10, "proc2.local_popL_LH":0, "proc2.local_popL_LHR":0, "proc2.local_popL_RH":0, "proc2.local_popR_LH":0, "proc2.local_popR_RH":0, "proc2.local_popR_RHL":0, "proc2.local_pushR_LH":0, "proc2.local_pushR_RH":0, "proc2.local_pushR_RHR":0} {"AllNodes_0_0":0, "AllNodes_0_1":0, "AllNodes_1_0":0, "AllNodes_1_1":0, "AllNodes_2_0":2, "AllNodes_2_1":2, "First_ProcID":1, "LeftHat":1, "RightHat":1, "Second_ProcID":2, "newnode":2, "popRightFAIL":false, "proc1.line":2, "proc1.local_popL_LH":0, "proc1.local_popL_LHR":0, "proc1.local_popL_RH":0, "proc1.local_popR_LH":0, "proc1.local_popR_RH":1, "proc1.local_popR_RHL":0, "proc1.local_pushR_LH":0, "proc1.local_pushR_RH":0, "proc1.local_pushR_RHR":0, "proc2.line":11, "proc2.local_popL_LH":0, "proc2.local_popL_LHR":0, "proc2.local_popL_RH":0, "proc2.local_popR_LH":0, "proc2.local_popR_RH":0, "proc2.local_popR_RHL":0, "proc2.local_pushR_LH":0, "proc2.local_pushR_RH":0, "proc2.local_pushR_RHR":0} {"AllNodes_0_0":0, "AllNodes_0_1":0, "AllNodes_1_0":0, "AllNodes_1_1":0, "AllNodes_2_0":2, "AllNodes_2_1":0, "First_ProcID":1, "LeftHat":1, "RightHat":1, "Second_ProcID":2, "newnode":2, "popRightFAIL":false, "proc1.line":2, "proc1.local_popL_LH":0, "proc1.local_popL_LHR":0, "proc1.local_popL_RH":0, "proc1.local_popR_LH":1, "proc1.local_popR_RH":1, "proc1.local_popR_RHL":0, "proc1.local_pushR_LH":0, "proc1.local_pushR_RH":0, "proc1.local_pushR_RHR":0, "proc2.line":12, "proc2.local_popL_LH":0, "proc2.local_popL_LHR":0, "proc2.local_popL_RH":0, "proc2.local_popR_LH":0, "proc2.local_popR_RH":0, "proc2.local_popR_RHL":0, "proc2.local_pushR_LH":0, "proc2.local_pushR_RH":0, "proc2.local_pushR_RHR":0} {"AllNodes_0_0":0, "AllNodes_0_1":0, "AllNodes_1_0":0, "AllNodes_1_1":0, "AllNodes_2_0":2, "AllNodes_2_1":0, "First_ProcID":1, "LeftHat":1, "RightHat":1, "Second_ProcID":2, "newnode":2, "popRightFAIL":false, "proc1.line":2, "proc1.local_popL_LH":0, "proc1.local_popL_LHR":0, "proc1.local_popL_RH":0, "proc1.local_popR_LH":1, "proc1.local_popR_RH":1, "proc1.local_popR_RHL":0, "proc1.local_pushR_LH":0, "proc1.local_pushR_RH":0, "proc1.local_pushR_RHR":0, "proc2.line":13, "proc2.local_popL_LH":0, "proc2.local_popL_LHR":0, "proc2.local_popL_RH":0, "proc2.local_popR_LH":0, "proc2.local_popR_RH":0, "proc2.local_popR_RHL":0, "proc2.local_pushR_LH":0, "proc2.local_pushR_RH":1, "proc2.local_pushR_RHR":0} {"AllNodes_0_0":0, "AllNodes_0_1":0, "AllNodes_1_0":0, "AllNodes_1_1":0, "AllNodes_2_0":2, "AllNodes_2_1":0, "First_ProcID":1, "LeftHat":1, "RightHat":1, "Second_ProcID":2, "newnode":2, "popRightFAIL":false, "proc1.line":2, "proc1.local_popL_LH":0, "proc1.local_popL_LHR":0, "proc1.local_popL_RH":0, "proc1.local_popR_LH":1, "proc1.local_popR_RH":1, "proc1.local_popR_RHL":0, "proc1.local_pushR_LH":0, "proc1.local_pushR_RH":0, "proc1.local_pushR_RHR":0, "proc2.line":16, "proc2.local_popL_LH":0, "proc2.local_popL_LHR":0, "proc2.local_popL_RH":0, "proc2.local_popR_LH":0, "proc2.local_popR_RH":0, "proc2.local_popR_RHL":0, "proc2.local_pushR_LH":0, "proc2.local_pushR_RH":1, "proc2.local_pushR_RHR":0} {"AllNodes_0_0":0, "AllNodes_0_1":0, "AllNodes_1_0":0, "AllNodes_1_1":0, "AllNodes_2_0":1, "AllNodes_2_1":0, "First_ProcID":1, "LeftHat":1, "RightHat":2, "Second_ProcID":2, "newnode":2, "popRightFAIL":false, "proc1.line":2, "proc1.local_popL_LH":0, "proc1.local_popL_LHR":0, "proc1.local_popL_RH":0, "proc1.local_popR_LH":1, "proc1.local_popR_RH":1, "proc1.local_popR_RHL":0, "proc1.local_pushR_LH":0, "proc1.local_pushR_RH":0, "proc1.local_pushR_RHR":0, "proc2.line":17, "proc2.local_popL_LH":0, "proc2.local_popL_LHR":0, "proc2.local_popL_RH":0, "proc2.local_popR_LH":0, "proc2.local_popR_RH":0, "proc2.local_popR_RHL":0, "proc2.local_pushR_LH":0, "proc2.local_pushR_RH":1, "proc2.local_pushR_RHR":0} {"AllNodes_0_0":0, "AllNodes_0_1":0, "AllNodes_1_0":0, "AllNodes_1_1":2, "AllNodes_2_0":1, "AllNodes_2_1":0, "First_ProcID":1, "LeftHat":1, "RightHat":2, "Second_ProcID":2, "newnode":2, "popRightFAIL":false, "proc1.line":2, "proc1.local_popL_LH":0, "proc1.local_popL_LHR":0, "proc1.local_popL_RH":0, "proc1.local_popR_LH":1, "proc1.local_popR_RH":1, "proc1.local_popR_RHL":0, "proc1.local_pushR_LH":0, "proc1.local_pushR_RH":0, "proc1.local_pushR_RHR":0, "proc2.line":18, "proc2.local_popL_LH":0, "proc2.local_popL_LHR":0, "proc2.local_popL_RH":0, "proc2.local_popR_LH":0, "proc2.local_popR_RH":0, "proc2.local_popR_RHL":0, "proc2.local_pushR_LH":0, "proc2.local_pushR_RH":1, "proc2.local_pushR_RHR":0} {"AllNodes_0_0":0, "AllNodes_0_1":0, "AllNodes_1_0":0, "AllNodes_1_1":2, "AllNodes_2_0":1, "AllNodes_2_1":0, "First_ProcID":1, "LeftHat":1, "RightHat":2, "Second_ProcID":2, "newnode":2, "popRightFAIL":false, "proc1.line":2, "proc1.local_popL_LH":0, "proc1.local_popL_LHR":0, "proc1.local_popL_RH":0, "proc1.local_popR_LH":1, "proc1.local_popR_RH":1, "proc1.local_popR_RHL":0, "proc1.local_pushR_LH":0, "proc1.local_pushR_RH":0, "proc1.local_pushR_RHR":0, "proc2.line":0, "proc2.local_popL_LH":0, "proc2.local_popL_LHR":0, "proc2.local_popL_RH":0, "proc2.local_popR_LH":0, "proc2.local_popR_RH":0, "proc2.local_popR_RHL":0, "proc2.local_pushR_LH":0, "proc2.local_pushR_RH":1, "proc2.local_pushR_RHR":0} {"AllNodes_0_0":0, "AllNodes_0_1":0, "AllNodes_1_0":0, "AllNodes_1_1":2, "AllNodes_2_0":1, "AllNodes_2_1":0, "First_ProcID":1, "LeftHat":1, "RightHat":2, "Second_ProcID":2, "newnode":2, "popRightFAIL":false, "proc1.line":2, "proc1.local_popL_LH":0, "proc1.local_popL_LHR":0, "proc1.local_popL_RH":0, "proc1.local_popR_LH":1, "proc1.local_popR_RH":1, "proc1.local_popR_RHL":0, "proc1.local_pushR_LH":0, "proc1.local_pushR_RH":0, "proc1.local_pushR_RHR":0, "proc2.line":19, "proc2.local_popL_LH":0, "proc2.local_popL_LHR":0, "proc2.local_popL_RH":0, "proc2.local_popR_LH":0, "proc2.local_popR_RH":0, "proc2.local_popR_RHL":0, "proc2.local_pushR_LH":0, "proc2.local_pushR_RH":1, "proc2.local_pushR_RHR":0} {"AllNodes_0_0":0, "AllNodes_0_1":0, "AllNodes_1_0":0, "AllNodes_1_1":2, "AllNodes_2_0":1, "AllNodes_2_1":0, "First_ProcID":1, "LeftHat":1, "RightHat":2, "Second_ProcID":2, "newnode":2, "popRightFAIL":false, "proc1.line":2, "proc1.local_popL_LH":0, "proc1.local_popL_LHR":0, "proc1.local_popL_RH":0, "proc1.local_popR_LH":1, "proc1.local_popR_RH":1, "proc1.local_popR_RHL":0, "proc1.local_pushR_LH":0, "proc1.local_pushR_RH":0, "proc1.local_pushR_RHR":0, "proc2.line":20, "proc2.local_popL_LH":1, "proc2.local_popL_LHR":0, "proc2.local_popL_RH":2, "proc2.local_popR_LH":0, "proc2.local_popR_RH":0, "proc2.local_popR_RHL":0, "proc2.local_pushR_LH":0, "proc2.local_pushR_RH":1, "proc2.local_pushR_RHR":0} {"AllNodes_0_0":0, "AllNodes_0_1":0, "AllNodes_1_0":0, "AllNodes_1_1":2, "AllNodes_2_0":1, "AllNodes_2_1":0, "First_ProcID":1, "LeftHat":1, "RightHat":2, "Second_ProcID":2, "newnode":2, "popRightFAIL":false, "proc1.line":2, "proc1.local_popL_LH":0, "proc1.local_popL_LHR":0, "proc1.local_popL_RH":0, "proc1.local_popR_LH":1, "proc1.local_popR_RH":1, "proc1.local_popR_RHL":0, "proc1.local_pushR_LH":0, "proc1.local_pushR_RH":0, "proc1.local_pushR_RHR":0, "proc2.line":21, "proc2.local_popL_LH":1, "proc2.local_popL_LHR":0, "proc2.local_popL_RH":2, "proc2.local_popR_LH":0, "proc2.local_popR_RH":0, "proc2.local_popR_RHL":0, "proc2.local_pushR_LH":0, "proc2.local_pushR_RH":1, "proc2.local_pushR_RHR":0} {"AllNodes_0_0":0, "AllNodes_0_1":0, "AllNodes_1_0":0, "AllNodes_1_1":2, "AllNodes_2_0":1, "AllNodes_2_1":0, "First_ProcID":1, "LeftHat":1, "RightHat":2, "Second_ProcID":2, "newnode":2, "popRightFAIL":false, "proc1.line":2, "proc1.local_popL_LH":0, "proc1.local_popL_LHR":0, "proc1.local_popL_RH":0, "proc1.local_popR_LH":1, "proc1.local_popR_RH":1, "proc1.local_popR_RHL":0, "proc1.local_pushR_LH":0, "proc1.local_pushR_RH":0, "proc1.local_pushR_RHR":0, "proc2.line":24, "proc2.local_popL_LH":1, "proc2.local_popL_LHR":0, "proc2.local_popL_RH":2, "proc2.local_popR_LH":0, "proc2.local_popR_RH":0, "proc2.local_popR_RHL":0, "proc2.local_pushR_LH":0, "proc2.local_pushR_RH":1, "proc2.local_pushR_RHR":0} {"AllNodes_0_0":0, "AllNodes_0_1":0, "AllNodes_1_0":0, "AllNodes_1_1":2, "AllNodes_2_0":1, "AllNodes_2_1":0, "First_ProcID":1, "LeftHat":1, "RightHat":2, "Second_ProcID":2, "newnode":2, "popRightFAIL":false, "proc1.line":2, "proc1.local_popL_LH":0, "proc1.local_popL_LHR":0, "proc1.local_popL_RH":0, "proc1.local_popR_LH":1, "proc1.local_popR_RH":1, "proc1.local_popR_RHL":0, "proc1.local_pushR_LH":0, "proc1.local_pushR_RH":0, "proc1.local_pushR_RHR":0, "proc2.line":25, "proc2.local_popL_LH":1, "proc2.local_popL_LHR":2, "proc2.local_popL_RH":2, "proc2.local_popR_LH":0, "proc2.local_popR_RH":0, "proc2.local_popR_RHL":0, "proc2.local_pushR_LH":0, "proc2.local_pushR_RH":1, "proc2.local_pushR_RHR":0} {"AllNodes_0_0":0, "AllNodes_0_1":0, "AllNodes_1_0":0, "AllNodes_1_1":1, "AllNodes_2_0":1, "AllNodes_2_1":0, "First_ProcID":1, "LeftHat":2, "RightHat":2, "Second_ProcID":2, "newnode":2, "popRightFAIL":false, "proc1.line":2, "proc1.local_popL_LH":0, "proc1.local_popL_LHR":0, "proc1.local_popL_RH":0, "proc1.local_popR_LH":1, "proc1.local_popR_RH":1, "proc1.local_popR_RHL":0, "proc1.local_pushR_LH":0, "proc1.local_pushR_RH":0, "proc1.local_pushR_RHR":0, "proc2.line":26, "proc2.local_popL_LH":1, "proc2.local_popL_LHR":2, "proc2.local_popL_RH":2, "proc2.local_popR_LH":0, "proc2.local_popR_RH":0, "proc2.local_popR_RHL":0, "proc2.local_pushR_LH":0, "proc2.local_pushR_RH":1, "proc2.local_pushR_RHR":0} {"AllNodes_0_0":0, "AllNodes_0_1":0, "AllNodes_1_0":0, "AllNodes_1_1":1, "AllNodes_2_0":1, "AllNodes_2_1":0, "First_ProcID":1, "LeftHat":2, "RightHat":2, "Second_ProcID":2, "newnode":2, "popRightFAIL":false, "proc1.line":2, "proc1.local_popL_LH":0, "proc1.local_popL_LHR":0, "proc1.local_popL_RH":0, "proc1.local_popR_LH":2, "proc1.local_popR_RH":1, "proc1.local_popR_RHL":0, "proc1.local_pushR_LH":0, "proc1.local_pushR_RH":0, "proc1.local_pushR_RHR":0, "proc2.line":0, "proc2.local_popL_LH":1, "proc2.local_popL_LHR":2, "proc2.local_popL_RH":2, "proc2.local_popR_LH":0, "proc2.local_popR_RH":0, "proc2.local_popR_RHL":0, "proc2.local_pushR_LH":0, "proc2.local_pushR_RH":1, "proc2.local_pushR_RHR":0} {"AllNodes_0_0":0, "AllNodes_0_1":0, "AllNodes_1_0":0, "AllNodes_1_1":1, "AllNodes_2_0":1, "AllNodes_2_1":0, "First_ProcID":1, "LeftHat":2, "RightHat":2, "Second_ProcID":2, "newnode":2, "popRightFAIL":false, "proc1.line":2, "proc1.local_popL_LH":0, "proc1.local_popL_LHR":0, "proc1.local_popL_RH":0, "proc1.local_popR_LH":2, "proc1.local_popR_RH":1, "proc1.local_popR_RHL":0, "proc1.local_pushR_LH":0, "proc1.local_pushR_RH":0, "proc1.local_pushR_RHR":0, "proc2.line":19, "proc2.local_popL_LH":1, "proc2.local_popL_LHR":2, "proc2.local_popL_RH":2, "proc2.local_popR_LH":0, "proc2.local_popR_RH":0, "proc2.local_popR_RHL":0, "proc2.local_pushR_LH":0, "proc2.local_pushR_RH":1, "proc2.local_pushR_RHR":0} {"AllNodes_0_0":0, "AllNodes_0_1":0, "AllNodes_1_0":0, "AllNodes_1_1":1, "AllNodes_2_0":1, "AllNodes_2_1":0, "First_ProcID":1, "LeftHat":2, "RightHat":2, "Second_ProcID":2, "newnode":2, "popRightFAIL":false, "proc1.line":2, "proc1.local_popL_LH":0, "proc1.local_popL_LHR":0, "proc1.local_popL_RH":0, "proc1.local_popR_LH":2, "proc1.local_popR_RH":1, "proc1.local_popR_RHL":0, "proc1.local_pushR_LH":0, "proc1.local_pushR_RH":0, "proc1.local_pushR_RHR":0, "proc2.line":20, "proc2.local_popL_LH":2, "proc2.local_popL_LHR":2, "proc2.local_popL_RH":2, "proc2.local_popR_LH":0, "proc2.local_popR_RH":0, "proc2.local_popR_RHL":0, "proc2.local_pushR_LH":0, "proc2.local_pushR_RH":1, "proc2.local_pushR_RHR":0} {"AllNodes_0_0":0, "AllNodes_0_1":0, "AllNodes_1_0":0, "AllNodes_1_1":1, "AllNodes_2_0":1, "AllNodes_2_1":0, "First_ProcID":1, "LeftHat":2, "RightHat":2, "Second_ProcID":2, "newnode":2, "popRightFAIL":false, "proc1.line":2, "proc1.local_popL_LH":0, "proc1.local_popL_LHR":0, "proc1.local_popL_RH":0, "proc1.local_popR_LH":2, "proc1.local_popR_RH":1, "proc1.local_popR_RHL":0, "proc1.local_pushR_LH":0, "proc1.local_pushR_RH":0, "proc1.local_pushR_RHR":0, "proc2.line":21, "proc2.local_popL_LH":2, "proc2.local_popL_LHR":2, "proc2.local_popL_RH":2, "proc2.local_popR_LH":0, "proc2.local_popR_RH":0, "proc2.local_popR_RHL":0, "proc2.local_pushR_LH":0, "proc2.local_pushR_RH":1, "proc2.local_pushR_RHR":0} {"AllNodes_0_0":0, "AllNodes_0_1":0, "AllNodes_1_0":0, "AllNodes_1_1":1, "AllNodes_2_0":1, "AllNodes_2_1":0, "First_ProcID":1, "LeftHat":2, "RightHat":2, "Second_ProcID":2, "newnode":2, "popRightFAIL":false, "proc1.line":2, "proc1.local_popL_LH":0, "proc1.local_popL_LHR":0, "proc1.local_popL_RH":0, "proc1.local_popR_LH":2, "proc1.local_popR_RH":1, "proc1.local_popR_RHL":0, "proc1.local_pushR_LH":0, "proc1.local_pushR_RH":0, "proc1.local_pushR_RHR":0, "proc2.line":22, "proc2.local_popL_LH":2, "proc2.local_popL_LHR":2, "proc2.local_popL_RH":2, "proc2.local_popR_LH":0, "proc2.local_popR_RH":0, "proc2.local_popR_RHL":0, "proc2.local_pushR_LH":0, "proc2.local_pushR_RH":1, "proc2.local_pushR_RHR":0} {"AllNodes_0_0":0, "AllNodes_0_1":0, "AllNodes_1_0":0, "AllNodes_1_1":1, "AllNodes_2_0":1, "AllNodes_2_1":0, "First_ProcID":1, "LeftHat":0, "RightHat":0, "Second_ProcID":2, "newnode":2, "popRightFAIL":false, "proc1.line":2, "proc1.local_popL_LH":0, "proc1.local_popL_LHR":0, "proc1.local_popL_RH":0, "proc1.local_popR_LH":2, "proc1.local_popR_RH":1, "proc1.local_popR_RHL":0, "proc1.local_pushR_LH":0, "proc1.local_pushR_RH":0, "proc1.local_pushR_RHR":0, "proc2.line":23, "proc2.local_popL_LH":2, "proc2.local_popL_LHR":2, "proc2.local_popL_RH":2, "proc2.local_popR_LH":0, "proc2.local_popR_RH":0, "proc2.local_popR_RHL":0, "proc2.local_pushR_LH":0, "proc2.local_pushR_RH":1, "proc2.local_pushR_RHR":0} {"AllNodes_0_0":0, "AllNodes_0_1":0, "AllNodes_1_0":0, "AllNodes_1_1":1, "AllNodes_2_0":1, "AllNodes_2_1":0, "First_ProcID":1, "LeftHat":0, "RightHat":0, "Second_ProcID":2, "newnode":2, "popRightFAIL":false, "proc1.line":2, "proc1.local_popL_LH":0, "proc1.local_popL_LHR":0, "proc1.local_popL_RH":0, "proc1.local_popR_LH":0, "proc1.local_popR_RH":1, "proc1.local_popR_RHL":0, "proc1.local_pushR_LH":0, "proc1.local_pushR_RH":0, "proc1.local_pushR_RHR":0, "proc2.line":0, "proc2.local_popL_LH":2, "proc2.local_popL_LHR":2, "proc2.local_popL_RH":2, "proc2.local_popR_LH":0, "proc2.local_popR_RH":0, "proc2.local_popR_RHL":0, "proc2.local_pushR_LH":0, "proc2.local_pushR_RH":1, "proc2.local_pushR_RHR":0} {"AllNodes_0_0":0, "AllNodes_0_1":0, "AllNodes_1_0":0, "AllNodes_1_1":1, "AllNodes_2_0":1, "AllNodes_2_1":0, "First_ProcID":1, "LeftHat":0, "RightHat":0, "Second_ProcID":2, "newnode":2, "popRightFAIL":false, "proc1.line":2, "proc1.local_popL_LH":0, "proc1.local_popL_LHR":0, "proc1.local_popL_RH":0, "proc1.local_popR_LH":0, "proc1.local_popR_RH":1, "proc1.local_popR_RHL":0, "proc1.local_pushR_LH":0, "proc1.local_pushR_RH":0, "proc1.local_pushR_RHR":0, "proc2.line":10, "proc2.local_popL_LH":2, "proc2.local_popL_LHR":2, "proc2.local_popL_RH":2, "proc2.local_popR_LH":0, "proc2.local_popR_RH":0, "proc2.local_popR_RHL":0, "proc2.local_pushR_LH":0, "proc2.local_pushR_RH":1, "proc2.local_pushR_RHR":0} {"AllNodes_0_0":0, "AllNodes_0_1":0, "AllNodes_1_0":0, "AllNodes_1_1":1, "AllNodes_2_0":1, "AllNodes_2_1":0, "First_ProcID":1, "LeftHat":0, "RightHat":0, "Second_ProcID":2, "newnode":2, "popRightFAIL":false, "proc1.line":2, "proc1.local_popL_LH":0, "proc1.local_popL_LHR":0, "proc1.local_popL_RH":0, "proc1.local_popR_LH":0, "proc1.local_popR_RH":1, "proc1.local_popR_RHL":0, "proc1.local_pushR_LH":0, "proc1.local_pushR_RH":0, "proc1.local_pushR_RHR":0, "proc2.line":11, "proc2.local_popL_LH":2, "proc2.local_popL_LHR":2, "proc2.local_popL_RH":2, "proc2.local_popR_LH":0, "proc2.local_popR_RH":0, "proc2.local_popR_RHL":0, "proc2.local_pushR_LH":0, "proc2.local_pushR_RH":1, "proc2.local_pushR_RHR":0} {"AllNodes_0_0":0, "AllNodes_0_1":0, "AllNodes_1_0":0, "AllNodes_1_1":1, "AllNodes_2_0":1, "AllNodes_2_1":0, "First_ProcID":1, "LeftHat":0, "RightHat":0, "Second_ProcID":2, "newnode":2, "popRightFAIL":false, "proc1.line":2, "proc1.local_popL_LH":0, "proc1.local_popL_LHR":0, "proc1.local_popL_RH":0, "proc1.local_popR_LH":0, "proc1.local_popR_RH":1, "proc1.local_popR_RHL":0, "proc1.local_pushR_LH":0, "proc1.local_pushR_RH":0, "proc1.local_pushR_RHR":0, "proc2.line":12, "proc2.local_popL_LH":2, "proc2.local_popL_LHR":2, "proc2.local_popL_RH":2, "proc2.local_popR_LH":0, "proc2.local_popR_RH":0, "proc2.local_popR_RHL":0, "proc2.local_pushR_LH":0, "proc2.local_pushR_RH":1, "proc2.local_pushR_RHR":0} {"AllNodes_0_0":0, "AllNodes_0_1":0, "AllNodes_1_0":0, "AllNodes_1_1":1, "AllNodes_2_0":1, "AllNodes_2_1":0, "First_ProcID":1, "LeftHat":0, "RightHat":0, "Second_ProcID":2, "newnode":2, "popRightFAIL":false, "proc1.line":2, "proc1.local_popL_LH":0, "proc1.local_popL_LHR":0, "proc1.local_popL_RH":0, "proc1.local_popR_LH":0, "proc1.local_popR_RH":1, "proc1.local_popR_RHL":0, "proc1.local_pushR_LH":0, "proc1.local_pushR_RH":0, "proc1.local_pushR_RHR":0, "proc2.line":13, "proc2.local_popL_LH":2, "proc2.local_popL_LHR":2, "proc2.local_popL_RH":2, "proc2.local_popR_LH":0, "proc2.local_popR_RH":0, "proc2.local_popR_RHL":0, "proc2.local_pushR_LH":0, "proc2.local_pushR_RH":0, "proc2.local_pushR_RHR":1} {"AllNodes_0_0":0, "AllNodes_0_1":0, "AllNodes_1_0":0, "AllNodes_1_1":1, "AllNodes_2_0":1, "AllNodes_2_1":0, "First_ProcID":1, "LeftHat":0, "RightHat":0, "Second_ProcID":2, "newnode":2, "popRightFAIL":false, "proc1.line":2, "proc1.local_popL_LH":0, "proc1.local_popL_LHR":0, "proc1.local_popL_RH":0, "proc1.local_popR_LH":0, "proc1.local_popR_RH":1, "proc1.local_popR_RHL":0, "proc1.local_pushR_LH":0, "proc1.local_pushR_RH":0, "proc1.local_pushR_RHR":0, "proc2.line":16, "proc2.local_popL_LH":2, "proc2.local_popL_LHR":2, "proc2.local_popL_RH":2, "proc2.local_popR_LH":0, "proc2.local_popR_RH":0, "proc2.local_popR_RHL":0, "proc2.local_pushR_LH":0, "proc2.local_pushR_RH":0, "proc2.local_pushR_RHR":1} {"AllNodes_0_0":0, "AllNodes_0_1":0, "AllNodes_1_0":0, "AllNodes_1_1":1, "AllNodes_2_0":1, "AllNodes_2_1":0, "First_ProcID":1, "LeftHat":0, "RightHat":2, "Second_ProcID":2, "newnode":2, "popRightFAIL":false, "proc1.line":2, "proc1.local_popL_LH":0, "proc1.local_popL_LHR":0, "proc1.local_popL_RH":0, "proc1.local_popR_LH":0, "proc1.local_popR_RH":1, "proc1.local_popR_RHL":0, "proc1.local_pushR_LH":0, "proc1.local_pushR_RH":0, "proc1.local_pushR_RHR":0, "proc2.line":17, "proc2.local_popL_LH":2, "proc2.local_popL_LHR":2, "proc2.local_popL_RH":2, "proc2.local_popR_LH":0, "proc2.local_popR_RH":0, "proc2.local_popR_RHL":0, "proc2.local_pushR_LH":0, "proc2.local_pushR_RH":0, "proc2.local_pushR_RHR":1} {"AllNodes_0_0":0, "AllNodes_0_1":0, "AllNodes_1_0":0, "AllNodes_1_1":1, "AllNodes_2_0":1, "AllNodes_2_1":0, "First_ProcID":1, "LeftHat":0, "RightHat":2, "Second_ProcID":2, "newnode":2, "popRightFAIL":false, "proc1.line":2, "proc1.local_popL_LH":0, "proc1.local_popL_LHR":0, "proc1.local_popL_RH":0, "proc1.local_popR_LH":0, "proc1.local_popR_RH":1, "proc1.local_popR_RHL":0, "proc1.local_pushR_LH":0, "proc1.local_pushR_RH":0, "proc1.local_pushR_RHR":0, "proc2.line":18, "proc2.local_popL_LH":2, "proc2.local_popL_LHR":2, "proc2.local_popL_RH":2, "proc2.local_popR_LH":0, "proc2.local_popR_RH":0, "proc2.local_popR_RHL":0, "proc2.local_pushR_LH":0, "proc2.local_pushR_RH":0, "proc2.local_pushR_RHR":1} {"AllNodes_0_0":0, "AllNodes_0_1":0, "AllNodes_1_0":0, "AllNodes_1_1":1, "AllNodes_2_0":1, "AllNodes_2_1":0, "First_ProcID":1, "LeftHat":0, "RightHat":2, "Second_ProcID":2, "newnode":2, "popRightFAIL":false, "proc1.line":2, "proc1.local_popL_LH":0, "proc1.local_popL_LHR":0, "proc1.local_popL_RH":0, "proc1.local_popR_LH":0, "proc1.local_popR_RH":1, "proc1.local_popR_RHL":0, "proc1.local_pushR_LH":0, "proc1.local_pushR_RH":0, "proc1.local_pushR_RHR":0, "proc2.line":0, "proc2.local_popL_LH":2, "proc2.local_popL_LHR":2, "proc2.local_popL_RH":2, "proc2.local_popR_LH":0, "proc2.local_popR_RH":0, "proc2.local_popR_RHL":0, "proc2.local_pushR_LH":0, "proc2.local_pushR_RH":0, "proc2.local_pushR_RHR":1} {"AllNodes_0_0":0, "AllNodes_0_1":0, "AllNodes_1_0":0, "AllNodes_1_1":1, "AllNodes_2_0":1, "AllNodes_2_1":0, "First_ProcID":1, "LeftHat":0, "RightHat":2, "Second_ProcID":2, "newnode":2, "popRightFAIL":false, "proc1.line":2, "proc1.local_popL_LH":0, "proc1.local_popL_LHR":0, "proc1.local_popL_RH":0, "proc1.local_popR_LH":0, "proc1.local_popR_RH":1, "proc1.local_popR_RHL":0, "proc1.local_pushR_LH":0, "proc1.local_pushR_RH":0, "proc1.local_pushR_RHR":0, "proc2.line":19, "proc2.local_popL_LH":2, "proc2.local_popL_LHR":2, "proc2.local_popL_RH":2, "proc2.local_popR_LH":0, "proc2.local_popR_RH":0, "proc2.local_popR_RHL":0, "proc2.local_pushR_LH":0, "proc2.local_pushR_RH":0, "proc2.local_pushR_RHR":1} {"AllNodes_0_0":0, "AllNodes_0_1":0, "AllNodes_1_0":0, "AllNodes_1_1":1, "AllNodes_2_0":1, "AllNodes_2_1":0, "First_ProcID":1, "LeftHat":0, "RightHat":2, "Second_ProcID":2, "newnode":2, "popRightFAIL":false, "proc1.line":2, "proc1.local_popL_LH":0, "proc1.local_popL_LHR":0, "proc1.local_popL_RH":0, "proc1.local_popR_LH":0, "proc1.local_popR_RH":1, "proc1.local_popR_RHL":0, "proc1.local_pushR_LH":0, "proc1.local_pushR_RH":0, "proc1.local_pushR_RHR":0, "proc2.line":20, "proc2.local_popL_LH":0, "proc2.local_popL_LHR":2, "proc2.local_popL_RH":2, "proc2.local_popR_LH":0, "proc2.local_popR_RH":0, "proc2.local_popR_RHL":0, "proc2.local_pushR_LH":0, "proc2.local_pushR_RH":0, "proc2.local_pushR_RHR":1} {"AllNodes_0_0":0, "AllNodes_0_1":0, "AllNodes_1_0":0, "AllNodes_1_1":1, "AllNodes_2_0":1, "AllNodes_2_1":0, "First_ProcID":1, "LeftHat":0, "RightHat":2, "Second_ProcID":2, "newnode":2, "popRightFAIL":false, "proc1.line":2, "proc1.local_popL_LH":0, "proc1.local_popL_LHR":0, "proc1.local_popL_RH":0, "proc1.local_popR_LH":0, "proc1.local_popR_RH":1, "proc1.local_popR_RHL":0, "proc1.local_pushR_LH":0, "proc1.local_pushR_RH":0, "proc1.local_pushR_RHR":0, "proc2.line":21, "proc2.local_popL_LH":0, "proc2.local_popL_LHR":2, "proc2.local_popL_RH":2, "proc2.local_popR_LH":0, "proc2.local_popR_RH":0, "proc2.local_popR_RHL":0, "proc2.local_pushR_LH":0, "proc2.local_pushR_RH":0, "proc2.local_pushR_RHR":1} {"AllNodes_0_0":0, "AllNodes_0_1":0, "AllNodes_1_0":0, "AllNodes_1_1":1, "AllNodes_2_0":1, "AllNodes_2_1":0, "First_ProcID":1, "LeftHat":0, "RightHat":2, "Second_ProcID":2, "newnode":2, "popRightFAIL":false, "proc1.line":2, "proc1.local_popL_LH":0, "proc1.local_popL_LHR":0, "proc1.local_popL_RH":0, "proc1.local_popR_LH":0, "proc1.local_popR_RH":1, "proc1.local_popR_RHL":0, "proc1.local_pushR_LH":0, "proc1.local_pushR_RH":0, "proc1.local_pushR_RHR":0, "proc2.line":24, "proc2.local_popL_LH":0, "proc2.local_popL_LHR":2, "proc2.local_popL_RH":2, "proc2.local_popR_LH":0, "proc2.local_popR_RH":0, "proc2.local_popR_RHL":0, "proc2.local_pushR_LH":0, "proc2.local_pushR_RH":0, "proc2.local_pushR_RHR":1} {"AllNodes_0_0":0, "AllNodes_0_1":0, "AllNodes_1_0":0, "AllNodes_1_1":1, "AllNodes_2_0":1, "AllNodes_2_1":0, "First_ProcID":1, "LeftHat":0, "RightHat":2, "Second_ProcID":2, "newnode":2, "popRightFAIL":false, "proc1.line":2, "proc1.local_popL_LH":0, "proc1.local_popL_LHR":0, "proc1.local_popL_RH":0, "proc1.local_popR_LH":0, "proc1.local_popR_RH":1, "proc1.local_popR_RHL":0, "proc1.local_pushR_LH":0, "proc1.local_pushR_RH":0, "proc1.local_pushR_RHR":0, "proc2.line":25, "proc2.local_popL_LH":0, "proc2.local_popL_LHR":0, "proc2.local_popL_RH":2, "proc2.local_popR_LH":0, "proc2.local_popR_RH":0, "proc2.local_popR_RHL":0, "proc2.local_pushR_LH":0, "proc2.local_pushR_RH":0, "proc2.local_pushR_RHR":1} {"AllNodes_0_0":0, "AllNodes_0_1":0, "AllNodes_1_0":0, "AllNodes_1_1":1, "AllNodes_2_0":1, "AllNodes_2_1":0, "First_ProcID":1, "LeftHat":0, "RightHat":2, "Second_ProcID":2, "newnode":2, "popRightFAIL":false, "proc1.line":2, "proc1.local_popL_LH":0, "proc1.local_popL_LHR":0, "proc1.local_popL_RH":0, "proc1.local_popR_LH":0, "proc1.local_popR_RH":1, "proc1.local_popR_RHL":0, "proc1.local_pushR_LH":0, "proc1.local_pushR_RH":0, "proc1.local_pushR_RHR":0, "proc2.line":26, "proc2.local_popL_LH":0, "proc2.local_popL_LHR":0, "proc2.local_popL_RH":2, "proc2.local_popR_LH":0, "proc2.local_popR_RH":0, "proc2.local_popR_RHL":0, "proc2.local_pushR_LH":0, "proc2.local_pushR_RH":0, "proc2.local_pushR_RHR":1} {"AllNodes_0_0":0, "AllNodes_0_1":0, "AllNodes_1_0":0, "AllNodes_1_1":1, "AllNodes_2_0":1, "AllNodes_2_1":0, "First_ProcID":1, "LeftHat":0, "RightHat":2, "Second_ProcID":2, "newnode":2, "popRightFAIL":false, "proc1.line":2, "proc1.local_popL_LH":0, "proc1.local_popL_LHR":0, "proc1.local_popL_RH":0, "proc1.local_popR_LH":0, "proc1.local_popR_RH":1, "proc1.local_popR_RHL":0, "proc1.local_pushR_LH":0, "proc1.local_pushR_RH":0, "proc1.local_pushR_RHR":0, "proc2.line":0, "proc2.local_popL_LH":0, "proc2.local_popL_LHR":0, "proc2.local_popL_RH":2, "proc2.local_popR_LH":0, "proc2.local_popR_RH":0, "proc2.local_popR_RHL":0, "proc2.local_pushR_LH":0, "proc2.local_pushR_RH":0, "proc2.local_pushR_RHR":1} {"AllNodes_0_0":0, "AllNodes_0_1":0, "AllNodes_1_0":0, "AllNodes_1_1":1, "AllNodes_2_0":1, "AllNodes_2_1":0, "First_ProcID":1, "LeftHat":0, "RightHat":2, "Second_ProcID":2, "newnode":2, "popRightFAIL":false, "proc1.line":2, "proc1.local_popL_LH":0, "proc1.local_popL_LHR":0, "proc1.local_popL_RH":0, "proc1.local_popR_LH":0, "proc1.local_popR_RH":1, "proc1.local_popR_RHL":0, "proc1.local_pushR_LH":0, "proc1.local_pushR_RH":0, "proc1.local_pushR_RHR":0, "proc2.line":19, "proc2.local_popL_LH":0, "proc2.local_popL_LHR":0, "proc2.local_popL_RH":2, "proc2.local_popR_LH":0, "proc2.local_popR_RH":0, "proc2.local_popR_RHL":0, "proc2.local_pushR_LH":0, "proc2.local_pushR_RH":0, "proc2.local_pushR_RHR":1} {"AllNodes_0_0":0, "AllNodes_0_1":0, "AllNodes_1_0":0, "AllNodes_1_1":1, "AllNodes_2_0":1, "AllNodes_2_1":0, "First_ProcID":1, "LeftHat":0, "RightHat":2, "Second_ProcID":2, "newnode":2, "popRightFAIL":false, "proc1.line":2, "proc1.local_popL_LH":0, "proc1.local_popL_LHR":0, "proc1.local_popL_RH":0, "proc1.local_popR_LH":0, "proc1.local_popR_RH":1, "proc1.local_popR_RHL":0, "proc1.local_pushR_LH":0, "proc1.local_pushR_RH":0, "proc1.local_pushR_RHR":0, "proc2.line":20, "proc2.local_popL_LH":0, "proc2.local_popL_LHR":0, "proc2.local_popL_RH":2, "proc2.local_popR_LH":0, "proc2.local_popR_RH":0, "proc2.local_popR_RHL":0, "proc2.local_pushR_LH":0, "proc2.local_pushR_RH":0, "proc2.local_pushR_RHR":1} {"AllNodes_0_0":0, "AllNodes_0_1":0, "AllNodes_1_0":0, "AllNodes_1_1":1, "AllNodes_2_0":1, "AllNodes_2_1":0, "First_ProcID":1, "LeftHat":0, "RightHat":2, "Second_ProcID":2, "newnode":2, "popRightFAIL":false, "proc1.line":2, "proc1.local_popL_LH":0, "proc1.local_popL_LHR":0, "proc1.local_popL_RH":0, "proc1.local_popR_LH":0, "proc1.local_popR_RH":1, "proc1.local_popR_RHL":0, "proc1.local_pushR_LH":0, "proc1.local_pushR_RH":0, "proc1.local_pushR_RHR":0, "proc2.line":21, "proc2.local_popL_LH":0, "proc2.local_popL_LHR":0, "proc2.local_popL_RH":2, "proc2.local_popR_LH":0, "proc2.local_popR_RH":0, "proc2.local_popR_RHL":0, "proc2.local_pushR_LH":0, "proc2.local_pushR_RH":0, "proc2.local_pushR_RHR":1}) ({"AllNodes_0_0":0, "AllNodes_0_1":0, "AllNodes_1_0":0, "AllNodes_1_1":1, "AllNodes_2_0":1, "AllNodes_2_1":0, "First_ProcID":1, "LeftHat":0, "RightHat":2, "Second_ProcID":2, "newnode":2, "popRightFAIL":false, "proc1.line":2, "proc1.local_popL_LH":0, "proc1.local_popL_LHR":0, "proc1.local_popL_RH":0, "proc1.local_popR_LH":0, "proc1.local_popR_RH":1, "proc1.local_popR_RHL":0, "proc1.local_pushR_LH":0, "proc1.local_pushR_RH":0, "proc1.local_pushR_RHR":0, "proc2.line":24, "proc2.local_popL_LH":0, "proc2.local_popL_LHR":0, "proc2.local_popL_RH":2, "proc2.local_popR_LH":0, "proc2.local_popR_RH":0, "proc2.local_popR_RHL":0, "proc2.local_pushR_LH":0, "proc2.local_pushR_RH":0, "proc2.local_pushR_RHR":1} {"AllNodes_0_0":0, "AllNodes_0_1":0, "AllNodes_1_0":0, "AllNodes_1_1":1, "AllNodes_2_0":1, "AllNodes_2_1":0, "First_ProcID":1, "LeftHat":0, "RightHat":2, "Second_ProcID":2, "newnode":2, "popRightFAIL":false, "proc1.line":2, "proc1.local_popL_LH":0, "proc1.local_popL_LHR":0, "proc1.local_popL_RH":0, "proc1.local_popR_LH":0, "proc1.local_popR_RH":1, "proc1.local_popR_RHL":0, "proc1.local_pushR_LH":0, "proc1.local_pushR_RH":0, "proc1.local_pushR_RHR":0, "proc2.line":25, "proc2.local_popL_LH":0, "proc2.local_popL_LHR":0, "proc2.local_popL_RH":2, "proc2.local_popR_LH":0, "proc2.local_popR_RH":0, "proc2.local_popR_RHL":0, "proc2.local_pushR_LH":0, "proc2.local_pushR_RH":0, "proc2.local_pushR_RHR":1} {"AllNodes_0_0":0, "AllNodes_0_1":0, "AllNodes_1_0":0, "AllNodes_1_1":1, "AllNodes_2_0":1, "AllNodes_2_1":0, "First_ProcID":1, "LeftHat":0, "RightHat":2, "Second_ProcID":2, "newnode":2, "popRightFAIL":false, "proc1.line":2, "proc1.local_popL_LH":0, "proc1.local_popL_LHR":0, "proc1.local_popL_RH":0, "proc1.local_popR_LH":0, "proc1.local_popR_RH":1, "proc1.local_popR_RHL":0, "proc1.local_pushR_LH":0, "proc1.local_pushR_RH":0, "proc1.local_pushR_RHR":0, "proc2.line":26, "proc2.local_popL_LH":0, "proc2.local_popL_LHR":0, "proc2.local_popL_RH":2, "proc2.local_popR_LH":0, "proc2.local_popR_RH":0, "proc2.local_popR_RHL":0, "proc2.local_pushR_LH":0, "proc2.local_pushR_RH":0, "proc2.local_pushR_RHR":1} {"AllNodes_0_0":0, "AllNodes_0_1":0, "AllNodes_1_0":0, "AllNodes_1_1":1, "AllNodes_2_0":1, "AllNodes_2_1":0, "First_ProcID":1, "LeftHat":0, "RightHat":2, "Second_ProcID":2, "newnode":2, "popRightFAIL":false, "proc1.line":2, "proc1.local_popL_LH":0, "proc1.local_popL_LHR":0, "proc1.local_popL_RH":0, "proc1.local_popR_LH":0, "proc1.local_popR_RH":1, "proc1.local_popR_RHL":0, "proc1.local_pushR_LH":0, "proc1.local_pushR_RH":0, "proc1.local_pushR_RHR":0, "proc2.line":0, "proc2.local_popL_LH":0, "proc2.local_popL_LHR":0, "proc2.local_popL_RH":2, "proc2.local_popR_LH":0, "proc2.local_popR_RH":0, "proc2.local_popR_RHL":0, "proc2.local_pushR_LH":0, "proc2.local_pushR_RH":0, "proc2.local_pushR_RHR":1} {"AllNodes_0_0":0, "AllNodes_0_1":0, "AllNodes_1_0":0, "AllNodes_1_1":1, "AllNodes_2_0":1, "AllNodes_2_1":0, "First_ProcID":1, "LeftHat":0, "RightHat":2, "Second_ProcID":2, "newnode":2, "popRightFAIL":false, "proc1.line":2, "proc1.local_popL_LH":0, "proc1.local_popL_LHR":0, "proc1.local_popL_RH":0, "proc1.local_popR_LH":0, "proc1.local_popR_RH":1, "proc1.local_popR_RHL":0, "proc1.local_pushR_LH":0, "proc1.local_pushR_RH":0, "proc1.local_pushR_RHR":0, "proc2.line":19, "proc2.local_popL_LH":0, "proc2.local_popL_LHR":0, "proc2.local_popL_RH":2, "proc2.local_popR_LH":0, "proc2.local_popR_RH":0, "proc2.local_popR_RHL":0, "proc2.local_pushR_LH":0, "proc2.local_pushR_RH":0, "proc2.local_pushR_RHR":1} {"AllNodes_0_0":0, "AllNodes_0_1":0, "AllNodes_1_0":0, "AllNodes_1_1":1, "AllNodes_2_0":1, "AllNodes_2_1":0, "First_ProcID":1, "LeftHat":0, "RightHat":2, "Second_ProcID":2, "newnode":2, "popRightFAIL":false, "proc1.line":2, "proc1.local_popL_LH":0, "proc1.local_popL_LHR":0, "proc1.local_popL_RH":0, "proc1.local_popR_LH":0, "proc1.local_popR_RH":1, "proc1.local_popR_RHL":0, "proc1.local_pushR_LH":0, "proc1.local_pushR_RH":0, "proc1.local_pushR_RHR":0, "proc2.line":20, "proc2.local_popL_LH":0, "proc2.local_popL_LHR":0, "proc2.local_popL_RH":2, "proc2.local_popR_LH":0, "proc2.local_popR_RH":0, "proc2.local_popR_RHL":0, "proc2.local_pushR_LH":0, "proc2.local_pushR_RH":0, "proc2.local_pushR_RHR":1} {"AllNodes_0_0":0, "AllNodes_0_1":0, "AllNodes_1_0":0, "AllNodes_1_1":1, "AllNodes_2_0":1, "AllNodes_2_1":0, "First_ProcID":1, "LeftHat":0, "RightHat":2, "Second_ProcID":2, "newnode":2, "popRightFAIL":false, "proc1.line":2, "proc1.local_popL_LH":0, "proc1.local_popL_LHR":0, "proc1.local_popL_RH":0, "proc1.local_popR_LH":0, "proc1.local_popR_RH":1, "proc1.local_popR_RHL":0, "proc1.local_pushR_LH":0, "proc1.local_pushR_RH":0, "proc1.local_pushR_RHR":0, "proc2.line":21, "proc2.local_popL_LH":0, "proc2.local_popL_LHR":0, "proc2.local_popL_RH":2, "proc2.local_popR_LH":0, "proc2.local_popR_RH":0, "proc2.local_popR_RHL":0, "proc2.local_pushR_LH":0, "proc2.local_pushR_RH":0, "proc2.local_pushR_RHR":1})
\end{MicroVerbatim}
\caption{Traces evidencing the SNARK bug for the $\text{SNARK}_{\text{lin}}$ benchmark provided by \tool{HyperQB} (top) and by \tool{AutoHyper} (bottom).}
\label{fig:low_snark}
\end{figure}

Within seconds, \tool{Alloy} identifies the SNARK bug and provides a minimal, 9-state relational graphical instance (Fig.~\ref{fig:snark}).
In comparison, the output from state-of-the-art tools (for the SNARK benchmark from \benchrev{}) places a significant cognitive burden on the user, as shown in Fig.~\ref{fig:low_snark}.\footnote{For clarity, the visualized instance contains a self-loop in the last state, which represents a valid trace under the model's stuttering semantics; the symbolic backend, however, produces only finite prefixes without loops.}
\tool{HyperQB}, for instance, directly reports the raw output of the \texttt{QuAbS} solver. This feedback consists of quantifier-indexed Boolean variables where truth values are indicated by the sign of the index. Because the tool converts all model integers to Booleans and duplicates every variable for each step of the $k$-unrolling, the resulting QBF-level witness is essentially inscrutable without extensive manual reverse-mapping.
Moving to \tool{AutoHyper}, it reports a trace at the level of SMV variables but does not guarantee minimal instances. For the same SNARK bug, it produces a sprawling 47-state witness containing a 7-state loop.
In contrast, the \tool{Alloy} visualizer abstracts away the low-level solver state, mapping results back to the original model context in a human-readable format. By ensuring both minimality\footnote{For the \tool{Sym} backend, any discovered witness is guaranteed to be of size $k$, where $k$ is the provided number of steps. Such a witness is inherently minimal, provided that the user follows an incremental search strategy by increasing $k$ sequentially until a solution is found.} and a high-level representation, our framework resolves the usability hindrances that previously made interpreting counter-examples in complex hyperproperties a challenging task.

\subsection{Quantitative evaluation}

We now report quantitative results addressing \textbf{RQ3} and \textbf{RQ4}, comparing the performance of \hpard{} and \hsmv{} against state-of-the-art HyperLTL model checkers.

\savebox{\oksym}{\usym{2714}}
\savebox{\noksym}{\usym{2717}}
\savebox{\inclsym}{\usym{2753}}
\savebox{\totsym}{\usym{1F551}}
\savebox{\oomsym}{\usym{2699}}
\newcommand{\ok}{\usebox{\oksym}}
\newcommand{\nok}{\usebox{\noksym}}
\newcommand{\incl}{\usebox{\inclsym}}
\newcommand{\tot}{\usebox{\totsym}}
\newcommand{\oom}{\usebox{\oomsym}}
\setlength{\tabcolsep}{.7pt}
\newcommand{\mytimes}{\hspace{-0.21em}\times\hspace{-0.21em}}

\begin{table*}
\centering
{\scriptsize
 \resizebox{\textwidth}{!}{
 \begin{tabular}{lccccc|cc|cc|cc|cc|cc|cc}
\toprule
\multirow{3}{*}{\bfseries Model (scope)} &
\multirow{3}{*}{\bfseries Q*} &
\multirow{3}{*}{\bfseries Res} &
\multirow{3}{*}{\bfseries solver} &
\multirow{3}{*}{\bfseries k} &
\multirow{3}{*}{\bfseries Sem} &
\multicolumn{4}{c|}{\tool{HyperPardinus}} &
\multicolumn{4}{c|}{\tool{HyperPardinus-}} &
\multicolumn{4}{c}{\tool{HyperPardinus}} \\
\cmidrule(lr){7-10}
\cmidrule(lr){11-14}
\cmidrule(lr){15-18}
&&&&&&
\multicolumn{2}{c|}{ \tool{AutoHyper}} &
\multicolumn{2}{c|}{ \tool{HyperQB}} &
\multicolumn{2}{c|}{ \tool{Exp}} &
\multicolumn{2}{c|}{ \tool{Sym}} &
\multicolumn{2}{c|}{ \tool{Exp}} &
\multicolumn{2}{c}{ \tool{Sym}} \\
\cmidrule(lr){1-6}
\cmidrule(lr){7-8}
\cmidrule(lr){9-10}
\cmidrule(lr){11-12}
\cmidrule(lr){13-14}
\cmidrule(lr){15-16}
\cmidrule(lr){17-18}
&&& {(exp)} &\multicolumn{2}{c|}{(sym)}&
\textbf{Size} & \textbf{t(s)} &
\textbf{Size} & \textbf{t(s)} &
\textbf{Size} & \textbf{t(s)} &
\textbf{Size} & \textbf{t(s)} &
\textbf{Size} & \textbf{t(s)} &
\textbf{Size} & \textbf{t(s)} \\
${\text{Mutant1}}_{\text{pk}} (3_Q)$ & $\boldsymbol{\exists}\forall$ & \ok{} & forklift & 6 & pes &
$21 \mytimes 21$ & 1.5 &
50KB &  2.2 &
$18 \mytimes 20$ & 1.0 &
87KB & 0.6 &
$18 \mytimes 20$ & 1.1 &
21KB & 0.5 \\
$\text{Mutant1}_{\text{dk}} (3_Q)$ & $\boldsymbol{\exists}\forall\forall$ & \ok{} & forklift &  6 & pes &
$21 \mytimes 82$ & 1.0 &
400KB &  ${\incl{}}$ &
$3 \mytimes 18 \mytimes 20$ & 1.1 &
141KB & 0.7 &
$3 \mytimes 30$ & 1.0 &
37KB & 0.6 \\
$\text{Mutant2}_{\text{pk}} (3_Q)$ & $\boldsymbol{\exists}\forall$ & \ok{} & forklift & 6 & pes &
$21 \mytimes 21$ & 1.0 &
50KB &  1.0 &
$21 \mytimes 20$ & 1.0 &
139KB & 0.6 &
$21 \mytimes 20$& 1.0 &
22KB & 0.6 \\
$\text{Mutant2}_{\text{dk}} (3_Q)$ & $\exists\forall\forall$ & \nok{} & forq & 6 & pes &
$21 \mytimes 105$ & 1.0 &
550KB & ${\incl{}}$ &
$3 \mytimes 21 \mytimes 20$ & 0.9 &
142KB & ${\incl{}}$ &
$3 \mytimes 45$& 0.9 &
40KB & ${\incl{}}$ \\
$\text{Mutant3}_{\text{pk}} (3_Q)$ & $\exists\forall$ & \nok{} & forq &  6 & pes &
$15 \mytimes 21$ & 0.9 &
50KB &  ${\incl{}}$ &
$14 \mytimes 20$ & 0.9 &
88KB & ${\incl{}}$ &
$14 \mytimes 20$ & 0.9 &
21KB & ${\incl{}}$ \\
$\text{Mutant3}_{\text{dk}} (3_Q)$ & $\exists\forall\forall$ & \nok{} & forq & 6 & pes &
$21 \mytimes 45$ & 1.0 &
200KB &  ${\incl{}}$ &
$3 \mytimes 14 \mytimes 20$ & 1.0 &
142KB & ${\incl{}}$ &
$3 \mytimes 21$ & 0.9 &
36KB & ${\incl{}}$ \\
$\text{Robot}_\text{two} (10_P)$ & $\boldsymbol{\exists\exists}$ & \ok{} & forklift & 20 & pes &
$310 \mytimes 310$ & 128.9 &
5MB &  3.0 &
$310 \mytimes 310$ & 45.1 &
7MB & 8.3 &
$310 \mytimes 310$ & 18.3 &
4MB & 4.8 \\
$\text{Robot}_\text{short} (10_P)$ & $\boldsymbol{\exists}\forall$ & \ok{} & forklift & 20 & opt &
$71 \mytimes 71$ & 2.2 &
750KB &  ${\incl{}}$ &
$71 \mytimes 71$ & 42.3 &
7MB & ${\incl{}}$ &
$71 \mytimes 71$ & 1.7 &
4MB & ${\incl{}}$ \\
$\text{Robot}_\text{score} (10_P \mytimes 6_Y)$ & $\boldsymbol{\exists}\forall$ & \ok{} & forklift &  20 & opt &
$570 \mytimes 570$ & 4.9 &
10MB &  ${\incl{}}$ &
\oom{} & \tot{} &
7MB & ${\incl{}}$ &
$570 \mytimes 570$ & 3.9 &
4MB & ${\incl{}}$ \\
$\text{Robot}_\text{robust} (10_P)$ & $\boldsymbol{\exists}\forall$ & \ok{} & forklift & 20 & opt &
$310 \mytimes 293$ & \tot{} &
5MB &  \tot{} &
$310 \mytimes 293$ & 25.0 &
6MB & ${\incl{}}$ &
$310 \mytimes 293$ & 2.4 &
4MB & ${\incl{}}$ \\
$\text{Robot}_\text{adv} (10_P)$ & $\boldsymbol{\exists}\forall$ & \ok{} & forklift &  20 & opt &
\oom{} & \tot{} &
7MB &  ${\incl{}}$ &
$100 \mytimes 36$ & \tot{} &
10MB & ${\incl{}}$ &
$100 \mytimes 36$ & 3.4 &
6MB & ${\incl{}}$ \\
$\text{CMSany}_{\text{NI}} (2_R \mytimes 2_A)$ & $\boldsymbol{\forall\forall}$ & \nok{} & forklift & 4 & opt &
7202 & 6.8 &
\oom{} & \tot{} &
$868 \mytimes 868$ & \tot{} &
192KB & ${\incl{}}$ &
67 & 1.8 &
114KB & 0.7 \\
$\text{CMSany}^{}_{\text{NI}} (3_R \mytimes 2_A)$ & $\boldsymbol{\forall\forall}$ & \nok{} & forklift & 6 & opt &
\oom{} & {\tot{}} &
\oom{} &  \tot{} &
$23836 \mytimes 23836$ & \tot{} &
322KB & ${\incl{}}$ &
\oom{} & \tot{} &
279KB & 0.9 \\
$\text{CMSany}^{}_{\text{GNI}} (2_R \mytimes 2_A)$ & $\boldsymbol{\forall\forall}\exists$ & \nok{} & forklift & 4 & opt &
$55037 \mytimes 868$ & \tot{} &
{\oom{}} &  {\tot{}} &
$355 \mytimes 77 \mytimes 646$ & {\tot{}} &
234KB & ${\incl{}}$ &
$689 \mytimes 433$ & 8.2 &
171KB & 0.8 \\
$\text{CMSany}^{}_{\text{GNI}} (3_R \mytimes 2_A)$ & $\boldsymbol{\forall\forall}\exists$ & \nok{} & forklift & 5 & opt &
\oom{} & {\tot{}} &
{\oom{}} &  {\tot{}} &
$1417 \mytimes 1712 \mytimes 11842$ & {\tot{}} &
392KB & ${\incl{}}$ &
\oom{} & {\tot{}} &
443KB & 1.0 \\
$\text{CMSndet}^{}_{\text{NI}} (2_R \mytimes 2_A)$ & $\boldsymbol{\forall\forall}$ & \nok{} & forklift &  6 & opt &
2450 & 1.6 &
{\oom{}} &  {\tot{}} &
$532 \mytimes 532$ & {\tot{}} &
192KB & ${\incl{}}$ &
63 & 1.3 &
172KB & 0.7 \\
$\text{CMSndet}^{}_{\text{NI}} (3_R \mytimes 2_A)$ & $\boldsymbol{\forall\forall}$ & \nok{} & forklift &  6 & opt &
356968 & 131.0 &
{\oom{}} & {\tot{}} &
$16036 \mytimes 16036$ & {\tot{}} &
265KB & ${\incl{}}$ &
160 & 115.0 &
268KB & 0.9 \\
$\text{CMSndet}^{}_{\text{GNI}} (2_R \mytimes 2_A)$ & $\forall\forall\exists$ & \ok{} & forq & 4 & opt &
$17957 \mytimes 532$ & \tot{} &
{\oom{}} & {\tot{}} &
$217 \mytimes 77 \mytimes 490$ & {\tot{}} &
236KB & ${\incl{}}$ &
$1631 \mytimes 490$ & 6.0 &
161KB & ${\incl{}}$ \\
$\text{CMSndet}^{}_{\text{GNI}} (3_R \mytimes 2_A)$ & $\forall\forall\exists$ & \ok{} & forq & 5 & opt &
\oom{} & {\tot{}} &
{\oom{}} &  {\tot{}} &
$16036 \mytimes 16036 \mytimes 16036$ & {\tot{}} &
356KB & ${\incl{}}$ &
\oom{} & {\tot{}} &
446KB & ${\incl{}}$ \\
$\text{CMSmax}^{}_{\text{NI}} (2_R \mytimes 2_A)$ & $\forall\forall$ & \ok{} & forq & 6 & opt &
1568 & 1.7 &
{\oom{}} &  {\tot{}} &
$448 \mytimes 448$ & {\tot{}} &
210KB & ${\incl{}}$ &
36 & 1.2 &
197KB & ${\incl{}}$ \\
$\text{CMSmax}^{}_{\text{NI}} (3_R \mytimes 2_A)$ & $\forall\forall$ & \ok{} & forq & 6 & opt &
127618 & 65.0 &
{\oom{}} &  {\tot{}} &
$10648 \mytimes 10648$ & {\tot{}} &
308KB & ${\incl{}}$ &
80 & 61.7 &
282KB & ${\incl{}}$ \\
$\text{CMSmax}^{}_{\text{GNI}} (2_R \mytimes 2_A)$ & $\forall\forall\exists$ & \ok{} & forq & 4 & opt &
$10901 \mytimes 448$ & \tot{} &
{\oom{}} &  {\tot{}} &
$224 \mytimes 77 \mytimes 406$ & {\tot{}} &
265KB & ${\incl{}}$ &
$1337 \mytimes 406$ & 4.8 &
206KB & ${\incl{}}$ \\
$\text{CMSmax}^{}_{\text{GNI}} (3_R \mytimes 2_A)$ & $\forall\forall\exists$ & \ok{} & forq & 5 & opt &
\oom{} & {\tot{}} &
{\oom{}} &  {\tot{}} &
$10648 \mytimes 10648 \mytimes 10648$ & {\tot{}} &
403KB & ${\incl{}}$ &
\oom{} & {\tot{}} &
449KB & ${\incl{}}$ \\
$\text{SNARK}^{}_{\text{bug}} (3_N \mytimes 2_V \mytimes 2_P)$ & $\boldsymbol{\forall}\exists$ & \nok{} & forklift & 8 & opt &
\oom{} & {\tot{}} &
{\oom{}} &  {\tot{}} &
\oom{} & {\tot{}} &
5MB & {\tot{}} &
\oom{} & {\tot{}} &
2MB & 19.3 \\
$\text{SNARK}^{}_{\text{fix}} (3_N \mytimes 2_V \mytimes 2_P)$ & $\forall\exists$ & \ok{} & forq & 8 & opt &
\oom{} & {\tot{}} &
{\oom{}} &  {\tot{}} &
\oom{} & {\tot{}} &
4MB & {\tot{}} &
\oom{} & {\tot{}} &
2MB & {\tot{}} \\
\bottomrule
\end{tabular}}}
\caption{Execution results for \benchnat{} examples for the \hpard{} pipeline with alternative SMV solvers.}
\label{tab:idiomatic}
\end{table*}

\begin{table*}[t]
\centering
{\scriptsize
\begin{tabular}{lccccc|cc|cc|cc|cc|cc|cc}
\toprule
\multirow{3}{*}{\bfseries Model} &
\multirow{3}{*}{\bfseries Q*} &
\multirow{3}{*}{\bfseries Res} &
\multirow{3}{*}{\bfseries solver} &
\multirow{3}{*}{\bfseries k} &
\multirow{3}{*}{\bfseries Sem} &
\multicolumn{8}{c|}{Original SMVs} &
\multicolumn{4}{c}{Translated \tool{Alloy} Models} \\
\cmidrule(lr){7-14} \cmidrule(lr){15-18}
&&&&&&
\multicolumn{2}{c|}{\tool{AutoHyper}} &
\multicolumn{2}{c|}{\tool{HyperQB}} &
\multicolumn{2}{c|}{ \tool{Exp}} &
\multicolumn{2}{c|}{ \tool{Sym}} &
\multicolumn{2}{c|}{ \tool{Exp}} &
\multicolumn{2}{c}{ \tool{Sym}} \\
\cmidrule(lr){1-6} \cmidrule(lr){7-8} \cmidrule(lr){9-10} \cmidrule(lr){11-12} \cmidrule(lr){13-14} \cmidrule(lr){15-16} \cmidrule(lr){17-18}
&&& {(exp)} &\multicolumn{2}{c|}{(sym)} &
\textbf{Size} & \textbf{t(s)} &
\textbf{Size} & \textbf{t(s)} &
\textbf{Size} & \textbf{t(s)} &
\textbf{Size} & \textbf{t(s)} &
\textbf{Size} & \textbf{t(s)} &
\textbf{Size} & \textbf{t(s)} \\
%
${\text{Bakery3}}_{\text{S1}}$ & $\exists\exists$ & \nok{} & forq & 7 & pes &
$167 \mytimes 167$ & 0.6 &
700KB & ${\incl{}}$ &
$112 \mytimes 112$ & 0.4 & 200KB & ${\incl{}}$ &
$112 \mytimes 112$ & 1.2 & 300KB & ${\incl{}}$ \\
$\text{Bakery3}_{\text{S2}}$ & $\boldsymbol{\forall}\exists$ & \nok{} & forklift & 12 & opt &
$167 \mytimes 167$ & 0.9 &
1MB & 1.1 &
$112 \mytimes 112$ & 1.6 & 350KB & 0.2 &
$112 \mytimes 112$ & 1.6 & 450KB & 0.8 \\
$\text{Bakery3}_{\text{S3}}$ & $\exists\forall$ & \nok{} & forq & 20 & opt &
$167 \mytimes 167$ & 1.1 &
2MB & 1.2 &
$112 \mytimes 112$ & 1.1 & 700KB &  0.3 &
$112 \mytimes 112$ & 1.3 & 750KB & 0.9 \\
$\text{Bakery3}_{\text{sym1}}$ & $\boldsymbol{\forall}\exists$ & \nok{} & forklift & 10 & opt &
$167 \mytimes 167$ & 0.7 &
1MB & 1.0 &
$112 \mytimes 112$ & 0.7 & 550KB & 0.2 &
$112 \mytimes 112$ & 1.3 & 350KB & 0.8 \\
$\text{Bakery3}_{\text{sym2}}$ & $\boldsymbol{\forall}\exists$ & \nok{} & forklift & 10 & opt &
$167 \mytimes 167$ & 0.6 &
1MB & 0.8 &
$1 \mytimes 1$ & 0.5 & 400KB & 0.2 &
$1 \mytimes 1$ & 1.1 & 350KB & 0.8 \\
$\text{Bakery5}_{\text{sym1}}$ & $\boldsymbol{\forall}\exists$ & \nok{} & forklift & 10 & opt &
$996 \mytimes 996$ & 0.9 &
11MB & 5.3 &
$478 \mytimes 478$ & 1.0 & 950KB & 0.5 &
$478 \mytimes 478$ & 1.8 & 600KB & 1.1 \\
$\text{Bakery5}_{\text{sym2}}$ & $\boldsymbol{\forall}\exists$ & \nok{} & forklift & 10 & opt &
$996 \mytimes 996$ &  0.8 &
11MB & 5.3 &
$130 \mytimes 130$ & 0.7 & 1MB & 0.5 &
$130 \mytimes 130$ & 1.6 & 550KB & 1.0 \\
$\text{Mutant}_{\text{mut}}$ & $\boldsymbol{\exists}\forall$ & \ok{} & forklift & 8 & pes &
$32 \mytimes 32$ & 0.8 &
200KB & 0.6 &
$21 \mytimes 21$ & 0.6 & 50KB & 0.1 &
$21 \mytimes 21$ & 1.1 & 100KB & 0.7 \\
$\text{\text{NI}}_{\text{correct}}$ & $\forall\exists$ & \ok{} & forq & 10 & opt &
$64 \mytimes 64$ & 0.8 &
350KB & ${\incl{}}$ &
$64 \mytimes 64 $ & 0.5 & 300KB & ${\incl{}}$ &
$64 \mytimes 64$ &  1.2 & 300KB & ${\incl{}}$ \\
$\text{\text{NI}}_{\text{incorrect}}$ & $\boldsymbol{\forall}\exists$ & \nok{} & forklift & 57 & pes &
$368 \mytimes 368$ &  1.2 &
12MB & ${\incl{}}$ &
$368 \mytimes 368$ & 1.2 & 2MB & ${\incl{}}$ &
$368 \mytimes 368$ & 2.4 & 2MB & ${\incl{}}$ \\
$\text{NRP}_{\text{fair}}$ & $\boldsymbol{\exists}\forall$ & \ok{} & forklift & 15 & opt &
$55 \mytimes 55$ & 0.7 &
600KB & ${\incl{}}$ &
$38 \mytimes 38$ & 0.6 & 100KB & ${\incl{}}$ &
$38 \mytimes 38$ & 1.1 & 150KB & ${\incl{}}$ \\
$\text{NRP}_{\text{unfair}}$ & $\boldsymbol{\exists}\forall$ & \ok{} & forklift & 15 & pes &
$54 \mytimes 54$ & 4.0 &
600KB &  0.8 &
$36 \mytimes 36$ & 0.7 & 100KB & 0.1 &
$36 \mytimes 36$ & 1.2 & 150KB & 0.7 \\
$\text{SNARK}_{\text{lin}}$ & $\boldsymbol{\forall}\exists$ & \nok{} & forklift & 26 & opt &
$4914 \mytimes 548$ & 3.6 &
132MB &  ${\incl{}}$ &
$71 \mytimes 137$ & 1.2 & 5MB & ${\incl{}}$ &
$71 \mytimes 137$ & 5.1 & 7MB & ${\incl{}}$ \\
$\text{Robot}_{\text{sp}}$ & $\boldsymbol{\exists}\forall$ & \ok{} & forklift & 20 & opt &
$146 \mytimes 146$ & 0.9 &
2MB & ${\incl{}}$ &
$124 \mytimes 124$ & 1.1 & 650KB & ${\incl{}}$ &
$124 \mytimes 124$ & 2.3 & 800KB & ${\incl{}}$ \\
$\text{Robot}_{\text{rb}}$ & $\boldsymbol{\exists}\forall$ & \ok{} & forklift & 20 & opt &
$266 \mytimes 266$ & 1.7 &
3MB & ${\incl{}}$ &
 $258 \mytimes 258$ & 1.3 & 950KB & ${\incl{}}$ &
 $258 \mytimes 258$ & 2.8 & 2MB & ${\incl{}}$ \\
\bottomrule
\end{tabular}}
\caption{Execution results for \benchrev{} examples for alternative SMV solvers.}
\label{tab:standard}
\end{table*}

\begin{table*}[t]
\centering
{\scriptsize
\begin{tabular}{l|cc|cc|cc|cc|cc|cc}
\toprule
\multirow{2}{*}{\bfseries Model} &
\multicolumn{2}{c|}{ \tool{C-}} &
\multicolumn{2}{c|}{ \tool{M-}}  &
\multicolumn{2}{c|}{ \tool{S-}}  &
\multicolumn{2}{c|}{ \tool{F-}}  &
\multicolumn{2}{c|}{ \tool{I-}}  &
\multicolumn{2}{c}{ \tool{B}}  \\
\cmidrule(lr){2-3} \cmidrule(lr){4-5} \cmidrule(lr){6-7} \cmidrule(lr){8-9} \cmidrule(lr){10-11} \cmidrule(lr){12-13}
&
\textbf{Size} & \textbf{t(s)} &
\textbf{Size} & \textbf{t(s)} &
\textbf{Size} & \textbf{t(s)} &
\textbf{Size} & \textbf{t(s)} &
\textbf{Size} & \textbf{t(s)} &
\textbf{Size} & \textbf{t(s)} \\
\midrule
\multicolumn{13}{c}{\tool{Exp}} \\
\midrule
${\text{Mutant1}}_{\text{pk}} (3_Q)$
& $18 \mytimes 20$ & 1.2 & $18 \mytimes 20$ & 1.0 & $18 \mytimes 20$ & 1.2 & $21 \mytimes 21$ & 1.6 & -- & -- & $18 \mytimes 20$ & 1.0 \\
${\text{Mutant1}}_{\text{dk}} (3_Q)$
& $3 \mytimes 18 \mytimes 20$ & 1.0 & $3 \mytimes 30$ & 1.0 & $3 \mytimes 30$ & 1.0 & $21 \mytimes 82$ & 1.0 & -- & -- & $3 \mytimes 30$ & 1.0 \\
${\text{Mutant2}}_{\text{pk}} (3_Q)$
& $21 \mytimes 20$ & 1.0 & $21 \mytimes 20$ & 1.0 & $21 \mytimes 20$ & 1.0 & $21 \mytimes 21$ & 1.1 & -- & -- & $21 \mytimes 20$ & 1.0 \\
${\text{Mutant2}}_{\text{dk}} (3_Q)$
& $3 \mytimes 21 \mytimes 20$ & 0.9 & $3 \mytimes 45$ & 0.9 & $3 \mytimes 45$ & 0.9 & $21 \mytimes 105$ & 1.1 & -- & -- & $3 \mytimes 45$ & 0.9 \\
${\text{Mutant3}}_{\text{pk}} (3_Q)$
& $14 \mytimes 20$ & 0.9 & $14 \mytimes 20$ & 0.9 & $14 \mytimes 20$ & 0.9 & $15 \mytimes 21$ & 1.0 & -- & -- & $14 \mytimes 20$ & 0.9 \\
${\text{Mutant3}}_{\text{dk}} (3_Q)$
& $3 \mytimes 14 \mytimes 20$ & 1.0 & $3 \mytimes 21$ & 0.9 & $3 \mytimes 21$ & 0.9 & $21 \mytimes 45$ & 0.9 & -- & -- & $3 \mytimes 21$ & 0.9 \\
${\text{Robot}}_{\text{two}} (10_P)$
& -- & -- & $310 \mytimes 310$ & 45.0 & $310 \mytimes 310$ & 18.8 & $310 \mytimes 310$ & 134.2 & -- & -- & $310 \mytimes 310$ & 18.4 \\
$\text{Robot}_\text{short} (10_P)$
& $71 \mytimes 71$ & 1.8 & $71 \mytimes 71$ & 41.4 & $71 \mytimes 71$ & 1.7 & $71 \mytimes 71$ & \tot{} & -- & -- & $71 \mytimes 71$ & 1.7 \\
$\text{Robot}_\text{score} (10_P \mytimes 6_Y)$
& $570 \mytimes 570$ & 4.0 & {\oom{}} & {\tot{}} & $570 \mytimes 570$ & 4.0 & $570 \mytimes 570$ & \tot{} & -- & -- & $570 \mytimes 570$ & 3.8 \\
$\text{Robot}_\text{robust} (10_P)$
& $310 \mytimes 293$ & 2.4 & $310 \mytimes 293$ & 24.8 & $310 \mytimes 293$ & 2.5 & $310 \mytimes 293$ & \tot{} & -- & -- & $310 \mytimes 293$ & 2.4 \\
$\text{Robot}_\text{adv} (10_P)$
& $100 \mytimes 36$ & 3.4 & {\oom{}} & {\tot{}} & $100 \mytimes 36$ & 3.5 & \oom{} & \tot{} & -- & -- & $100 \mytimes 36$ & 3.3 \\
$\text{CMSany}_{\text{NI}} (2_R \mytimes 2_A)$
& $434 \mytimes 518$ & 2.8 & 67 & 8.6 & 67 & 2.6 & 7202 & 2.7 & -- & -- & 67 & 1.7 \\
$\text{CMSany}_{\text{NI}} (3_R \mytimes 2_A)$
& {\oom{}} & {\tot{}} & {\oom{}} & {\tot{}} & {\oom{}} & {\tot{}} & {\oom{}} & {\tot{}} & -- & -- & {\oom{}} & {\tot{}} \\
$\text{CMSany}_{\text{GNI}} (2_R \mytimes 2_A)$
& $434 \mytimes 77 \mytimes 646$ & {\tot{}} & {\oom{}} & {\tot{}} & $2485 \mytimes 646$ & 66.6 & $110 \mytimes 868$ & \tot{} & $3365 \mytimes 646$ & 14.9 & $689 \mytimes 433$ & 8.5 \\
$\text{CMSany}_{\text{GNI}} (3_R \mytimes 2_A)$
& {\oom{}} & {\tot{}} & {\oom{}} & {\tot{}} & {\oom{}} & {\tot{}} & {\oom{}} & {\tot{}} & {\oom{}} & {\tot{}} & {\oom{}} & {\tot{}} \\
$\text{CMSndet}_{\text{NI}} (2_R \mytimes 2_A)$
& $266 \mytimes 298$ & 2.0 & 63 & 5.0 & 63 & 1.9 & 2450 & 1.5 & -- & -- & 63 & 1.4 \\
$\text{CMSndet}_{\text{NI}} (3_R \mytimes 2_A)$
& $5993 \mytimes 3630$ & {\tot{}} & {\oom{}} & {\tot{}} & {\oom{}} & {\tot{}} & 356968 & 134.7 & -- & -- & 160 & 113.9 \\
$\text{CMSndet}_{\text{GNI}} (2_R \mytimes 2_A)$
& $266 \mytimes 77 \mytimes 490$ & {\tot{}} & {\oom{}} & {\tot{}} & $1519 \mytimes 490$ & {\tot{}} & $110 \mytimes 532$ & \tot{} & $2141 \mytimes 490$ & 6.8 & $1631 \mytimes 490$ & 5.9 \\
$\text{CMSndet}_{\text{GNI}} (3_R \mytimes 2_A)$
& {\oom{}} & {\tot{}} & {\oom{}} & {\tot{}} & {\oom{}} & {\tot{}} & {\oom{}} & {\tot{}} & {\oom{}} & {\tot{}} & {\oom{}} & {\tot{}} \\
$\text{CMSmax}_{\text{NI}} (2_R \mytimes 2_A)$
& $224 \mytimes 262$ & 1.7 & 36 & 3.3 & 36 & 1.4 & 1568 & 1.3 & -- & -- & 36 & 1.1 \\
$\text{CMSmax}_{\text{NI}} (3_R \mytimes 2_A)$
& $3887 \mytimes 2994$ & {\tot{}} & {\oom{}} & {\tot{}} & 80 & 178.4 & 127618 & 62.6 & -- & -- & 80 & 63.1 \\
$\text{CMSmax}_{\text{GNI}} (2_R \mytimes 2_A)$
& $224 \mytimes 77 \mytimes 406$ & {\tot{}} & {\oom{}} & {\tot{}} & $1225 \mytimes 406$ & 159.9 & $110 \mytimes 448$ & \tot{} & $1847 \mytimes 406$ & 14.9 & $1337 \mytimes 406$ & 4.7 \\
$\text{CMSmax}_{\text{GNI}} (3_R \mytimes 2_A)$
& {\oom{}} & {\tot{}} & {\oom{}} & {\tot{}} & {\oom{}} & {\tot{}} & {\oom{}} & {\tot{}} & {\oom{}} & {\tot{}} & {\oom{}} & {\tot{}} \\
$\text{SNARK}_{\text{bug}} (3_N \mytimes 2_V \mytimes 2_P)$
& {\oom{}} & {\tot{}} & {\oom{}} & {\tot{}} & {\oom{}} & {\tot{}} & {\oom{}} & {\tot{}} & -- & -- & {\oom{}} & {\tot{}} \\
$\text{SNARK}_{\text{fix}} (3_N \mytimes 2_V \mytimes 2_P)$
& {\oom{}} & {\tot{}} & {\oom{}} & {\tot{}} & {\oom{}} & {\tot{}} & {\oom{}} & {\tot{}} & -- & -- & {\tot{}} & {\tot{}} \\
\midrule
\multicolumn{13}{c}{\tool{Sym}} \\
\midrule
${\text{Mutant1}}_{\text{pk}} (3_Q)$
& 25KB & 0.6 & 100KB & 0.6 & 25KB & 0.5 & 25KB & 0.6 & -- & -- & 25KB & 0.6 \\
${\text{Mutant1}}_{\text{dk}} (3_Q)$
& 50KB & 0.6 & 200KB & 0.8 & 50KB & 0.6 & 50KB & 0.6 & -- & -- & 50KB & 0.6 \\
${\text{Mutant2}}_{\text{pk}} (3_Q)$
& 25KB & 0.6 & 150KB & 0.7 & 25KB & 0.6 & 25KB & 0.6 & -- & -- & 25KB & 0.6 \\
${\text{Mutant2}}_{\text{dk}} (3_Q)$
& 50KB & ${\incl{}}$ & 200KB & ${\incl{}}$ & 50KB & ${\incl{}}$ & 50KB & ${\incl{}}$ & -- & -- & 50KB & ${\incl{}}$ \\
${\text{Mutant3}}_{\text{pk}} (3_Q)$
& 25KB & ${\incl{}}$ & 100KB & ${\incl{}}$ & 25KB & ${\incl{}}$ & 25KB & ${\incl{}}$ & -- & -- & 25KB & ${\incl{}}$ \\
${\text{Mutant3}}_{\text{dk}} (3_Q)$
& 50KB & ${\incl{}}$ & 200KB & ${\incl{}}$ & 50KB & ${\incl{}}$ & 50KB & ${\incl{}}$ & -- & -- & 50KB & ${\incl{}}$ \\
$\text{Robot}_\text{two} (10_P)$
& 4MB & 4.8 & 7MB & 8.3 & 4MB & 4.7 & 4MB & 4.4 & -- & -- & 4MB & 4.7 \\
$\text{Robot}_\text{short} (10_P)$
& 4MB & ${\incl{}}$ & 7MB & ${\incl{}}$ & 4MB & ${\incl{}}$ & 4MB & ${\incl{}}$ & -- & -- & 4MB & ${\incl{}}$ \\
$\text{Robot}_\text{score} (10_P \mytimes 6_Y)$
& 4MB & ${\incl{}}$ & 7MB & ${\incl{}}$ & 4MB & ${\incl{}}$ & 4MB & ${\incl{}}$ & -- & -- & 4MB & ${\incl{}}$ \\
$\text{Robot}_\text{robust} (10_P)$
& 4MB & ${\incl{}}$ & 6MB & ${\incl{}}$ & 4MB & ${\incl{}}$ & 4MB & ${\incl{}}$ & -- & -- & 4MB & ${\incl{}}$ \\
$\text{Robot}_\text{adv} (10_P)$
& 7MB & ${\incl{}}$ & 10MB & ${\incl{}}$ & 7MB & ${\incl{}}$ & 7MB & ${\incl{}}$ & -- & -- & 7MB & ${\incl{}}$ \\
$\text{CMSany}_{\text{NI}} (2_R \mytimes 2_A)$
& 150KB & ${\incl{}}$ & 150KB & 0.7 & 100KB & 0.8 & 150KB & 0.7 & -- & -- & 100KB & 0.7 \\
$\text{CMSany}_{\text{NI}} (3_R \mytimes 2_A)$
& 350KB & ${\incl{}}$ & 250KB & 0.9 & 250KB & 0.9 & 300KB & 0.9 & -- & -- & 300KB & 0.8 \\
$\text{CMSany}_{\text{GNI}} (2_R \mytimes 2_A)$
& 150KB & ${\incl{}}$ & 300KB & 0.8 & 250KB & 0.8 & 150KB & 0.8 & -- & -- & 150KB & 0.8 \\
$\text{CMSany}_{\text{GNI}} (3_R \mytimes 2_A)$
& 350KB & ${\incl{}}$ & 400KB & 0.9 & 450KB & 1.0 & 450KB & 1.1 & -- & -- & 450KB & 1.1 \\
$\text{CMSndet}_{\text{NI}} (2_R \mytimes 2_A)$
& 200KB & ${\incl{}}$ & 200KB & 0.6 & 200KB & 0.7 & 150KB & 0.7 & -- & -- & 150KB & 0.6 \\
$\text{CMSndet}_{\text{NI}} (3_R \mytimes 2_A)$
& 350KB & ${\incl{}}$ & 250KB & 0.8 & 250KB & 0.8 & 300KB & 0.8 & -- & -- & 250KB & 0.8 \\
$\text{CMSndet}_{\text{GNI}} (2_R \mytimes 2_A)$
& 200KB & ${\incl{}}$ & 250KB & ${\incl{}}$ & 200KB & ${\incl{}}$ & 200KB & ${\incl{}}$ & -- & -- & 150KB & ${\incl{}}$ \\
$\text{CMSndet}_{\text{GNI}} (3_R \mytimes 2_A)$
& 550KB & ${\incl{}}$ & 450KB & ${\incl{}}$ & 400KB & ${\incl{}}$ & 400KB & ${\incl{}}$ & -- & -- & 450KB & ${\incl{}}$ \\
$\text{CMSmax}_{\text{NI}} (2_R \mytimes 2_A)$
& 150KB & ${\incl{}}$ & 200KB & ${\incl{}}$ & 150KB & ${\incl{}}$ & 150KB & ${\incl{}}$ & -- & -- & 200KB & ${\incl{}}$ \\
$\text{CMSmax}_{\text{NI}} (3_R \mytimes 2_A)$
& 350KB & ${\incl{}}$ & 250KB & ${\incl{}}$ & 250KB & ${\incl{}}$ & 250KB & ${\incl{}}$ & -- & -- & 300KB & ${\incl{}}$ \\
$\text{CMSmax}_{\text{GNI}} (2_R \mytimes 2_A)$
& 200KB & ${\incl{}}$ & 300KB & ${\incl{}}$ & 200KB & ${\incl{}}$ & 200KB & ${\incl{}}$ & -- & -- & 200KB & ${\incl{}}$ \\
$\text{CMSmax}_{\text{GNI}} (3_R \mytimes 2_A)$
& 450KB & ${\incl{}}$ & 350KB & ${\incl{}}$ & 400KB & ${\incl{}}$ & 400KB & ${\incl{}}$ & -- & -- & 450KB & ${\incl{}}$ \\
$\text{SNARK}_{\text{bug}} (3_N \mytimes 2_V \mytimes 2_P)$
& 2MB & 19.5 & 6MB & 58.9 & 3MB & {\tot{}} & 2MB & 20.2 & -- & -- & 2MB & 19.0 \\
$\text{SNARK}_{\text{fix}} (3_N \mytimes 2_V \mytimes 2_P)$
& 3MB & {\tot{}} & 6MB & {\tot{}} & 2MB & {\tot{}} & 3MB & {\tot{}} & -- & -- & 3MB & {\tot{}} \\
\bottomrule
\end{tabular}}
\caption{Ablation tests for \benchnat{} examples using the two \tool{HyperPardinus} backends.}
\label{tab:ablations}
\end{table*}

\paragraph{How to read the data}
The results relevant for \textbf{RQ3} and \textbf{RQ4} are reported in Tables~\ref{tab:idiomatic} and~\ref{tab:standard} for \benchnat{} and \benchrev{}, respectively. An entry corresponds to an \tool{Alloy} command for an example, and shows the quantifier alternation (\textbf{Q*}, with bold quantifiers denoting when witnesses are generated), and the validity of the property (\textbf{Res}, \ok{} if it is valid, \nok{} otherwise). 
For explicit backends, we additionally show the used automaton \textbf{solver} (namely, \tool{forklift} when witnesses are expected and \tool{forq} when not).
For symbolic backends, the considered prefix size (\textbf{k}) and finite semantics (\textbf{Sem}, pes(simistic) or opt(imistic)) are shown. The execution results report the \textbf{Size} of the intermediate model representations (when model splitting is enabled, only the largest sub-model is being reported), in number of states in explicit backends (with \hpard{} model composition, there may be less models than initially in \textbf{Q*}) and consumed memory in symbolic ones (\oom{} denotes inability to build these within the defined time period), and execution \textbf{t}ime (\tot{} denotes timeouts). Inconclusive results due to finite semantics are signaled with \incl{} (note that they may also be wrong, but that is meaningless in an inconclusive result).
In both tables, the \tool{AutoHyper} and \tool{HyperQB} columns considers the off-the-shelf tools, with minor improvements.
Table~\ref{tab:standard} shows the performance of \hsmv{} (with \tool{Exp} and \tool{Sym}), \tool{AutoHyper} and \tool{HyperQB} when given original SMVs from the \benchrev{} benchmark. Additionally, it shows the performance of the complete \hpard{} pipeline (also with \tool{Exp} and \tool{Sym}) starting with the \tool{Alloy} version of \benchrev{} models.
Table~\ref{tab:idiomatic} considers only the full \hpard{} pipeline. Here, each entry additionally shows the used scope (due to their origin, examples from Table~\ref{tab:standard} have a fixed universe). To assess the impact of the \hpard{} optimizations, we also show the results with quantifier composition, symmetry breaking, and multiplicity constraints disabled (\tool{HyperPardinus-}); and with the state-of-the-art model checkers for HyperLTL instead of \hsmv{}.
For a finer analysis of the optimizations employed along the pipeline, Table~\ref{tab:ablations} compares the performance of \benchnat{} examples with only one optimization selectively disabled. For \hpard{}, \tool{C-} stands for quantifier composition, \tool{M-} for multiplicity constraints, \tool{S-} for symmetry breaking. For \hsmv{}, \tool{F-} stands for the new model constructions (formula sub-expressions, bisimulation quotients and AP compaction) and \tool{I-} for splitting initial configurations; \hpard{} is the baseline with all optimizations enabled, the same values presented in Table~\ref{tab:idiomatic}; \text{--} indicates that an optimization was not used for a benchmark.

\paragraph{\tool{HyperPardinus} and \benchnat{}}
Focusing on Table~\ref{tab:idiomatic} and addressing \textbf{RQ3}, we can see that \hpard{} was able to analyze models written at a much higher level of abstraction than those in \benchrev{}, with \tool{Sym} performing considerably better (though only considering finite prefixes and often inconclusive). For instance, \tool{Sym} can verify all CMS entries below the timeout, while \tool{Exp} reaches scope $3_R\times2_A$ only for some NI properties.
Regarding the impact of the \hpard{} optimizations, \tool{Exp-} is able to generate the state machines and terminate for most entries, $\text{SNARK}$ being a notable exception. $\text{CMS}$ is the demonstrative class for which the state machines can be generated, but their size becomes prohibitive; the worst case is $\text{CMSndet}_{NI}(3_R\times2_A)$, with $16036 \times 16036$ versus $160$ states.
The gains from \tool{Sym-} are less evident and harder to compare: the finite semantics is only conclusive when it finds a (counter-)example, and many temporal sub-expressions reduce to trivial statements when inconclusive, affecting the overall complexity; to avoid misleading comparisons, we therefore refrain from reporting those times; sizes are independent of the finite semantics. A notable exception is $\text{Robot}_\text{two}$, the only conclusive Robot entry, where disabling the optimizations still yields an answer but roughly doubles both time and size.
The SNARK model exhibits too much non-determinism for any \tool{Exp} backend to produce a state machine: the state space explodes because every distinct sequence of the 4 concurrent operations, combined with the full history required to model linearizability, yields a unique state. The $\text{SNARK}_\text{lin}$ model from \benchrev{} is therefore far more constrained, disallowing operation reorderings (and thus requiring no history) and restricting the allowed sequences of operations. 
Nonetheless, for \tool{Sym}, \hpard{} finds the counter-example in $\text{SNARK}_\text{bug}$ for the much more complex \benchnat{} model in comparable time to the \benchrev{} model, while \tool{HyperQB} cannot even produce the QBF problem. 
More importantly, the \benchnat{} SNARK counter-example is conclusive while the \benchrev{} one is not; we express linearizability as a safety property over explicit histories, so a violation is observed at a concrete state of the finite prefix and cannot be repaired by extending the trace.
$\text{SNARK}_\text{fix}$ has no bugs, so it is intrinsically inconclusive for \tool{Sym} regardless of the chosen semantics; for \tool{opt} it ultimately concludes in 5.5H -- demonstrating that \tool{Sym} is far more suitable for bug finding, as expected.

\paragraph{\tool{HyperSMV} and \benchnat{}} 
Turning to \textbf{RQ4}, it is interesting to see that the \hsmv{} optimizations are also essential to the pipeline: even with \hpard{} optimizations enabled, there are many Robot and CMS examples where \tool{AutoHyper} and \tool{HyperQB} time out but the corresponding \hsmv{} procedure returns an answer in a few seconds. It is worth noting that more declarative modeling seems to have a toll on the state-of-the-art tools. For instance, for $\text{Robot}_\text{robust}$ natively developed for \tool{Alloy} both state-of-the-art tools timed out (unlike \hsmv{}), but in the original low-level version $\text{Robot}_\text{rb}$ (Table~\ref{tab:standard}), they finished in few seconds. This is likely due to the level of non-determinism, as for instance robot movements in $\text{Robot}_\text{rb}$ cannot happen freely in all directions, unlike in $\text{Robot}_\text{robust}$.

\paragraph{\tool{HyperSMV} and \benchrev{}}
Focusing on Table~\ref{tab:standard} and addressing \textbf{RQ4}, \hsmv{} consistently replicates the expected outcomes for \benchrev{}, and has competitive performance with state-of-the-art tools.
Focusing on \tool{Exp}, the worst scenario was $\text{Bakery3}_{\text{S2}}$, becoming $2\times$ slower. \tool{Exp} is particularly successful in $\text{SNARK}_{\text{lin}}$, becoming $3\times$ faster; the size of the state machine implies that optimizations were key. \tool{Sym} was equivalent or better in all executions, with an evident improvement on the size of the models, spending $26\times$ less memory in $\text{SNARK}_{\text{lin}}$.
Nonetheless, the symbolic approaches (\tool{HyperQB} and \tool{Sym}) are naturally inconclusive for many of the examples, hindering the significance of the comparisons; this is the case for $\text{SNARK}_{\text{lin}}$, whose property is of the form $F \phi_0 \wedge G \phi_1$ and $F \phi_0$ becomes trivially true under the optimistic semantics.

\paragraph{\tool{HyperPardinus} and \benchrev{}}
Regarding \textbf{RQ3}, the results of the full \hpard{} pipeline show an expected overhead, but remain well competitive with the low-level backends. 
The overhead is near-constant, and mostly visible on the sub-second entries; in the worst case, $\text{SNARK}_\text{lin}$, it becomes $4\times$ slower but still reports in a few seconds.
Recall that these executions start from a relational model translated from SMVs; \hpard{} then translates them back into SMV and calls \hsmv{}; while this has the overhead of the \proc{RL2LTL} translation, it also benefits by the optimizations implemented in \hpard{}. On the other hand, the translations break the structure of the original formulas, which may affect performance. Explicit backends are more sensitive to this because the original benchmarks are annotated with APs, but there are no such annotations at the level of \tool{HyperPardinus}, and \tool{HyperSMV} must infer APs from the translated formula. 

\paragraph{Optimizations in the pipeline}
Focusing on Table~\ref{tab:ablations}, we can support a more detailed discussion of \textbf{RQ3} and reveal how the optimizations have proven crucial for \benchnat{} examples\footnote{We don't show the same results for \benchrev{} examples, as optimizations are only significant for  $\text{SNARK}_{\text{lin}}$.}. For the smaller $\text{Mutant}$ examples, the optimizations were naturally not significant.
For $\text{Robot}$ examples, we can see that \tool{M-} and \tool{F-} optimizations were key for the viability of \tool{Exp}. The same examples are inconclusive for \tool{Sym}, with the exception of $\text{Robot}_\text{two}$ for which \tool{M-} is also crucial.
For $\text{CMS}$ examples, we can attest that all optimizations had a significant impact on the result for the \tool{Exp} backend. For instance, for $\text{CMSmax}_\text{GNI}$ only \tool{S-} and \tool{I-} still produce a result, at respectively $34\times$ and $3\times$ the baseline cost. For the \tool{Sym} backend, the most interesting result is that without the \tool{C-} optimization all $\text{CMS}$ results are inconclusive.
For the $\forall$ properties that are valid, and thus require an exploration of all possible models, we can see that \tool{S-} trades slightly larger models for a narrowing of the search space.
This can be evidenced clearly for $\text{CMSmax}_\text{GNI}$ under \tool{Exp}, where \tool{S-} induces a $34\times$ larger time, and for $\text{SNARK}_\text{bug}$ under \tool{Sym}, where \tool{S-} concludes above the timeout.

\subsection{Threats to validity}

\paragraph{Benchmark selection and construction} The \benchnat{} idiomatic models were developed by the authors who are well-versed in \tool{Alloy} and \hpard{}, which may introduce a bias. Nonetheless, these examples and hyperproperties were selected from the literature: CMS is a popular example in the security verification community, and some of the Robot and Mutant examples are higher-level re-implementations from \benchrev{}.

\paragraph{Model translation and optimization} The \benchrev{} \tool{Alloy} models were generated automatically from SMV files. The outcome of all executions was consistent with the expected result, giving us confidence in this translation. However, note that we do not aim to validate this SMV-to-\tool{Alloy} translation, but only to show that \hpard{} is able to deal with models at a low level of abstraction with little performance overhead.

In BMC semantics, moving expressions from the formula to the models can impact the result. Note that in high-level \tool{Alloy} the same declarative specification defines models and properties, and there is no separation between hyper formula and non-hyper state machines.
For the \tool{Exp} backend, there is no concern on this regard, and we effectively try to push most of the formula to the state machines (and eventually do all the solving in this phase, what happens for examples whose model size becomes $1$ after optimizations).
For the \tool{Sym} backend, however, we must exercise additional caution to ensure that our translation and optimizations are sound. This is because, for the particular cases when the finite semantics is conclusive, sound infinite inference assumes that the state machines are total, while moving restrictions from the formula to the state machines may break totality. All our \benchnat{} examples have stuttering and are thus also total; in our examples, we can safety push invariants from the formula, preserving stuttering and totality. The state machines of all \benchrev{} examples are total, as can be verified with a simple syntactic check; we have verified that all \hpard{}-generated SMV models preserve such totality, and have disabled further formula optimizations in \hsmv{} for these examples. As a result, in all our benchmarks, moving expressions from the formula to the models can only improve the conclusiveness of BMC.

\paragraph{Toolchain and implementation}
Regarding the trusted computing base, our pipeline is long and contains many non-validated components, both developed by us and by third-parties. At the bottom of the pipeline, both \hsmv{} and state-of-the-art tools either rely on robust QBF solvers (which support certificate generation) or mature libraries for automata manipulation. The fact that 4 different backends provided consistent results (except for some inconclusive symbolic instances) gives us further confidence.

\section{Conclusion}
\label{sec:conclusions}

This paper proposes a model finding approach for hyper relational temporal logic, backed by automated analysis procedures. Through a minimal extension to \tool{Alloy}, we show that the approach allows model checking hyperproperties for design models written in a high-level specification language. 
This makes it more expressive than state-of-the-art model checkers for HyperLTL, and more suitable for end users as models are specified at a higher-level of abstraction and feedback provided in the \tool{Alloy Analyzer}. Models are automatically analyzed in a two-phased backend that has showed to be competitive with the state of the art for low-level models, and outperform it in high-level ones. 

There are many open challenges in the verification of HyperLTL properties, such as finding looping traces in BMC. By relying on standard formats, our approach will benefit from improvements to low-level model checkers for HyperLTL. At the modeling language level, we relied on a simple extension to \tool{Alloy} to demonstrate the feasibility of the approach, but further studies are needed to evaluate it from the end-user perspective.

\begin{acks}
The authors would like to thank David Chemouil for support on the \tool{Alloy 6} to SMV translation, and Rui Gonçalves and Miguel Montes for their implementation efforts on previous prototypes.

This work is financed by National Funds through the FCT -- Fundação para a Ciência e a Tecnologia, I.P. (Portuguese Foundation for Science and Technology) within the project POISE, with reference 2024.15783.PEX (\url{https://doi.org/10.54499/2024.15783.PEX}).
\end{acks}

\bibliographystyle{ACM-Reference-Format}
\bibliography{bib}

\end{document}